%% file: ms.tex
\documentclass{aastex631}
\usepackage{CJK}
\usepackage{graphicx}
\usepackage{tabularx}
\usepackage{booktabs}
\usepackage{multirow}
\usepackage{makecell}
\usepackage{rotating}
\usepackage{subfigure}
\usepackage{gensymb}
\usepackage{placeins}  
\usepackage{etoolbox}
\makeatletter
\patchcmd{\@outputpage@head}{\maxdeadcycles=10}{\maxdeadcycles=100}{}{}
\makeatother

\shorttitle{Velocity range comparison of circumstellar OH masers}
\shortauthors{}
\graphicspath{{./}{figures/}}

\begin{document}
\begin{CJK*}{UTF8}{gkai}
\title{A Search for Asymmetric Kinematic Components in Circumstellar Envelopes \\Using OH Main Line Masers}

\correspondingauthor{Jun-ichi Nakashima}
\email{junichin@mail.sysu.edu.cu, nakashima.junichi@gmail.com}

\author[0000-0003-1015-2967]{Jia-Yong Xie (谢嘉泳)}
\affiliation{School of Physics and Astronomy, Sun Yat-sen University, Tang Jia Wan, Zhuhai, 519082, P. R. China}

\author[0000-0003-3324-9462]{Jun-ichi Nakashima(中岛淳一)}
\affiliation{School of Physics and Astronomy, Sun Yat-sen University, Tang Jia Wan, Zhuhai, 519082, P. R. China}
\affiliation{CSST Science Center for the Guangdong-Hong Kong-Macau Greater Bay Area, \\Sun Yat-Sen University, 2 Duxue Road, Zhuhai 519082, Guangdong Province, PR China}

\author[0000-0002-1086-7922]{Yong Zhang (张泳)}
\affiliation{School of Physics and Astronomy, Sun Yat-sen University, Tang Jia Wan, Zhuhai, 519082, P. R. China}
\affiliation{CSST Science Center for the Guangdong-Hong Kong-Macau Greater Bay Area, \\Sun Yat-Sen University, 2 Duxue Road, Zhuhai 519082, Guangdong Province, PR China}

\begin{abstract}

Circumstellar OH maser lines are useful for studying the dynamics of the circumstellar envelope (CSE) around evolved stars. This study aims to identify CSEs around cold stars, which exhibit deviations from the spherical expansion, by comparing the velocity ranges of the OH main lines (1665/1667 MHz) with those of the satellite line (1612 MHz), using a database of circumstellar OH maser sources. We performed this comparison for 377 circumstellar OH maser sources. In addition, using infrared two-color diagrams, we examined the evolutionary stages and infrared properties of objects showing velocity excess (velocity excess means the detection of the main lines outside the velocity range of the satellite line). A periodicity analysis of the WISE light curves was also carried out. As a result of the velocity range comparison, eight circumstellar OH maser sources were found to exhibit velocity excess. The infrared colors of these objects match those of post-AGB stars. Periodic variations were observed in the WISE light curves of five of these eight objects. The results suggest that  examining velocity excess of the main lines relative to the satellite line is scientifically significant because mainline masers probe the CSE dynamics over a broader range of evolutionary stages compared to the 22.235 GHz H${_2}$O maser line. Additionally, during the post-AGB phase, the emission regions of the mainline and 22.235 GHz H${_2}$O masers may overlap in a CSE, whereas they originate from different regions during the AGB phase.

\end{abstract}


\section{Introduction} \label{sec: Intro}

Maser lines are frequently detected in the circumstellar envelopes (CSEs) of evolved stars undergoing active mass loss, such as Asymptotic Giant Branch (AGB) stars, post-AGB stars, and red supergiants (RSGs). Specifically, maser lines of OH, $\rm H_{2}O$, and SiO molecules are commonly found in the oxygen-rich CSEs of these stars, where the number of oxygen atoms exceeds that of carbon atoms. Recently, it has been suggested that these circumstellar masers may also be observed in red nova remnants (RNRs), which result from stellar mergers between two main-sequence stars or between a main-sequence star and an evolved low- or intermediate-mass star \citep{2005PASJ...57L..25D,2020A&A...638A..17O}. Observations of circumstellar maser lines provide valuable insights into the morphology and kinematics of the circumstellar molecular gas.

Despite extensive observations of circumstellar maser sources since the late 1970s, efforts to compile the observed data into a comprehensive database have been relatively limited until recently. Early works, such as those by \citet{1989A&AS...78..399T} and \citet{1990ApJS...74..911B}, have diminished in scientific value over time, partly due to their limited scope and the lack of updates. Recently, \citet{2015A&A...582A..68E} compiled and organized observational data on the three circumstellar OH maser lines (1612 MHz, 1665 MHz, and 1667 MHz) and released an online database\footnote{https://hsweb.hs.uni-hamburg.de/projects/maserdb//}. This database has been updated from time-to-time since its initial release in 2015 and contains nearly up-to-date information as of 2024. The same group is also compiling a database of circumstellar $\rm H_{2}O$ maser sources, with some of the collected data (excluding non-detections) available as an appendix in \citet{2024ApJS..270...13F}.

\citet{2024ApJS..270...13F} analyzed the database of circumstellar $\rm H_{2}O$ and OH maser sources to identify water fountain (WF) candidates by comparing the velocity ranges of the 1612 MHz OH and 22.235 GHz $\rm H_{2}O$ maser lines. A WF is a low- to intermediate-mass evolved star with a small-scale, high-velocity, highly collimated molecular jet at the center of the CSE \citep{2012PASJ...64...98I}. Evolved stars with low to intermediate mass undergo a transformation in the morphology of their CSE from spherical to non-spherical as they progress from the AGB to the planetary nebula (PN) phase. However, the physical processes driving this morphological evolution are not yet fully understood. WFs are considered as key objects for studying this evolution, as they represent a phase immediately following the onset of the morphological transition from spherical to non-spherical.  WF is classically defined as a circumstellar 22.235 GHz H$_2$O maser source, with a velocity range exceeding 100 km~s$^{-1}$. However, from an astrophysical perspective, WF describes an object where a bipolar molecular outflow has just begun to form near the center of a spherically symmetric, expanding envelope developed during the AGB phase. To identify such objects, \citet{2024ApJS..270...13F} established a selection criterion wherein the velocity range of the 22.235 GHz H$_2$O maser exceeds that of the 1612 MHz OH maser line. In this paper, we refer to cases where the velocity range of a maser line surpasses that of the 1612 MHz OH maser line as a "velocity excess." WF candidates selected based on this H$_2$O maser velocity excess may not strictly conform to the classical definition of WF. Rather, as \citet{2024ApJS..270...13F} discussed, these candidates may represent WFs at an earlier evolutionary stage than those traditionally classified as WFs. \citet{2024ApJS..270...13F} identified 11 sources based on the velocity excess of the 22.235 GHz H$_2$O maser line. The selection criteria set by \citet{2024ApJS..270...13F} were designed to detect high-velocity gas components, such as bipolar jets, within the spherically expanding envelope formed during the AGB phase. However, the criteria proposed by \citet{2024ApJS..270...13F} may select not only WFs but also CSEs with kinematic components deviating from spherical expansion, such as the RNRs mentioned earlier. Whether the object is a WF, an RNR, or another type, CSEs with non-spherical kinematic components are considered to contain many astrophysically interesting features.

For a subset of the sample, \citet{2024ApJS..270...13F} compared the velocity ranges of the OH satellite line (1612 MHz) with those of the OH main lines (1665 MHz and 1667 MHz). Interestingly, they found that in two sources, the velocity ranges of the OH main lines exceeded those of the OH satellite line (IRAS 18251--1048 and IRAS 22516+0838). In the case of these two sources, the 1665 MHz OH and/or 1667 MHz OH lines were detected outside the double peak of the 1612 MHz OH maser line (i.e., they find the velocity excess of the 1665/1667 MHz OH lines). In a spherically expanding CSE, the 1612 MHz OH maser line typically exhibits a double-peaked profile, where the velocity separation between the peaks is known to be twice the expansion velocity of the spherical CSE. Therefore, this result suggests (1) the presence of velocity components within the CSE that deviate from spherical expansion, and (2) that these irregular kinematic components can be detected in the 1665 MHz OH and/or 1667 MHz OH lines. Since the main objective of \citet{2024ApJS..270...13F} was to compare the velocity ranges of the 1612 MHz OH and 22.235 GHz $\rm H_{2}O$ lines, the comparison between the OH satellite (1612 MHz) and main lines (1665 MHz and 1667 MHz) was limited to a small sample.

In the present study, we performed a thorough comparison of the velocity ranges of the OH satellite and main lines using the latest version of the database of circumstellar OH maser sources \citep{2015A&A...582A..68E}. Our goal is to identify circumstellar OH maser sources where the velocity range of the OH main lines exceeds that of the OH satellite line. Additionally, we utilized catalogs of known AGB and post-AGB stars to establish evolutionary tracks on infrared two-color diagrams. We then investigated the infrared properties of the selected sources exhibiting kinematic irregularities in the OH maser lines by comparing them with known samples of AGB and post-AGB stars.

 For simplicity, we will refer to the 1612 MHz OH maser line as the satellite line and the 1665 MHz and 1667 MHz OH maser lines as the main lines. We will refer to the 22.235 GHz H$_2$O maser line simply as the H$_2$O maser line. As there are two main lines, the rest frequency is mentioned when it is necessary to distinguish between them.

\section{Methodology and Data Analysis}
	\label{sect: Methodology and Data Analysis}
\subsection{Data and Preprocessing}
    \label{sect: data}

\begin{figure}
    \centering
    \includegraphics[width=\textwidth]{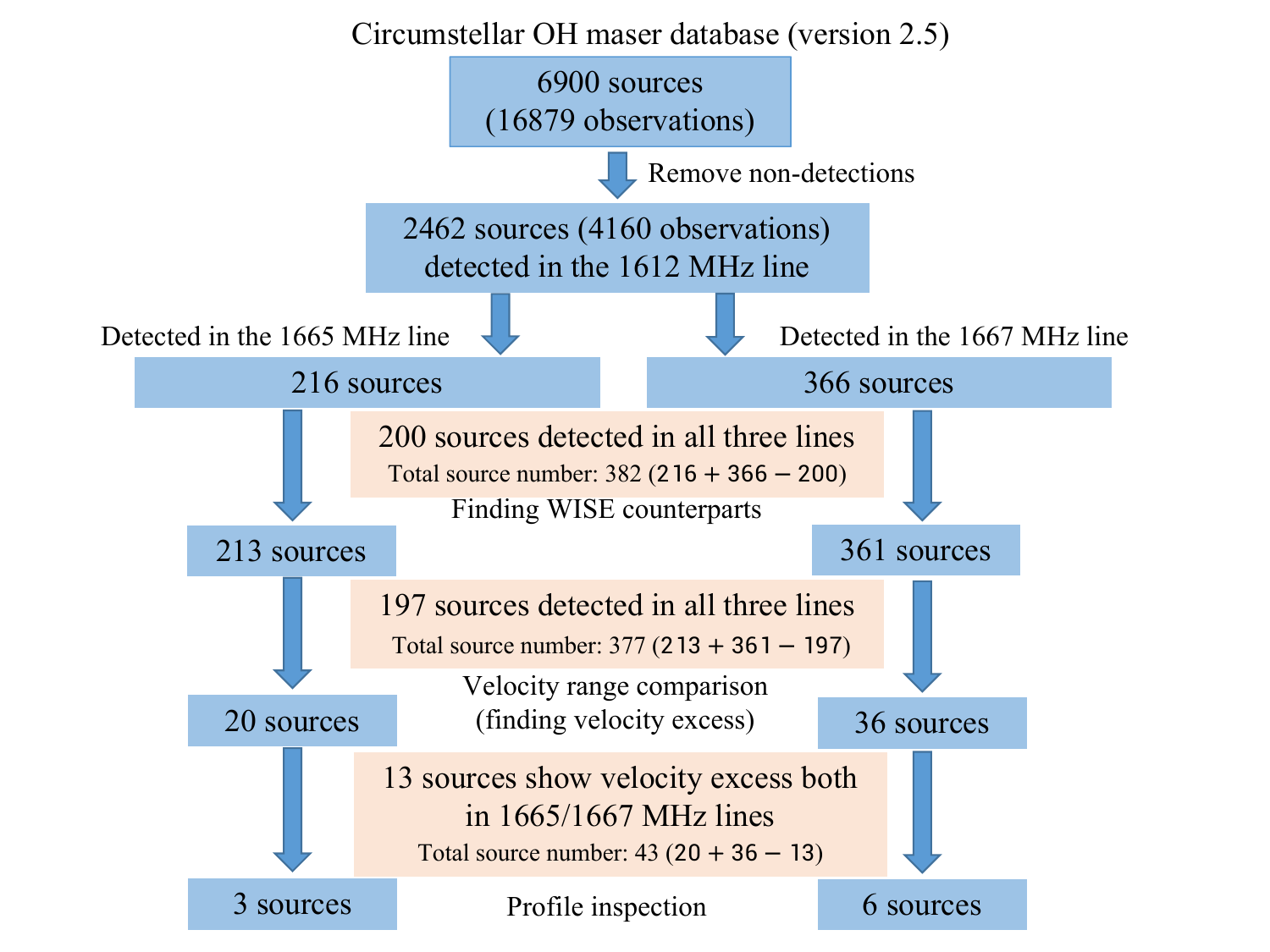}
    \caption{Flowchart of the source selection process in the present study}.
    \label{Fig: [Flowchart]}			
\end{figure}

We utilized the circumstellar OH maser database \citep[][version 2.5; hereafter referred to as the OH database]{2015A&A...582A..68E}, accessible via CDS/VizieR\footnote{https://cdsarc.cds.unistra.fr/viz-bin/cat/J/A+A/582/A68}. This database contains 16,879 observations of 6,900 sources. The OH database was initially released in 2015 and has been updated from time-to-time since then. Version 2.5 includes data released after 2015, such as ATCA follow-up of SPLASH \citep{2020ApJS..247....5Q} and THOR \citep{2019A&A...628A..90B}. The OH database includes flux densities and velocities for both blue-shifted and red-shifted peaks ($V_{\rm blue}$ and $V_{\rm red}$), the expansion velocity of the envelope, and the radial velocity of the source. Specifically, the database contains 9,720 observations for 2,462 observed sources and 4,160 detections for the satellite line. For the main line at 1665 MHz, there are 3,491 observations for 248 sources and 315 detections, and for the main line at 1667 MHz, 3,667 observations for 446 sources and 673 detections. Among these OH maser sources, 216 were detected in both the satellite and main (1665 MHz) lines, and 366 in both the satellite and main (1667 MHz) lines, with 200 sources detected in all three OH lines (1612 MHz, 1665 MHz, and 1667 MHz). The number of sources detected in each OH maser line is summarized in Figure~\ref{Fig: [Flowchart]}. The line parameters of the OH maser sources used for velocity comparison, along with their coordinates, are provided in Table~\ref{Tab: [WISE position of the selected sources]} and Table~\ref{Tab: [Velocity information of the selected the OH maser sources with WISE counterparts]}.

The positional accuracy of circumstellar maser sources in the OH database is sometimes suboptimal, occasionally worse than 10 arcseconds. This is primarily because many OH maser observations were conducted using single-dish radio telescopes, often relying on coordinates cited from the IRAS Point Source Catalog (PSC) as the observation position, which can introduce relatively large  positional uncertainties. To mitigate the impact of unreliable coordinate information in the OH database, it is crucial to first obtain more accurate positional information. To this end, we compared the sample of circumstellar OH maser sources with the WISE PSC \citep{2010AJ....140.1868W,2011ApJ...731...53M}\footnote{https://wise2.ipac.caltech.edu/docs/release/allwise/}, and identified WISE counterparts for the circumstellar OH maser sources. \citet{2024ApJS..270...13F} demonstrated that, compared to other infrared PSC (such as 2MASS and Spitzer/GLIMPSE), the WISE PSC offers a higher probability of identifying counterparts for circumstellar maser sources due to its wavelength coverage and sensitivity. Given that WISE's angular resolution  (6.1$''$, 6.4$''$, 6.5$''$, 12.0$''$ in the W1, W2, W3, and W4 bands respectively) is significantly higher than that of IRAS, we can expect the positional information from the WISE PSC to be more reliable than that from the IRAS PSC and the OH database. Therefore, we used the WISE PSC to obtain the infrared positions of circumstellar OH maser sources. Additionally, other information available from the WISE PSC, such as infrared photometric data, was also utilized in our analysis.

\begin{figure}
\centering
\includegraphics[width=0.3\textwidth, angle=0]{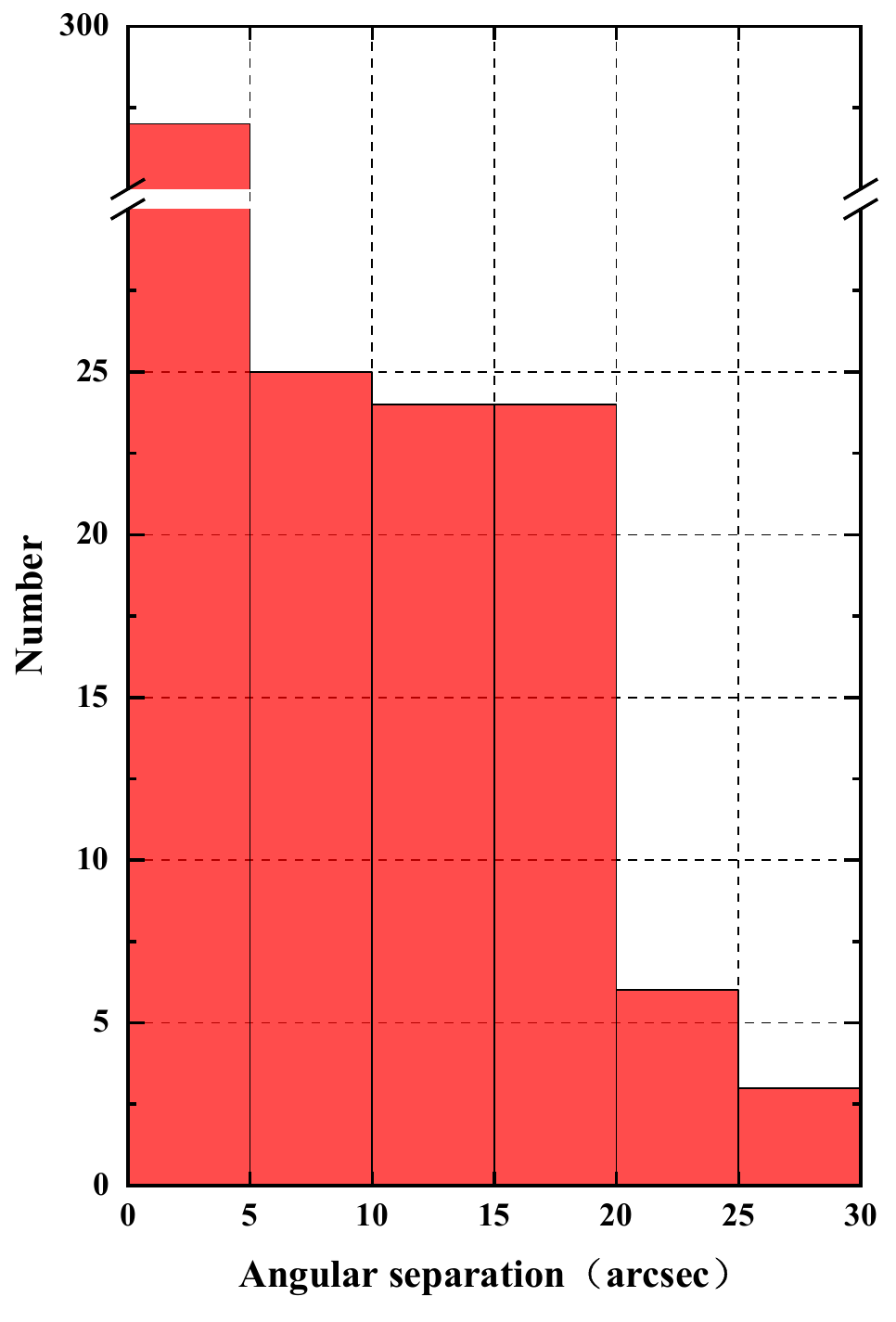}
\caption{Histogram showing the angular separations between the positions of circumstellar OH maser sources in the database and their corresponding AllWISE counterparts. Approximately 80\% of the OH maser sources have a WISE counterpart within 5$''$.}
\label{Fig: [Histogram of the angDist distribution]}
\end{figure}

The cross-checking procedure we applied closely followed that described by \citet{2024ApJS..270...13F}, who first automatically listed WISE sources within 20$''$ of circumstellar maser sources. In most cases (about 80\%), a bright infrared counterpart was found within 5$''$ (see Figure~\ref{Fig: [Histogram of the angDist distribution]}), leaving no doubt unique identification of the infrared counterpart. In cases where identifying the counterpart is not straightforward—such as when more than one bright WISE objects are within a few arcseconds of the maser source—we used color indices to make the determination. Since circumstellar maser sources have thick dust/molecular envelopes, we expect to see a flux excess in the W4 (22 $\mu {\rm m}$) band, making color indices good indicators for identifying circumstellar maser sources. However, many sources in the current OH maser sample were too bright in the mid-infrared, leading to saturation and unreliable color indices (quality flags for the photometric measurements are provided in Table~\ref{Tab: [WISE, IRAS and 2MASSS information of the selected sources]}). In some cases, we had to forgo checking the color indices due to such saturation, particularly in the W4 and W3 (12 $\mu {\rm m}$) bands. On the other hand, even if the photometric data quality is low due to saturation, it is reasonable to assume a large flux density. In such cases, we can infer that the color index is large (above a certain threshold), even if the exact value cannot be accurately calculated. To ensure accuracy, we visually inspected the WISE images for all sources with poor quality flags and confirmed that a very bright infrared source could be found at the location of the OH maser source. By comparing this with the positional information from other infrared archives, such as 2MASS and Spitzer/GLIMPSE, we confirmed that the positional information provided by the WISE catalog is reliable (with uncertainties within 2$''$--3$''$), even when the objects are too bright and the photometric quality flag is low. Coordinate information for the identified WISE counterparts is given in Table~\ref{Tab: [WISE position of the selected sources]}, and photometric parameters for the WISE counterparts and their quality flags are given in Table~\ref{Tab: [WISE, IRAS and 2MASSS information of the selected sources]}).

During the inspection of the WISE images, we confirmed the presence of an extended infrared nebula towards IRAS 05506+2414. Previous studies listed in SIMBAD suggest that this object is a young stellar object \citep[YSO,][]{1990A&AS...84..517M,2008ApJ...680..483S, 2017ApJ...850..158S}, so we excluded it from further analysis. Additionally, for four sources (i.e., IRAS 19059--2219, IRAS 19244+1115, IRAS 20077--0625, and IRAS 22177+5936), no clear counterparts were found in the WISE point source catalog within 20$''$, and even extending the search scale to 30$''$ did not yield any identifications. 
However, the WISE images show a point source, which is thought to be a counterpart, near the OH maser source.
The reasons for the lack of WISE counterparts in the point source catalog for these sources are unclear \citep[a similar case exists where no IRAS point source is found for circumstellar SiO maser sources,][]{ 2005PASJ...57..933D}. Since the locations and infrared fluxes are unknown, these sources were excluded from the present analysis (see Table~\ref{Tab: [Sources without the WISE counterparts]}).

\subsection{Velocity comparison and line profile inspection}
\label{sect: Velocity comparison and line profile inspection}

\begin{figure}
\centering
\includegraphics[width=0.8\textwidth, angle=0]{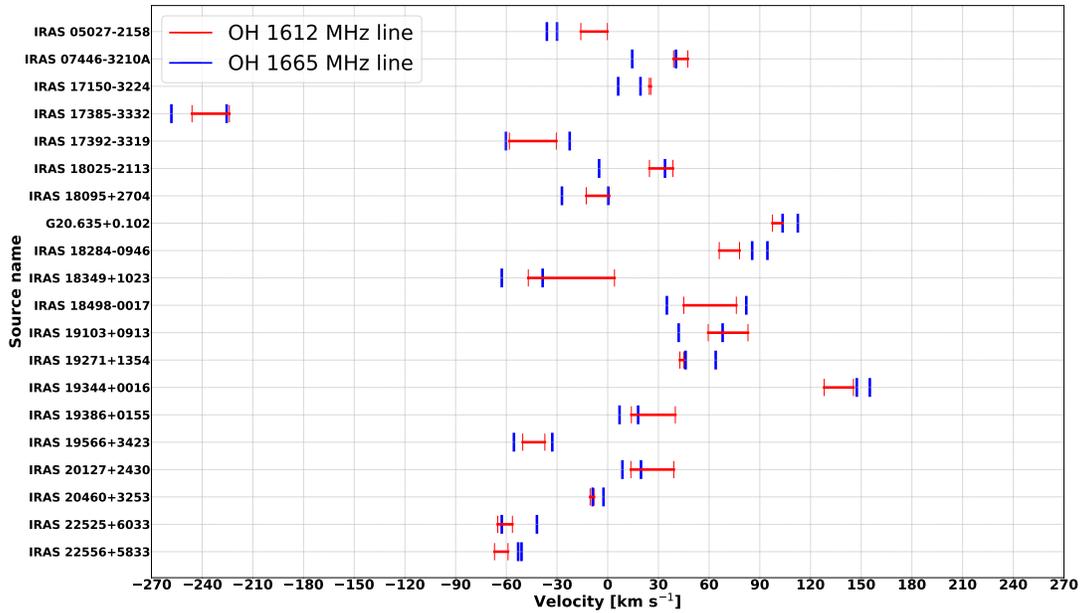}
		\caption{Comparison of the velocity ranges between the satellite line and the main (1665 MHz) lines for the 20 selected sources that exhibit velocity excess in the 1665 MHz OH line. The red line represents the velocity range of the satellite line, while the two blue vertical lines indicate the velocity range of the main line (1665 MHz). The sources are arranged in ascending order by the right ascension.}
		\label{Fig: [V1665_morethan_V1612]}
	\end{figure}

\begin{figure}
\centering
\includegraphics[width=0.8\textwidth, angle=0]{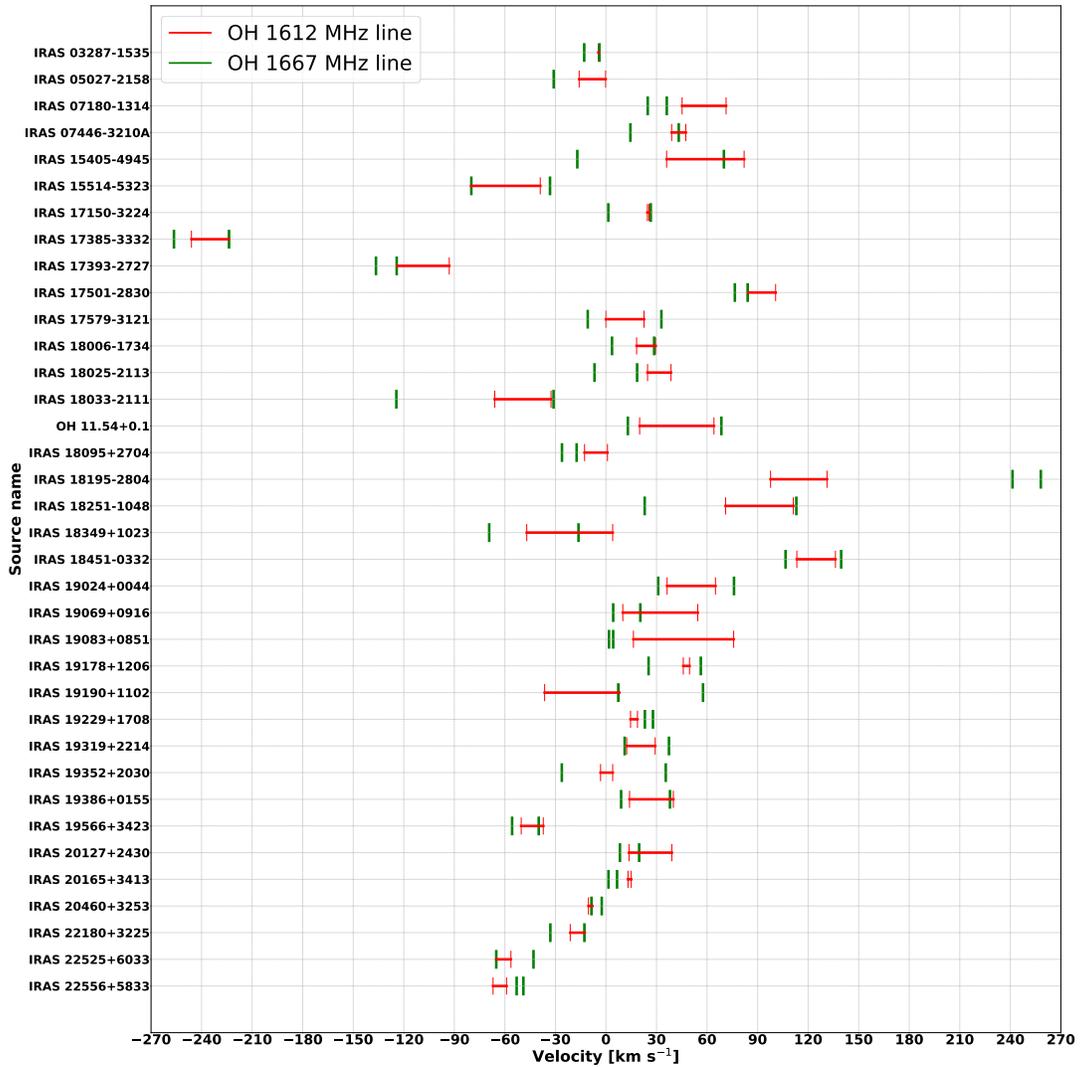}
		\caption{Comparison of the velocity ranges between the satellite and main (1667 MHz) lines for the 36 selected sources that exhibit velocity excess in the main line (1667 MHz). The red line represents the velocity range of the satellite line, while the green vertical lines indicate the velocity range of the main line (1667 MHz). The sources are arranged in ascending order by the right ascension.}
		\label{Fig: [V1667_morethan_V1612]}
	\end{figure}

The next step was to select OH maser sources from the database where the velocity range of either the two main lines exceeds that of the satellite line. In this paper, the velocity is defined as the line-of-sight velocity with respect to the local standard of rest (LSR). We conducted a velocity comparison for 377 OH maser sources with available data (see Section~\ref{sect: data}) and identified those with velocity excess. Initially, we selected sources where the velocity range of the main lines exceeds that of the satellite line by more than 5 km~s$^{-1}$ on either the red-shifted edge, the blue-shifted edge, or both, based on the velocity ranges provided in the database. Figures \ref{Fig: [V1665_morethan_V1612]} and \ref{Fig: [V1667_morethan_V1612]} compare the velocity ranges of the main lines with the satellite line for OH maser sources exhibiting velocity excess. The same figures for sources without velocity excess are provided in Appendix~\ref{appendix: Velocity range comparison of all the selected sources}. From this comparison, we identified 20 sources where the velocity range of the main line (1665 MHz) exceeds that of the satellite MHz line and 36 sources for the other main line (1667 MHz). Thirteen of these sources were common to both categories, exhibiting velocity excess in both the two main lines.

As discussed in Section~\ref{sec: Intro}, comparison of the velocity ranges of the satellite line with those of the main lines enables us to utilize the satellite line as a probe for the expansion velocity of a spherically expanding envelope, typically observed in AGB stars. The satellite line profile from such a spherically expanding envelope (i.e., the AGB wind) usually exhibits a double-peak profile \citep{1996A&ARv...7...97H}. In this profile, the red-shifted peak corresponds to the receding part of the spherically expanding envelope, while the blue-shifted peak corresponds to the approaching part. In fact, the satellite line emission originates from the relatively outer regions of such a spherical AGB envelope, where the material has reached its terminal expansion velocity. Therefore, detecting the main lines at velocities beyond the range of the satellite line suggests the presence of kinematic components within the envelope that deviate from spherical expansion.

However, it is important to note that a detection of a double-peaked profile is not always the case. Variations in maser intensity over time and the inhomogeneous distribution of material within the CSE can result in the detection of OH maser lines from only one side of the spherically expanding envelope \citep[i.e., either the receding or approaching side, see, e.g.,][]{1996A&ARv...7...97H,2024ApJS..270...13F}. In such a case, the value of $(V_{\rm red} - V_{\rm blue})/2$ does not represent the expansion velocity of the spherically expanding envelope but rather the velocity width of the red-shifted or blue-shifted component. If the value of $(V_{\rm red} - V_{\rm blue})/2$ is smaller than the typical expansion velocities of AGB stars or RSGs (approximately 20~km~s$^{-1}$--40 km~s$^{-1}$), we must be particularly cautious, as the intensity peaks corresponding to $V_{\rm red}$ and $V_{\rm blue}$ might both lie on the receding (red-shifted) side or the approaching (blue-shifted) side.

To address this issue, we collected spectra from the original papers reporting observations of the OH maser lines for all sources showing velocity excess in the main lines and visually inspected the line profiles of the satellite line. The line profile generated from a spherically expanding envelope not only display a double-peak profile but also often feature a characteristic profile that gradually approaches zero intensity near the systemic velocity of the star while steeply declining on the other side \citep{1996A&ARv...7...97H}. This characteristic can be a clue to identifying the true double-peak profile of a spherically expanding envelope. The results of this visual inspection are summarized in Section~\ref{sec: Results} and Appendix~\ref{appendix: Maser information of the other sources exhibiting velocity deviation}. The inspection revealed that many of the sources shown in Figures~\ref{Fig: [V1665_morethan_V1612]} and \ref{Fig: [V1667_morethan_V1612]} do not actually exhibit true velocity excess. Detailed results are provided in Section~\ref{sec: Results}.

\subsection{Infrared two-color diagrams}
\label{sect: Infrared two-color diagrams}

As a preliminary step for the discussions in the following sections, we plotted the regions of oxygen-rich AGB stars and post-AGB stars on various infrared two-color diagrams. \citet{1988A&A...194..125V} proposed regions on the IRAS two-color diagram where different types of evolved stars are located. Our approach here is similar to their classification. Specifically, we plotted samples of 3,828 oxygen-rich AGB stars \citep{2017JKAS...50..131S} and 89 oxygen-rich post-AGB stars \citep{2007ApJ...658.1096D,2019arXiv190105866M, 2022ApJ...927L..13K} on two-color diagrams using photometric data from the IRAS, WISE, and 2MASS archives. We then identified regions containing the majority (80\%-–90\%) of these oxygen-rich AGB and post-AGB star samples. Objects beyond the post-AGB phase (i.e., pre-PN and PN) were excluded due to the insufficient number of samples available to define regions on the two-color diagrams, as well as the rarity of detected circumstellar masers.

For the oxygen-rich AGB stars, since \citet{2017JKAS...50..131S} already identified IRAS and WISE counterparts, we used their coordinates to retrieve photometric data from the infrared archives. In other cases, following the method described in Section~\ref{sect: data}, we first identified WISE counterparts and then matched them with counterparts in other infrared archives using the coordinates provided in the WISE PSC. To ensure the reliability of our analysis, we used only photometric data with high-quality flags (q = 3 for IRAS; q = A, B, or C for AllWISE; q $\neq$ U for 2MASS). The number of objects with high-quality flags was 2,492 (AllWISE), 2,056 (IRAS) and 3,272 (2MASS) for oxygen-rich AGB stars, and 81 (AllWISE), 52 (IRAS) and 55 (2MASS) for oxygen-rich post-AGB stars.

The photometric data from the WISE PSC are provided in magnitudes, based on the Vega magnitude system. Therefore, following the method used by \citet{2024ApJS..270...13F}, we first converted the magnitudes to flux densities before calculating the color indices. The color indices were then calculated as [W1]$-$[W4]$=2.5\log(F_{\text{W4}}/F_{\text{W1}})$ or [W2]$-$[W4]$=2.5\log(F_{\text{W4}}/F_{\text{W2}})$, where $F_{\text{W1}}$, $F_{\text{W2}}$, and $F_{\text{W4}}$ are the flux densities at the W1, W2, and W4 bands, respectively, derived from the WISE magnitudes. All other infrared color indices mentioned below are defined in a similar manner.

\begin{figure}
\centering
\includegraphics[width=0.8\textwidth, angle=0]{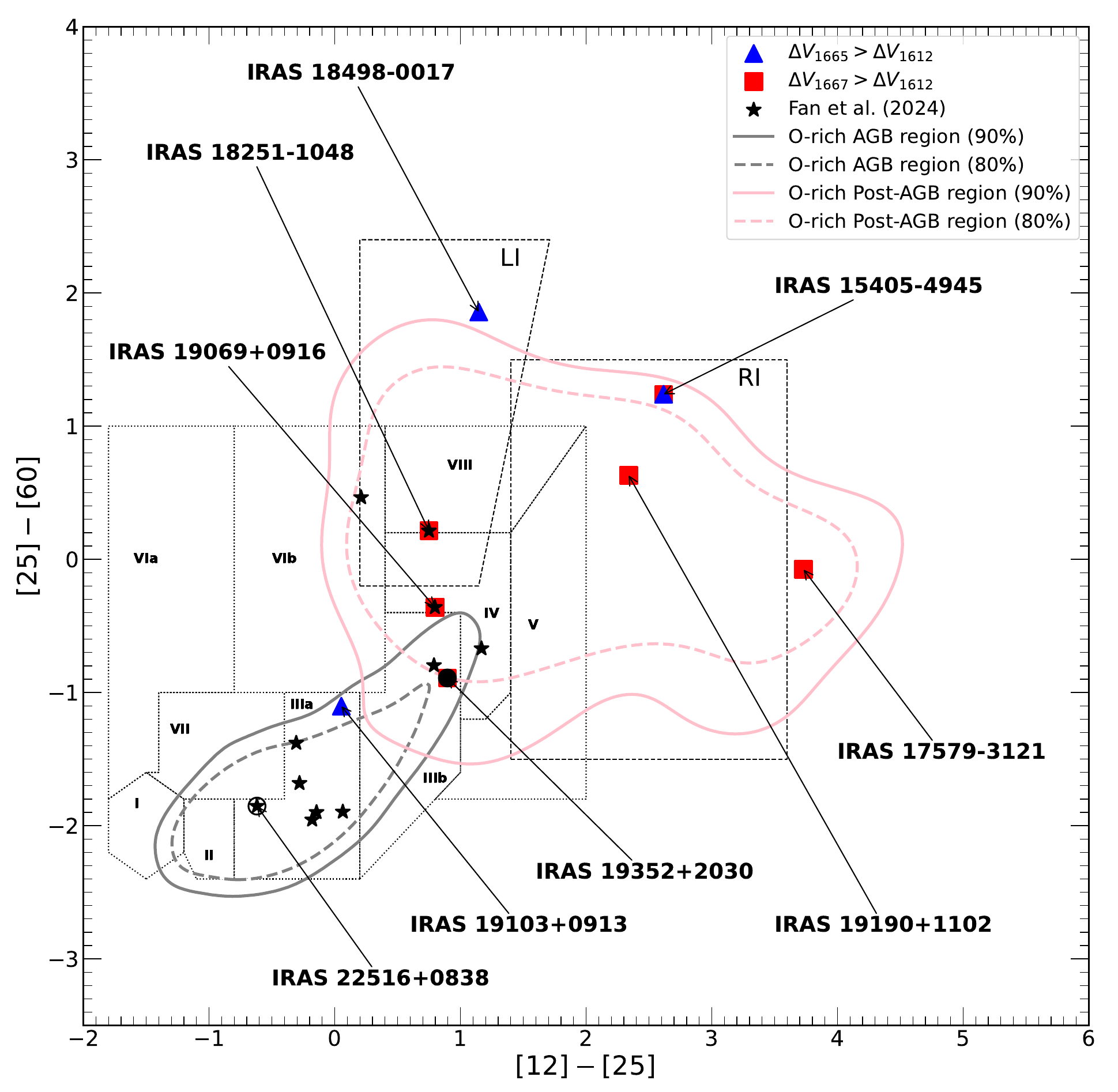}
\caption{IRAS [12]$-$[25] versus [25]$-$[60] two-color diagram of objects exhibiting high-velocity components traced by the main lines. The boxes (black dotted lines) are defined by \citet{1988A&A...194..125V}, and the regions LI and RI (black dashed lines) represent two post-AGB sequences \citep{2002AJ....123.2772S,2002AJ....123.2788S}. The area enclosed by the gray solid and dashed lines contains 90\% and 80\% of the oxygen-rich AGB star samples, respectively, while the area enclosed by the red solid and dashed lines contains 90\% and 80\% of the oxygen-rich post-AGB star samples. The blue triangles represent objects with velocity excess in the 1665 MHz OH line ($\triangle V_{\rm 1665} > \triangle V_{\rm 1612}$), and the red squares represent objects with velocity excess in the main line (1667 MHz, $\triangle V_{\rm 1667} > \triangle V_{\rm 1612}$). The black stars indicate the 11 WF candidates identified by \citet{2024ApJS..270...13F}. The black open circle (IRAS 22516+0838) represents another object  with velocity deviation in the main line (1667 MHz) from \citet{2024ApJS..270...13F}, while the black filled circle (IRAS 19352+2030) represents the object with velocity deviation in the  H$_2$O maser line confirmed in this work.}
\label{Fig: [New_IRAS_two_color_diagram_of_AGB_and_postAGB]}
\end{figure}

To verify the reliability of the samples, we first examined their distribution on the IRAS two-color diagram, where regions for different types of evolved stars were proposed by \citet{1988A&A...194..125V}. Figure~\ref{Fig: [New_IRAS_two_color_diagram_of_AGB_and_postAGB]} shows the regions where oxygen-rich AGB stars and oxygen-rich post-AGB stars are predominantly distributed. The region enclosed by the gray solid line contains 90\% of the oxygen-rich AGB star samples, while the region enclosed by the red solid line contains 90\% of the oxygen-rich post-AGB star samples. The boxed regions (I to VIII) in Figure~\ref{Fig: [New_IRAS_two_color_diagram_of_AGB_and_postAGB]} are those proposed by \citet{1988A&A...194..125V}. According to their work, region II contains relatively young pulsating variables with oxygen-rich chemistry, which evolve through regions IIIa, IIIb, and IV as their CSE becomes thicker. Region V is mainly occupied by PNe. Region VIb contains stars with hot dust near the central star and cold dust farther out, many of which are thought to have oxygen-rich chemistry. Region VIII is believed to contain a mixture of objects at different evolutionary stages with extreme properties. Regions IV, V, and VIII are recognized as areas where post-AGB stars and/or pre-PNe are mainly distributed, according to \citet{1998A&AS..127..185N} and \citet{2002A&A...388..252E}. Additionally, the LI and RI regions proposed by \citet{2002AJ....123.2772S} and \citet{2002AJ....123.2788S} are also shown in Figure~\ref{Fig: [New_IRAS_two_color_diagram_of_AGB_and_postAGB]} for comparison. The LI region contains AGB stars with relatively high expansion velocities ($>$ 15 km s$^{-1}$) at the late AGB phase or early post-AGB phase. Stars in this region are expected to have a higher average initial mass \citep[around 4 ${\rm M}_\odot$,][]{2005MNRAS.357.1189C,2006MNRAS.369..189C}. On the other hand, the RI region is thought to contain post-AGB stars, pre-PNe, and PNe with relatively low expansion velocities ($<$ 15 km s$^{-1}$). The average initial mass of objects in this region is about 1.7 ${\rm M}_\odot$. \citet{2004ApJS..155..595D} suggests that objects in the RI region often exhibit stronger OH main line emission and more irregular OH maser line profiles than those in the LI region. Comparing the regions proposed by \citet{1988A&A...194..125V}, \citet{2002AJ....123.2772S}, and \citet{2002AJ....123.2788S} with our proposed regions, we confirmed that there is no inconsistency.

Next, we performed a similar task on the WISE two-color diagram. Figure~\ref{Fig: [New_WISE_two_color_diagram_of_AGB_and_postAGB]} shows the two-color diagram between the color indices [W1]$-$[W4] and [W2]$-$[W4]. Similarly to Figure~\ref{Fig: [New_IRAS_two_color_diagram_of_AGB_and_postAGB]}, the region enclosed by the gray solid line contains 90\% of the oxygen-rich AGB star samples, and the region enclosed by the red solid line contains 90\% of the oxygen-rich post-AGB star samples. Similarly, Figure~\ref{Fig: [New_2MASS_two_color_diagram_of_AGB_and_postAGB]} presents the 2MASS two-color diagram between [J]$-$[H] and [H]$-$[K] colors, showing the distribution of oxygen-rich AGB and post-AGB star samples. As with the other two-color diagrams, only data with a high-quality flag (i.e., q $\neq$ U) were used. The 2MASS PSC provides fluxes in the J (1.25 $\mu {\rm m}$), H (1.65 $\mu {\rm m}$), and K (2.17 $\mu {\rm m}$) bands and has an angular resolution of 2.0$''$ in all three bands, corresponding to sensitivities of 0.4, 0.5, and 0.6 mJy, respectively \citep{2003tmc..book.....C,2006AJ....131.1163S}. The conversion of 2MASS magnitudes to Jy is based on \citet{2003AJ....126.1090C}. In the 2MASS wavelength range, interstellar extinction cannot be ignored. Therefore, we applied the same correction method for interstellar extinction as that of \citet{2017MNRAS.465.4482Y} before plotting the data. In addition to IRAS and WISE, we also checked the data from AKARI and Spitzer/GLIMPSE, but since there was not enough data available to plot the sample in the present study, the results are omitted.

\section{Results} \label{sec: Results}

\subsection{Sources with True Velocity Excess} \label{sec: Sources with true velocity deviation}

Table~\ref{Tab: [Sources showing true velocity deviations in the OH 1665/1667 MHz line]} summarizes the sources identified with genuine velocity excess in the main line, as discussed in Section~\ref{sect: Velocity comparison and line profile inspection}. The second column of this table presents the spectral types (i.e., profile shapes) of the satellite line.

The spectral type D represents a double-peak profile, characteristic of spherically expanding envelopes (e.g., AGB winds). The spectral type $\rm D_{+}$ denotes a D-type profile with an additional weak emission peak detected outside the double peak. The spectral type Irr indicates an irregular line profile that differs from both D-type and $\rm D_{+}$-type profiles.

For sources classified as type D, if the velocity of the main line exceeds the velocity range of the satellite line by 5 km~s$^{-1}$ or more on either the redshifted or blueshifted side, we consider it as a genuine velocity excess. Four sources—IRAS 18251--1048, IRAS 18498--0017, IRAS 19069+0916, and IRAS 19103+0913—fall into this category. In the case of type $\rm D_{+}$, $(V_{\rm red} - V_{\rm blue})/2$ exceeds the velocity interval between the two main peaks. Although it is unclear whether the additional peaks detected outside the double peak belong to the spherically expanding envelope or another kinematic component, we assume here that they are part of the spherically expanding envelope. We applied the same criterion as for D-type sources: if the velocity range of the OH main lines exceeds the velocity range of the OH satellite line by 5 km~s$^{-1}$ or more, we consider it as a genuine velocity excess. Two sources—IRAS 17579--3121 and IRAS 19190+1102—fit this classification. For type Irr sources, similarly to D-type and $\rm D_{+}$-type sources, if the velocity range of the OH main lines exceeds that of the OH satellite line by 5 km~s$^{-1}$ or more, we consider it as a genuine velocity excess. Two sources—IRAS 15405--4945 and IRAS 19352+2030—fall into this group. If the satellite line shows an Irr-type profile, it raises the question of whether it truly traces AGB spherically expanding winds. This issue will be discussed further in Section~\ref{sec: Discussion}.

In Section~\ref{sect: Velocity comparison and line profile inspection}, we also noted cases where only one of the redshifted or blueshifted components of the spherically expanding envelope is detected as a single-peak profile in the satellite line. In such instances, it is suggested that if the velocity of the OH main line deviates beyond the expansion velocity of the circumstellar envelope around evolved stars relative to the peak velocity of the satellite line, it may indicate the presence of kinematic components different from the spherically expanding component. However, there were no sources in the present sample that met this condition.

\subsection{Comparison with Fan et al. (2024)} \label{sec: Comparison with Fan et al. (2024)}

\citet{2024ApJS..270...13F} identified 11 sources that show a velocity excess in the H$_2$O maser line compared to the velocity range of the satellite line.  Among these 11 sources, two (IRAS 18251--1048 and IRAS 19069+0916) are also listed in Table~\ref{Tab: [Sources showing true velocity deviations in the OH 1665/1667 MHz line]}. Of these sources, IRAS 19069+0916 is considered by \citet{2024ApJS..270...13F} to have a high probability of being a WF source among the 11 with confirmed velocity excess. \citet{2024ApJS..270...13F} based their findings on velocity data from the H$_2$O maser database (Refer to their paper for details about the H$_2$O maser database). Therefore, if the H$_2$O maser database information is incomplete, there is a possibility that some sources with velocity excess were overlooked. In the present study, we collected spectral data for the H$_2$O and OH satellite lines from the literature for the sources listed in Figures \ref{Fig: [V1665_morethan_V1612]} and \ref{Fig: [V1667_morethan_V1612]} (i.e., sources where our initial analysis using the OH maser database detected velocity excess) and visually inspected their velocity ranges. As a result, we confirmed that the H$_2$O maser line of IRAS 19352+2030 \citep{2001PASJ...53..517T} shows a clear velocity excess relative to the satellite line \citep{1978A&AS...31..127B,1979A&AS...36..193B,1989ApJ...344..350L,2019A&A...628A..90B}. This velocity excess in the H$_2$O maser emission for IRAS 19352+2030 was missed by \citet{2024ApJS..270...13F}. Notably, \citet{2013ApJ...769...20Y} identified this object as a potential WF candidate. The satellite line OH maser of IRAS 19352+2030 exhibits a single strong peak near $-$3 km~s$^{-1}$, along with several additional peaks. According to the profile classification mentioned in Section~\ref{sec: Sources with true velocity deviation}, this corresponds to the Irr type. Detailed maser characteristics of this object are summarized in Section~\ref{sec: IRAS 19352+2030}.

\citet{2024ApJS..270...13F} also noted that the main line (1667 MHz) of IRAS 18251--1048 and IRAS 22516+0838 shows a velocity excess relative to the satellite OH line, based on a visual inspection of the spectra. Among these, IRAS 18251--1048 was also detected with a velocity excess in the present study (see Table~\ref{Tab: [Sources showing true velocity deviations in the OH 1665/1667 MHz line]}), but IRAS 22516+0838 was not. This suggests that the velocity excess of IRAS 22516+0838 could not be identified using the velocity information provided in the OH maser database. These findings indicate that some sources with velocity excess can only be confirmed through visual inspection of the spectra.

\begin{figure}
\centering
\includegraphics[width=0.8\textwidth, angle=0]{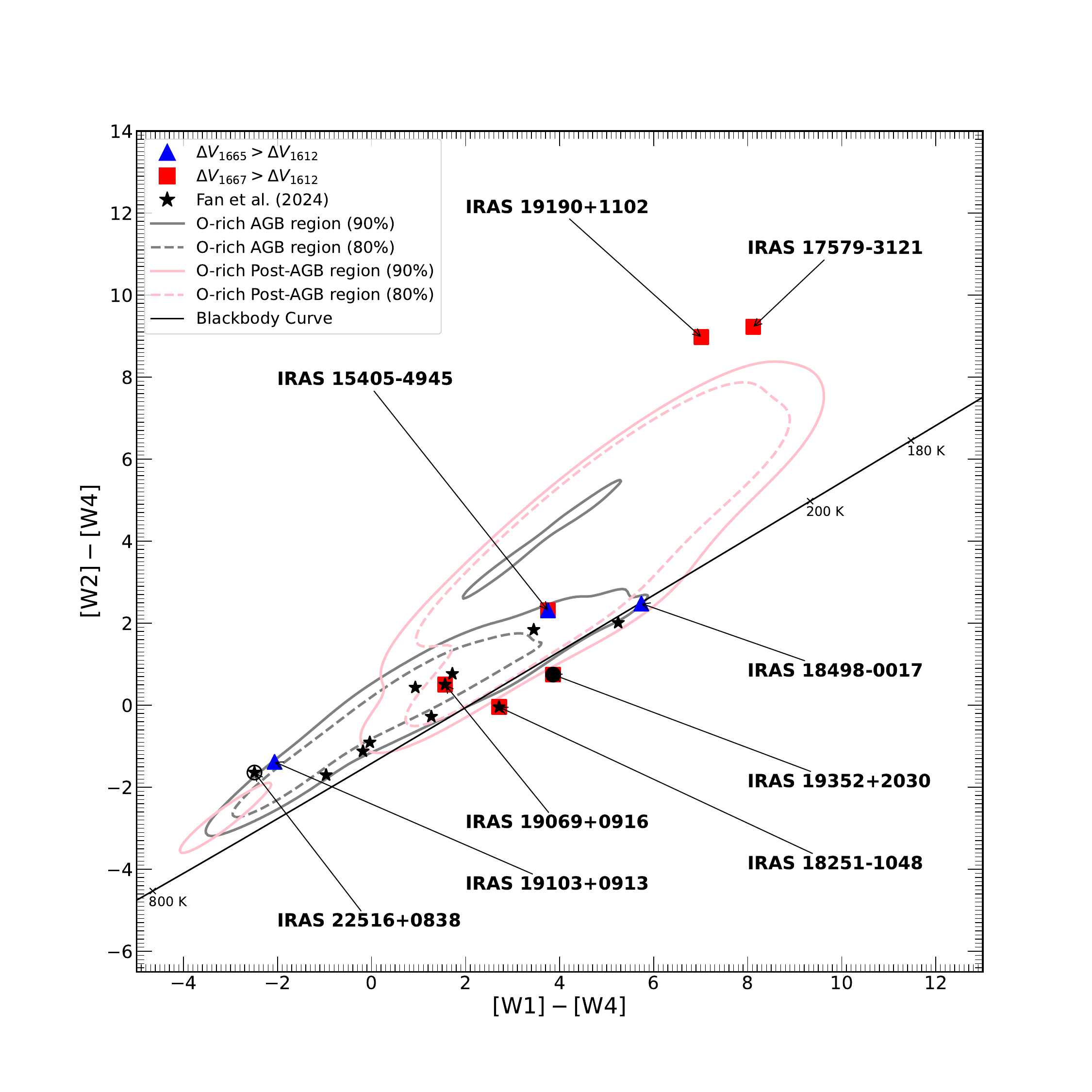}
\caption{AllWISE [W1]$-$[W4] versus [W2]$-$[W4] two-color diagram shows objects with high-velocity components traced by the main  lines. The black line represents the black body radiation curve. Other notations are consistent with those in Figure~\ref{Fig: [New_IRAS_two_color_diagram_of_AGB_and_postAGB]}.}
\label{Fig: [New_WISE_two_color_diagram_of_AGB_and_postAGB]}
\end{figure}

\begin{figure}
\centering
\includegraphics[width=0.8\textwidth, angle=0]{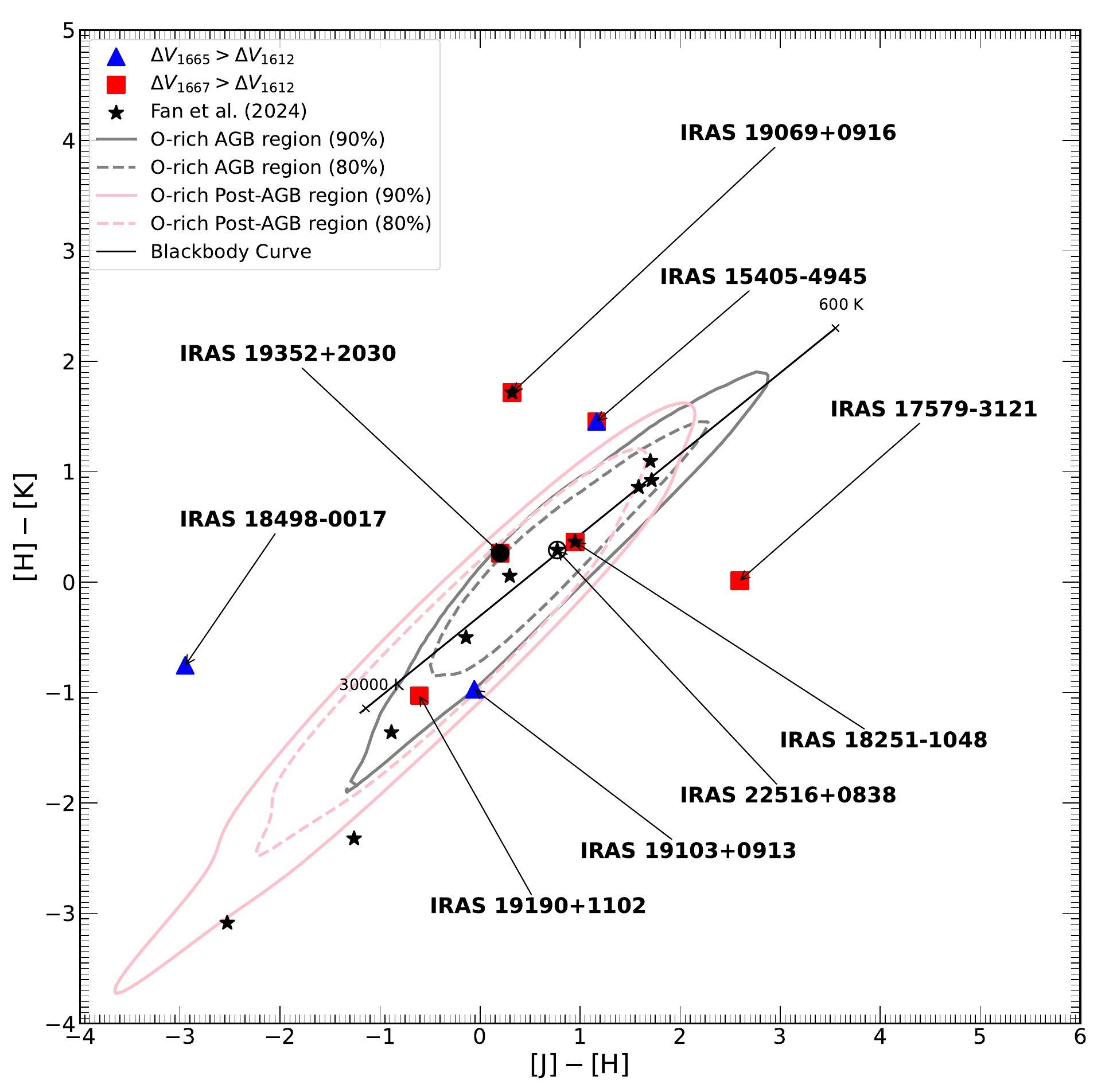}
\caption{2MASS [J]$-$[H] versus [H]$-$[K] two-color diagram shows objects with high-velocity components traced by the main lines. The black line represents the black body radiation curve. Other notations are consistent with those in Figure~\ref{Fig: [New_IRAS_two_color_diagram_of_AGB_and_postAGB]}.}
\label{Fig: [New_2MASS_two_color_diagram_of_AGB_and_postAGB]}
\end{figure}

When comparing the distribution of the 11 sources selected by \citet{2024ApJS..270...13F} with the 8 sources where velocity excess was confirmed in  the present study on the infrared two-color diagram, it is clear that their distributions differ significantly. In Figures~\ref{Fig: [New_IRAS_two_color_diagram_of_AGB_and_postAGB]} and \ref{Fig: [New_WISE_two_color_diagram_of_AGB_and_postAGB]}, many sources selected by \citet{2024ApJS..270...13F} are located in the AGB star region. In contrast, none of the 8 sources with confirmed velocity excess in the present study are located in the AGB star color region; all are found in the post-AGB region. In Figure~\ref{Fig: [New_2MASS_two_color_diagram_of_AGB_and_postAGB]}, it is observed that the sources selected by \citet{2024ApJS..270...13F} tend to be positioned near the line representing blackbody radiation, while the sources selected in the present study are distributed away from the blackbody radiation line. This deviation is presumably due to the appearance of multiple temperature components within the circumstellar envelope as stellar evolution progresses. For comparison, IRAS 19352+2030 is plotted as a black filled circle, and IRAS 22516+0838 as a black open circle, in the infrared two-color diagrams (Figures~\ref{Fig: [New_IRAS_two_color_diagram_of_AGB_and_postAGB]}, \ref{Fig: [New_WISE_two_color_diagram_of_AGB_and_postAGB]}, and \ref{Fig: [New_2MASS_two_color_diagram_of_AGB_and_postAGB]}).

\subsection{Comparison with Known WFs and RNRs}\label{sec: Comparison with Known WFs and RNRs}

To compare the infrared colors of the eight sources with confirmed velocity excess in the present study with those of known WFs and RNRs, we plotted the data points of the known WFs and RNRs on the IRAS two-color diagram (Figure~\ref{Fig: [New_IRAS_two_color_diagram_of_known_WFs_RNRs]}) and the WISE two-color diagram (Figure~\ref{Fig: [New_WISE_two_color_diagram_of_known_WFs_RNRs]}). The infrared photometric data for the known WFs and RNRs are summarized in Table~\ref{Tab: [WISE, IRAS and 2MASS Infrared photometric data of the known WFs and RNRs]} (see Appendix~\ref{appendix: Infrared photometric data for known WFs and RNRs}). In Figure~\ref{Fig: [New_IRAS_two_color_diagram_of_known_WFs_RNRs]}, the known WFs are distributed in the post-AGB regions (i.e., IIIb, IV, V, VIII, LI, RI), and it was found that the eight sources with velocity excess are also distributed in nearly the same regions as the known WFs. This contrasts with the results of \citet{2024ApJS..270...13F}, where many of the 11 sources they selected were located in the AGB region (see, Figures~\ref{Fig: [New_IRAS_two_color_diagram_of_AGB_and_postAGB]} and \ref{Fig: [New_WISE_two_color_diagram_of_AGB_and_postAGB]}). Similarly, on the WISE two-color diagram (Figure~\ref{Fig: [New_WISE_two_color_diagram_of_known_WFs_RNRs]}), the eight sources selected in the present study are distributed in almost the same regions as the known WFs. For RNRs, since there are only five known objects in total, the number of objects that can be plotted on Figures~\ref{Fig: [New_IRAS_two_color_diagram_of_known_WFs_RNRs]} and \ref{Fig: [New_WISE_two_color_diagram_of_known_WFs_RNRs]}
 is very limited. However, for comparison purposes, the known objects have been plotted (one RNR is plotted in Figure~\ref{Fig: [New_IRAS_two_color_diagram_of_known_WFs_RNRs]}, and four RNRs are plotted in Figure~\ref{Fig: [New_WISE_two_color_diagram_of_known_WFs_RNRs]}). Although RNRs are fundamentally different from AGB and post-AGB stars, they are lying, from the perspective of IRAS and WISE colors, in regions that overlap with the AGB and post-AGB star regions.

\subsection{Maser Properties of Individual Sources}\label{sec: Maser Properties of Individual Sources} For each source listed in Table~\ref{Tab: [Sources showing true velocity deviations in the OH 1665/1667 MHz line]}, we summarize the maser line properties derived from spectra published in previous studies and related information.

\subsubsection{IRAS 15405--4945}\label{sec: IRAS 15405-4945} The spectrum of the satellite line is classified as Irr type, showing multiple emission peaks in the velocity range from 36 km~s$^{-1}$ to 82 km~s$^{-1}$ \citep{1988A&A...202L..19T}. The spectrum of the main line (1665 MHz) exhibits multiple peaks in the velocity range from 10 km~s$^{-1}$ to 65 km~s$^{-1}$ \citep{1988A&A...202L..19T}, while the main line (1667 MHz) also shows multiple peaks but in the velocity range from $-17$ km~s$^{-1}$ to 70 km~s$^{-1}$ \citep{1988A&A...202L..19T}. These velocity ranges are consistent with the results of \citet{1996A&AS..119..459T}. The blue-shifted components of the main lines clearly deviate from the velocity range of the satellite line, with the velocity excess of the main line (1667 MHz) exceeding 50 km~s$^{-1}$.  There are no reports of the detection of any transition lines of H$_2$O or SiO masers.

This source is classified as a pre-PN \citep{1988A&A...202L..19T,1998A&AS..127..185N}. Located in the Galactic plane, it exhibits IRAS colors indicative of a low effective temperature, leading to its initial classification as a possible ultra-compact HII region. However, there is no clear evidence to support this classification \citep{1997MNRAS.291..261W}. \citet{2001MNRAS.322..280Z} suggested that hydrodynamical interactions between two stellar winds with different velocities could form a shell-like structure. Additionally, \citet{2001MNRAS.322..280Z} reported significant temporal variations in the line profiles of the satellite and main (1667 MHz) lines \citep[see Figures 1 and 2 in][]{1988A&A...202L..19T}. Using the Hubble Space Telescope (HST), \citet{2007AJ....134.2200S} confirmed the presence of a complex bipolar nebula associated with this source. The size of the high-velocity outflows traced by the OH masers corresponds to the size of the bipolar nebula observed with the HST.

\subsubsection{IRAS 17579--3121}\label{sec: IRAS 17579-3121} The spectrum of the satellite line is of the $\rm D_{+}$ type. According to \citet{2012A&A...537A...5W}, the line profile of the satellite line has two velocity components around 1 km~s$^{-1}$ and 23 km~s$^{-1}$, both showing finer velocity structures. This result generally agrees with \citet{2008AJ....135.2074G}, though they also reported another weak peak around 32 km~s$^{-1}$. The profile of the main line (1667 MHz) also shows multiple peaks in the velocity range from $-11$ km~s$^{-1}$ to 33 km~s$^{-1}$ \citep{1997A&AS..124...85S}. The blue-shifted components of the main line (1667 MHz) clearly deviate from the velocity range of the satellite line. The main line (1665 MHz)  was observed by \citet{1997A&AS..124...85S}, but with a negative detection. This source was classified as a post-AGB star by \citet{2006A&A...458..173S} based on its optical spectrum. Initially, it was classified as a PN by \citet{1990A&A...233..181R} due to the detection of centimeter-wave continuum emission, but later \citet{2008AJ....135.2074G} argued that the continuum source is physically unrelated to this object, rejecting the PN classification and identifying it as a post-AGB star.

\subsubsection{IRAS 18251--1048}\label{sec: IRAS 18251-1048} The spectrum of the satellite line is of the D type. The line profile of the satellite line shows a typical double-peak profile with peaks around 70 km~s$^{-1}$ and 110 km~s$^{-1}$ \citep{2014ApJ...794...81Y}. The main line (1667 MHz) has intensity peaks around 20 km~s$^{-1}$ and 115 km~s$^{-1}$, clearly deviating from the velocity range of the satellite line \citep{2014ApJ...794...81Y}. Notably, the blue-shifted side has a large velocity excess of about 50 km~s$^{-1}$. The H$_2$O maser was detected by \citet{2013ApJ...769...20Y}. The H$_2$O maser shows two emission peaks around 75 km~s$^{-1}$ and 108 km~s$^{-1}$. \citet{2017ApJS..232...13C} also reported a peak of the H$_2$O maser line around 74 km~s$^{-1}$. This source was listed as a WF candidate by \citet{2024ApJS..270...13F} based on the velocity information in the H$_2$O and OH databases. The SiO maser lines show a single peak around 88 km~s$^{-1}$ in both the \textit{v} = 1 and \textit{v} = 2, \textit{J} = 1$-$0 transitions \citep{2017ApJS..232...13C}.

\subsubsection{IRAS 18498--0017} The spectrum of the satellite line is of the D type, with peaks around 42 km~s$^{-1}$ and 75 km~s$^{-1}$ \citep{2019A&A...628A..90B}. The main line (1667 MHz)  shows a profile similar to the satellite line \citep{1977A&AS...28..199J, 1991ApJS...75.1323D}. Additionally, the main line (1665 MHz)  shows a peak around 82 km~s$^{-1}$, clearly outside the velocity range of the satellite line \citep{1991ApJS...75.1323D}. For the H$_2$O maser line, \citet{1976ApJ...208...87C} confirmed emission peaks at 60 km~s$^{-1}$, 62 km~s$^{-1}$, and 78 km~s$^{-1}$. Subsequently, \citet{1986A&A...167..129E} detected emission peaks at 47.8 km~s$^{-1}$ and 71.7 km~s$^{-1}$. Furthermore, \citet{2017ApJS..232...13C} detected emission peaks at 73 km~s$^{-1}$ and 78.5 km~s$^{-1}$. The maximum velocity of the previously observed H$_2$O maser line deviates by 3.5 km~s$^{-1}$ from the velocity range of the satellite line. Based on this, \citet{2013ApJ...769...20Y} listed this source as a WF candidate. The SiO maser lines (\textit{v} = 0 and \textit{v} = 2, \textit{J} = 1$-$0) were detected by \citet{2017ApJS..232...13C}. Both transitions show a single peak profile with a peak around 62 km~s$^{-1}$.

\subsubsection{IRAS 19069+0916}\label{sec: IRAS 19069+0916} The spectrum of the satellite line is of the D type, showing a typical double-peak profile with peaks around 10 km~s$^{-1}$ and 53 km~s$^{-1}$ \citep{1988ApJS...66..183E}. The spectrum of the main line (1665 MHz)  shows multiple emission peaks in the velocity range from 12 km~s$^{-1}$ to 23 km~s$^{-1}$. For the main line (1667 MHz), multiple peaks are seen in the velocity range from 2 km~s$^{-1}$ to 18 km~s$^{-1}$ \citep{1997ApJS..109..489L}. The blue-shifted end of the main line (1667 MHz) deviates from the velocity range of the satellite line. For the H$_2$O maser line, \citet{2010ApJS..188..209K} and \citet{2014A&A...569A..92V} detected two emission peaks around 10 km~s$^{-1}$ and 45 km~s$^{-1}$. Additionally, \citet{2001PASJ...53..517T} detected peaks around $-35$ km~s$^{-1}$ and 10 km~s$^{-1}$. Notably, the peak at $-35$ km~s$^{-1}$ deviates significantly from the velocity range of the satellite line, with a velocity excess reaching 45 km~s$^{-1}$. Due to this large velocity excess, this source was listed as a WF candidate by \citet{2024ApJS..270...13F}. The infrared colors of this source place it within the post-AGB star color region \citep{2024ApJS..270...13F}. Additionally, the SiO maser lines (\textit{v} = 1 and \textit{v} = 2, \textit{J} = 1$-$0) have been detected \citep{2010ApJS..188..209K}. Both transitions show a single peak profile with a peak around 30 km~s$^{-1}$.

\subsubsection{IRAS 19103+0913}\label{sec: IRAS 19103+0913} The spectrum of the satellite line is of the D type, showing peaks around 60 km~s$^{-1}$ and 83 km~s$^{-1}$ \citep{2019A&A...628A..90B}. For the main line (1667 MHz), a strong single peak is detected around 65 km~s$^{-1}$ \citep{1991ApJS...75.1323D}. For the main line (1665 MHz), emission peaks are detected around 42 km~s$^{-1}$ and 65 km~s$^{-1}$ \citep{1991ApJS...75.1323D}. The blue-shifted peak is outside the velocity range of the satellite line. To date, there are no reports of observations of H$_2$O and SiO maser lines for this source.

\subsubsection{IRAS 19190+1102}\label{sec: IRAS 19190+1102} The spectrum of the satellite line is of the $\rm D_{+}$ type, showing peaks at $-36.4$ km~s$^{-1}$, $-8.2$ km~s$^{-1}$, and 7.9 km~s$^{-1}$ \citep{1989ApJ...344..350L}. The main line (1667 MHz)  shows peaks around 7 km~s$^{-1}$, 23 km~s$^{-1}$, and 58 km~s$^{-1}$ \citep{1989ApJ...344..350L}, with the red-shifted components clearly deviating from the velocity range of the satellite line. This velocity excess reaches 50 km~s$^{-1}$. The H$_2$O maser line shows multiple peaks in the velocity range from $-4.9$ km~s$^{-1}$ to 65 km~s$^{-1}$ \citep{1989ApJ...344..350L}, significantly deviating from the velocity range of the 1612 MHz maser line. Similarly, \citet{2014ApJS..211...15Y} detected multiple peaks in the velocity range from 0 km~s$^{-1}$ to 65 km~s$^{-1}$. Subsequent observations by \citet{2015A&A...578A.119G} confirmed that the velocity range of the detected H$_2$O maser line extends from $-55$ km~s$^{-1}$ to 87.9 km~s$^{-1}$. This change in velocity range was also confirmed by the FLASHING project by \citet{2023PASJ...75.1183I}. \citet{2010ApJ...713..986D} observed the H$_2$O maser line over two epochs and noted changes. The velocity width of the detected H$_2$O maser was about 90 km~s$^{-1}$ in epoch 1 (2004 March 19) and about 130 km~s$^{-1}$ in epoch 2 (2006 May 31). To date, there have been no observational reports of SiO masers. \citet{2010ApJ...713..986D} suggested that the wide velocity range of the H$_2$O maser (about 130 km~s$^{-1}$) and its expansion over time indicate that this source is in the pre-PN stage. From the proper motion of the H$_2$O masers obtained by VLBI observations, the dynamical age of the molecular jet was calculated to be about 59 years \citep{2010ApJ...713..986D}.

\subsubsection{IRAS 19352+2030}\label{sec: IRAS 19352+2030} The spectrum of the satellite line is of the Irr type, showing two broad (width about 5 km~s$^{-1}$) velocity components around $-3$ km~s$^{-1}$ and 0 km~s$^{-1}$. Multiple peaks are seen within each velocity component \citep{1978A&AS...31..127B,1979A&AS...36..193B,1989ApJ...344..350L,2019A&A...628A..90B}. The main line (1665 MHz) shows a single peak profile with a peak around $-2$ km~s$^{-1}$. For the main line (1667 MHz), peaks are observed around $-2$ km~s$^{-1}$ and 7 km~s$^{-1}$ \citep{2004A&A...420..217E,2019A&A...628A..90B}. The main line (1667 MHz)  has also been reported by \citet{1989ApJ...344..350L}, showing emission peaks at $-1.4$ km~s$^{-1}$ and 35.5 km~s$^{-1}$. Furthermore, \citet{1993A&AS...98..245D} suggested the presence of peaks around $-26$ km~s$^{-1}$ and 0 km~s$^{-1}$. The  H$_2$O maser line shows a strong peak around 65 km~s$^{-1}$ \citep{2001PASJ...53..517T}. The peak velocity of this H$_2$O maser line is within the velocity range of the satellite line \citep{2001PASJ...53..517T}. The velocity excess reaches 65 km~s$^{-1}$. There are no reported observations of SiO maser lines. \citet{2013ApJ...769...20Y} listed this source as a WF candidate. Additionally, the main line (1667 MHz)  shows a relatively wide velocity range exceeding 60 km~s$^{-1}$. Comparing the observations from \citet{1978A&AS...31..127B} and \citet{2019A&A...628A..90B}, which have significantly different observation periods, the OH maser line profiles have not changed significantly. Moreover, due to the similarity of the satellite line profile to the peculiar type II OH/IR star U Ori, \citet{1977ApJ...214...60R} classified this source as a type II OH/IR star (Note: Type II OH/IR sources have stronger OH satellite line masers than OH main line masers).

\section{Discussion}\label{sec: Discussion}

\subsection{Significance of Using the Mainline OH Maser for Finding the Velocity Excess}\label{sec: significance of main line} 
\citet{2024ApJS..270...13F} compared the velocity range of the H$_2$O maser line with that of the satellite line. In the present study, we compared the velocity range of the main lines with that of the satellite line. In both cases, the aim of the velocity comparison was to identify circumstellar maser sources associated with CSEs whose velocity components deviate from spherically symmetric expansion. In principle, both types of comparison can achieve the same goal, as explained in Section~\ref{sect: Velocity comparison and line profile inspection}. However, the samples of objects detected in the  H$_2$O maser line and the satellite line, as well as those detected in the main lines and the satellite line, are overall clearly different in nature, despite some overlap. As shown in Figure~\ref{Fig: [OH maser distribution]}, the black data points (representing objects detected in both the H$_2$O maser line and the satellite line) are clearly concentrated in color regions where AGB stars are abundant. In contrast, the red, blue, and green data points (representing objects detected in either or both the two main lines and the satellite line are more broadly distributed across a wider color range, with less concentration in the regions where AGB stars are abundant. This difference suggests that using the H$_2$O maser line covers evolutionary stages closer to the AGB phase, while the main lines can cover a broader range of evolutionary stages within a single sample. 

It is important to note that the H$_2$O maser line and the main lines are not necessarily emitted from the same region within the CSE. Figure~\ref{Fig: Spectra_IRAS18251-1048_and_IRAS19069+0916} presents the spectra of the H$_2$O maser line and the main lines for IRAS 18251--1048 and IRAS 19069+0916, which are listed in Table~\ref{Tab: [Sources showing true velocity deviations in the OH 1665/1667 MHz line]}. These spectra, compiled from literature with observation dates as close as possible, reveal that in both sources, the velocity of the main lines and the H$_2$O line exceeds the velocity range of the satellite line. However, when examining the spectra, it is evident that for IRAS 19069+0916, the intensity peaks of the mainline OH maser and the H$_2$O maser are located at relatively close velocities, whereas for IRAS 18251--1048, the velocity of the intensity peaks differs significantly. This suggests that the emission regions of these two maser lines within the CSE are not necessarily closely located. This speculation is supported by VLBI observations. For instance, \citet{1987A&A...174...95D} and \citet{1999NewAR..43..563M} confirmed through VLBI observations that the main lines in S Per are emitted from locations close to the H$_2$O maser emission region. However, the mainline OH maser can also be emitted from regions corresponding to the satellite line (which is usually emitted from a location distinct from the H$_2$O maser emission region), as seen in OH 127.8--0.0 \citep{1985MNRAS.216P...1D}, VY CMa \citep{1982ApJ...253..199B}, and VX Sgr \citep{1986MNRAS.220..513C}.

Let us look at Figure~\ref{Fig: [OH maser distribution]} in a little more detail. The red, blue, and green data points are located within the color regions associated with AGB stars (i.e., the IIIa region), but as seen in  Figure~\ref{Fig: [New_IRAS_two_color_diagram_of_AGB_and_postAGB]}, few objects with confirmed velocity excess are found in this region. In contrast, in Figure~\ref{Fig: [OH maser distribution]}, the black data points are abundant in the AGB region, and as seen in Figure~\ref{Fig: [New_IRAS_two_color_diagram_of_AGB_and_postAGB]}, many of the velocity excess sources identified by \citet{2024ApJS..270...13F} are also found in this region. If the main lines and the H$_2$O maser line are emitted from closely located regions within the CSE, more cases of velocity excess in the main line should be found in the IIIa region. Table~\ref{Tab: [Detection rate of velocity excess sources]} summarizes the detection rates of velocity excess in the color regions of AGB and post-AGB phases. Due to the small sample size for the main line (1665 MHz), it is statistically difficult to make definitive statements based on the characteristics of the main line (1665 MHz). However, when comparing the detection rate of velocity excess using the main line (1667 MHz, see ``1612--1667'' in  Table~\ref{Tab: [Detection rate of velocity excess sources]}) with that using the  H$_2$O line (see ``1612--H$_2$O'' in Table~\ref{Tab: [Detection rate of velocity excess sources]}), it is clear that during the AGB phase, there is a significant difference in the detection rates of velocity excess, but this difference narrows in the post-AGB phase. As mentioned above, this result seems to suggest that in the CSE, the emission regions of the main line (1667 MHz) and the  H$_2$O line are distinctly different during the AGB phase, whereas in the post-AGB phase, they are located closer to each other.

The properties of OH mainline masers, as described above (particularly those of the 1667 MHz main line), indicate that the OH mainline maser could be an effective tool for investigating the dynamical characteristics of the CSEs at various evolutionary stages. Moreover, in the case of objects with poorly understood characteristics, such as RNRs, the detectability of OH mainline masers over a wide range of effective temperatures could serve as a valuable tool for investigation. Regarding OH masers, it is generally believed that radiation at wavelengths of 35~$\mu$m and 53~$\mu$m is closely related to the population inversion \citep[see, e.g.,][]{Gray_2012}. However, the pumping mechanism of the mainline OH maser is not yet fully understood, as complex physical mechanisms beyond a simple radiative excitation are involved. Investigating the relationship between the evolutionary stage of an object and the proximity of the emission regions of the H$_2$O maser and the mainline OH maser within the CSE also may be valuable for understanding maser pumping in the future.

\subsection{Comparison with Known WF Samples}\label{sec: comparison with known WFs} 

In this section, we examine the rate of detecting velocity excess in the OH main lines among known WF samples. There are 16 known WFs, of which the satellite line has been detected from 12 sources. The OH main lines have been detected from 11 sources (see Table~\ref{Tab: [OH maser status of the known WF samples]}). In all the WFs where the main lines are detected, the satellite line is also detected. Among the 11 sources, only one source (IRAS 19190+1102) shows a velocity excess in the main line. Therefore, the detection rate of velocity excess is 9\%. In other words, using the main lines, there is a high probability of missing known WFs.

In \citet{2024ApJS..270...13F}, velocity excess was searched for using H$_2$O masers, and among the 16 known WFs, 7 sources showed velocity excess in the H$_2$O maser line. As stated in Section~\ref{sec: Intro}1, note that the classical definition of WF is "a circumstellar H$_2$O maser source with a velocity range of over 100 km~s$^{-1}$," whereas the WF candidates selected by \citet{2024ApJS..270...13F} are described as "H$_2$O maser sources exhibiting a velocity excess relative to the OH satellite lines." By the definition of classical WF, the  H$_2$O maser is detected from all known WFs. Since the satellite line is detected from 12 sources (refer to Table~\ref{Tab: [OH maser status of the known WF samples]} for information on which sources exhibit the satellite line detection), the detection rate of velocity excess in the sample where both the satellite line and H$_2$O masers are detected is 58\%. This value is significantly higher compared to the detection rate using the main lines.

What is the cause of this difference? The low detection rate of velocity excess in the main line from known WFs does not necessarily mean that the general detection rate of velocity excess from post-AGB stars is low. According to Table~\ref{Tab: [Detection rate of velocity excess sources]}, there is no statistically significant difference in the detection rates when compared with the case of H$_2$O masers. As shown in Figure~\ref{Fig: [New_IRAS_two_color_diagram_of_known_WFs_RNRs]}, there does not appear to be a significant difference in the distribution between the sources with detected velocity excess in the main line maser and the known WF sources on the IRAS two-color diagram. If we consider that the distribution on the IRAS two-color diagram reflects the evolutionary stages of evolved stars, we should consider that the evolutionary stages of these two samples are similar.

The known WF must be considered a special sample with a selection bias. This is because all known WFs are sources where the H$_2$O maser line is bright and the velocity range exceeds 100 km~s$^{-1}$. Additionally, for such a velocity range to be observed, there are constraints on the angle of the jet relative to the line of sight. WF candidates selected from the perspective of velocity excess can be detected even with a narrower velocity range. From this perspective, it should be considered that the conditions differ from the circumstellar OH maser sources analyzed in this study.


Within the CSE of known WFs, there are bipolar jets that have already undergone sufficient acceleration. Therefore, it is possible that the physical conditions in the emission region of the OH satellite line change due to interactions with the bipolar jets, making it difficult to detect the double-peak profile that serves as the standard to determine the CSE expansion velocity. In fact, out of the eight sources where velocity excess was detected in this study, the OH satellite line profile was of Irr type in two sources (25\%), while in the known WFs, five out of the 12 sources with detected the satellite line were of Irr type (42\%), indicating a slightly higher rate of Irr type.

Moreover, as pointed out by \citet{2024ApJS..270...13F}, it is highly likely that the known WFs are observing the CSE from a direction close to the jet direction (i.e., polar direction). Such geometrically special conditions may also be related to the low detection rate of velocity excess in the OH main line. In any case, it is noteworthy that there is little overlap between the known WF sample and the sources with the velocity excess in the main lines. So far, there have been very few case studies on sources with the velocity excess in the main lines. In the future, high angular resolution mapping observations using radio interferometers to map the satellite and main lines and the H$_2$O maser in the samples from this study could help clarify the differences compared to the known WFs.

\subsection{Can the Source with an Irregular Profile be an AGB Star?}\label{sec: Irr type}
As mentioned in Section~\ref{sec: Sources with true velocity deviation}, among the eight sources where velocity excess was detected (see Table~\ref{Tab: [Sources showing true velocity deviations in the OH 1665/1667 MHz line]}), two objects (IRAS 15405--4945 and IRAS 19352+2030) exhibit an irregular (Irr) type profile in the satellite line. These sources may be in a more advanced evolutionary stage than the AGB phase in which the satellite line generates from a spherically expanding CSE and exhibits a double-peaked profile \citep[see, e.g.,][]{2004ApJS..155..595D}. As evolution progresses and asymmetric motions (e.g., bipolar outflows) occur, it is possible that irregular profiles deviating from a double-peaked profile may form (e.g., \citet{2001MNRAS.322..280Z}). However, it is noteworthy that among the Irr-type sources listed in Table~\ref{Tab: [Sources showing true velocity deviations in the OH 1665/1667 MHz line]}, one has an IRAS variability index of 99 (specifically, the IRAS variability index of IRAS 19352+2030 is 99). Typically, such an index indicates a  Mira-type pulsating variable star in the AGB phase. However, in Figure~\ref{Fig: [New_IRAS_two_color_diagram_of_AGB_and_postAGB]}, IRAS 19352+2030 is located near the boundary between regions IIIb and IV and appears post-AGB-like in terms of infrared color. Given these characteristics, sources like IRAS 19352+2030 might be objects transitioning from the AGB phase to the post-AGB phase.

To further investigate the infrared variability characteristics of the sources listed in Table~\ref{Tab: [Sources showing true velocity deviations in the OH 1665/1667 MHz line]} aside from the IRAS variability index, a simple analysis of photometric data in the W1 and W2 bands from the past 10 years, provided by the NEOWISE-R mission \citep{2014ApJ...792...30M}, was conducted. In this analysis, only photometric data of good quality—where the signal-to-noise ratio in each band exceeded 20, the frame image quality score was greater than 0, and the frame quality score was greater than 0—were used to plot the light curves (data from one of the bands, W1 or W2, that provided high-quality light curves were used; see Figure~\ref{Fig: [WISE light curves with periodicity]} and Figure~\ref{Fig: [WISE light curves without periodicity}. In these light curves, data observed within a 48-hour period are averaged and plotted as a single data point. The standard deviation of multiple data points collected within this period is shown as the error bar. Additionally, a period analysis was performed using the Lomb–Scargle periodogram to detect periodicity in the light curves. This is a widely used statistical algorithm for period analysis, suitable for detecting and characterizing periodic signals in observations irregularly spaced in time \citep[see, e.g.,][]{2021ApJS..256...43S}. The periodograms were created using the AstroPy library\footnote{\url{https://docs.astropy.org/en/stable/timeseries/lombscargle.html}}, specifically utilizing the "chi2" method with the auto power option. The resulting light curves and period analysis results are shown in Figure~\ref{Fig: [WISE light curves with periodicity]} for cases where periodicity was found. The pulsation periods for IRAS 18251--1048, IRAS 18498--0017, IRAS 19069+0916, IRAS 19103+0913, and IRAS 19352+2030 were found to be 1396.5, 1583.2, 1384.4, 505.6, and 517.5 days, respectively. When the positions of these objects were checked on Figure~\ref{Fig: [New_IRAS_two_color_diagram_of_AGB_and_postAGB]}, it was found that IRAS 18251--1048 and IRAS 18498--0017 are located in the post-AGB region (LI region), and IRAS 19103+0913 (IIIa region) and IRAS 19352+2030 are located in the AGB region. It is intriguing that stars exhibiting both velocity excess and pulsating variability are found among sources significantly separated from the AGB star color region, such as IRAS 18251--1048 and IRAS 18498--0017. These stars may be on the verge of leaving the AGB stage.

\subsection{Additional Note}\label{sec: additional note}
As shown in Figure~\ref{Fig: [OH maser distribution]}, when comparing the sample of objects for which both the main (1667 MHz) and satellite line observations are available (red data points) with that but for other main line (1665 MHz) and satellite line observations are available (blue data points), the red data points are overwhelmingly more numerous. From the perspective of the detection rate of velocity excess, due to the smaller sample size for the main line (1665 MHz), it is unclear whether there is a difference in the detection rates of the velocity excess for the main line (1667 MHz, see Table~\ref{Tab: [Detection rate of velocity excess sources]}). To draw statistically significant conclusions, it is desirable to increase the sample size of main line OH maser sources using more sensitive observational instruments such as FAST and SKA in the future. Identifying differences in the properties of the two main lines from the perspectives of evolutionary stage and infrared color could be helpful for understanding the pumping mechanism.

Additionally, as confirmed by \citet{2024ApJS..270...13F} and in the present study, when searching for sources showing velocity excess using only the line parameters of maser lines listed in the circumstellar maser source database, there is a possibility of overlooking a non-negligible number of velocity excess sources (see Section~\ref{sect: Velocity comparison and line profile inspection}). As the circumstellar maser source database expands in the future, it would be beneficial and necessary to collect spectral data and include them in the database.

\section{Summary} \label{sec: summary}

In the present study, we conducted a comprehensive analysis of circumstellar OH maser sources, focusing on identifying CSEs of cold stars with velocity components deviating from spherically symmetric expansion. The main findings and conclusions are summarized as follows:

\begin{enumerate} 
\item We compared the velocity ranges of the main lines and satellite line for 377 circumstellar OH maser sources, identifying 8 circumstellar OH maser sources where the main lines exhibited significant velocity excess compared to the satellite lines. 

\item The infrared colors of these 8 circumstellar OH maser sources were consistent with post-AGB star-like colors. This result differs from previous studies using H$_2$O maser lines, where many of the objects with detected velocity excesses exhibit AGB star-like colors. 

\item It is suggested that the main lines, compared to the H$_2$O maser line, are detected from a wider range of infrared colors and are effective probes for investigating the dynamics of CSEs (circumstellar envelopes) across a broader range of evolutionary stages. 

\item Two of the objects with detected velocity excesses (IRAS 15405--4945 and IRAS 19352+2030) showed irregularities in the satellite line profiles, suggesting that these objects may be in the transition phase from the AGB to post-AGB stages. 

\item Analysis of WISE light curves revealed periodic variability in 5 of these 8 objects. This suggests that pulsating variable AGB stars may be included among objects located in the post-AGB star color region. 

\item It is recognized that during the transition from the AGB to post-AGB stages, the emission regions of OH main lines and H$_2$O maser lines within the CSE may change. 

\item  Of the 16 known WFs, 11 have been detected in the satellite and main lines, but only one WF exhibits the velocity excess, with a detection rate of 9\%, which is relatively low.

\end{enumerate}

This study demonstrates the usefulness of circumstellar OH main line masers in investigating the complex dynamics of envelopes of evolved stars and emphasizes the need for further investigation of maser properties during the transition period from the AGB to post-AGB stages. Additionally, it is suggested that the database of circumstellar OH maser sources may also be useful for exploring objects that cannot be explained by the conventional schemes of stellar evolution, such as RNR.

\begin{acknowledgments}
We acknowledge the science research grants from the China Manned Space Project with No. CMS-CSST-2021-A03, No.CMS-CSST-2021-B01. 
JN acknowledges financial support from the `One hundred top talent program of Sun Yat-sen University' grant no. 71000-18841229. YZ thanks the financial supports from the National Natural Science Foundation of China (NSFC, No.12473027 and No.12333005) and the Guangdong Basic and Applied Basic Research Funding (No.2024A1515010798)
\end{acknowledgments}

\appendix

\section{Results of velocity range comparison: Sources without confirmed velocity excess}\label{appendix: Velocity range comparison of all the selected sources}

In this section, we present the results of the velocity comparison for sources where no velocity excess was confirmed, as discussed in Section~\ref{sect: Velocity comparison and line profile inspection}. Figures~\ref{Fig: a[comparison of 1612 OH maser with that of 1665 MHz]}, \ref{Fig: b[comparison of 1612 OH maser with that of 1665 MHz-2]}, \ref{Fig: c[comparison of 1612 OH maser with that of 1665 MHz-3]}, and \ref{Fig: d[comparison of 1612 OH maser with that of 1665 MHz-4]} show the comparison between the velocity range of the satellite line and the main line (1665 MHz). Similarly, Figures~\ref{Fig: a[comparison of 1612 OH maser with that of 1667 MHz]}, \ref{Fig: b[comparison of 1612 OH maser with that of 1667 MHz-2]}, \ref{Fig: c[comparison of 1612 OH maser with that of 1667 MHz-3]}, \ref{Fig: d[comparison of 1612 OH maser with that of 1667 MHz-4]}, \ref{Fig: e[comparison of 1612 OH maser with that of 1667 MHz-5]}, \ref{Fig: f[comparison of 1612 OH maser with that of 1667 MHz-6]}, and \ref{Fig: g[comparison of 1612 OH maser with that of 1667 MHz-7]} illustrate the comparison between the velocity range of the satellite line and the main line (1667 MHz).

\section{Maser properties of sources where velocity excess was not confirmed}\label{appendix: Maser information of the other sources exhibiting velocity deviation}

As discussed in Section~\ref{sect: Velocity comparison and line profile inspection}, for sources with velocity excesses identified from database values, we visually examined the spectral diagrams published in previous studies to confirm whether the velocity excess was genuine. For sources where the velocity excess could not be definitively determined, this section summarizes the maser properties reported in previous studies.

\subsection{IRAS 03287--1535}\label{sec: IRAS 03287-1535} The satellite line spectrum primarily shows a single peak at $-4.5$ km~s$^{-1}$ \citep{1992A&A...254..133L}, with an additional peak at $-4.3$ km~s$^{-1}$ reported in the database. The main line (1667 MHz) spectrum reveals two peaks at $-12.9$ km~s$^{-1}$ and $-3.9$ km~s$^{-1}$ \citep{1993A&AS...98..245D}, with the blue-shifted component deviating from the velocity range of the satellite line. The  $\rm H_{2}O$ maser spectrum shows a single peak around $-5$ km~s$^{-1}$ \citep{2013AJ....145...22K}, and the deviation of the $\rm H_{2}O$ maser line from the satellite line velocity range is minimal (0.5 km~s$^{-1}$). No SiO maser spectra have been detected for this source.

\subsection{IRAS 05027--2158}\label{sec: IRAS 05027-2158} The OH maser lines were observed four times by \citet{2010ARep...54..400R}. A faint emission peak at $-$6.9 km~s$^{-1}$ may be present in the satellite line. The main line (1667 MHz) shows a single emission peak around $-$31 km~s$^{-1}$ in all observations, which is close to the systemic velocity ($-$30 km~s$^{-1}$). Similarly, the main line (1665 MHz) exhibits an emission feature around $-$30 km~s$^{-1}$ in all observed profiles. The satellite line profile was not detected by \citet{1988A&A...206..285S}, but the database records indicate two peaks at $-$15.9 km~s$^{-1}$ and $-$0.2 km~s$^{-1}$. The  $\rm H_{2}O$ maser spectrum reported by \citet{2010ApJS..188..209K} shows an emission peak at $-$30.0 km~s$^{-1}$. Both the SiO \textit{v} = 1 and \textit{v} = 2, \textit{J} = 1$-$0 maser lines show a single peak at $-$30.0 km~s$^{-1}$ \citep{2010ApJS..188..209K}.

\subsection{IRAS 07180--1314}\label{sec: IRAS 07180-1314} The satellite line exhibits a typical double peak around 45 km~s$^{-1}$ and 71 km~s$^{-1}$ \citep{1992A&A...254..133L}. \citet{1995A&AS..111..237L} detected double-peaked emission in the main line (1667 MHz) at 24.8 km~s$^{-1}$ and 36.1 km~s$^{-1}$; however, this detection is questionable as the velocities do not align with those of the satellite line, and no corresponding spectra have been reported. The main line (1665 MHz) spectrum was not observed. The $\rm H_{2}O$ maser spectrum shows two peaks at 47.5 km~s$^{-1}$ and 57.5 km~s$^{-1}$ \citep{2014ApJS..211...15Y}, whereas another spectrum shows a single peak at 48.7 km~s$^{-1}$ \citep{1993A&A...275..163K}. \citet{1993A&A...275..163K} also suggested two SiO maser (\textit{v} = 1, \textit{J} = 1$-$0) features at 57.4 km~s$^{-1}$ and 62.3 km~s$^{-1}$, consistent with the findings of \citet{1998A&AS..127..185N}. However, later studies by \citet{2007PASJ...59..559D, 2014ApJS..211...15Y} detected the SiO \textit{v} = 1 and \textit{v} = 2, \textit{J} = 1$-$0 maser lines, both showing typical single peaks around 60 km~s$^{-1}$. Interestingly, the blue-shifted peak of the SiO \textit{v} = 1, \textit{J} = 1$-$0 maser line appears to fade over time \citep{1993A&A...275..163K, 1998A&AS..127..185N, 2007PASJ...59..559D, 2014ApJS..211...15Y}. \citet{1995A&A...299..453L} classified this source as an AGB star (i.e., OH/IR star).

\subsection{IRAS 07446--3210A}\label{sec: IRAS 07446-3210A} The satellite line shows standard two peaks at 39.0 km~s$^{-1}$ and 47.4 km~s$^{-1}$ \citep{1991A&AS...90..327T}, though no corresponding spectra have been reported in the literature. The main line (1665 MHz)spectrum shows two emission features around 15 km~s$^{-1}$ and 40 km~s$^{-1}$, similar to those seen in the main line (1667 MHz) \citep{1995A&AS..111..237L}. The  $\rm H_{2}O$ maser spectrum displays peaks spanning a velocity range from 20.0 km~s$^{-1}$ to 35.0 km~s$^{-1}$ \citep{2010ApJS..188..209K}. SiO \textit{v} = 1 and \textit{v} = 2, \textit{J} = 1$-$0 maser lines were also detected by \citet{2010ApJS..188..209K}.

\subsection{IRAS 15514--5323}\label{sec: IRAS 15514-5323} 
The satellite line spectrum shows typical double-peaked profiles of AGB stars around $-80$ km~s$^{-1}$ and $-40$ km~s$^{-1}$, and the redshifted component exhibits a somewhat complex line profile, consisting of multiple peaks rather than a single peak \citep{2004ApJS..155..595D}. The main line (1667 MHz) shows two peaks around $-80$ km~s$^{-1}$ and $-34$ km~s$^{-1}$ \citep{2004ApJS..155..595D}. No previous  $\rm H_{2}O$ and SiO maser spectra have been reported.

\subsection{IRAS 17150--3224}\label{sec: IRAS 17150-3224} A single emission peak at 25 km~s$^{-1}$ was detected in the satellite line \citep{2004ApJS..155..595D,2012A&A...537A...5W}. The main line (1665 MHz)  shows a single emission peak around 10 km~s$^{-1}$, with weak feature close proximity to the single peak and a large linewidth of about 20 km~s$^{-1}$, as reported by \citet{2004ApJS..155..595D}. The main line (1667 MHz) displays a typical double-peaked profile at 0 km~s$^{-1}$ and 25 km~s$^{-1}$ \citep{2004ApJS..155..595D}, consistent with \citet{2012A&A...537A...5W}. The 10 km~s$^{-1}$ component in the main line (1665 MHz)  and the 0 km~s$^{-1}$ component in the main line (1667 MHz) deviate from the peak in the satellite line. This star is classified as a bipolar pre-PN by \citet{1998ApJ...501L.117K}. \citet{2020ApJ...889...85H} suggested that the AGB wind from this object ended about 1300 years ago, with quadrupolar outflows ejected around 350 years ago, and two additional bipolar outflows ejected about 280 and 200 years ago, respectively.

\subsection{IRAS 17385--3332}\label{sec: IRAS 17385-3332} This object exhibits a double-peaked profile around $-$247 km~s$^{-1}$ and $-$225 km~s$^{-1}$ in the satellite line spectrum \citep{1994A&AS..103..301H,2002AJ....123.2772S, 2004ApJS..155..595D,2012A&A...537A...5W}. The main line (1665 MHz) shows multiple peaks spanning from $-$258 km~s$^{-1}$ to $-$228 km~s$^{-1}$ in both left-hand-circular (LHC) and right-hand-circular (RHC) polarization spectra. The most blue-shifted component lies outside the velocity range of the satellite line. The main line (1667 MHz) shows one emission peak at $-$225 km~s$^{-1}$, which was not clearly detected by \citet{2004ApJS..155..595D}, but both LHC and RHC spectra detected this peak as reported by \citet{2004A&A...420..217E} and \citet{2012A&A...537A...5W}.

\subsection{IRAS 17392--3319}\label{sec: IRAS 17392-3319}
The satellite line exhibits a typical double-peaked profile with velocities at $-$57.5 km~s$^{-1}$ and $-$30.5 km~s$^{-1}$ \citep{1999MNRAS.304..622S}. The main line (1667 MHz) also shows two emission peaks at $-$59.5 km~s$^{-1}$ and $-$30.5 km~s$^{-1}$ \citep{1999MNRAS.304..622S}. Compared to the satellite line, the profile of the main line (1667 MHz) displays a slight shift (about 2 km~s$^{-1}$) on the blue-shifted side. \citet{2004A&A...420..217E} detected a similar double-peaked profile in both the left-hand and right-hand circular polarization spectra of the main line (1667 MHz) at the same velocities. The main line (1665 MHz) was not detected by \citet{2004A&A...420..217E}.

\subsection{IRAS 17393--2727}\label{sec: IRAS 17393-2727}
The satellite line exhibits two peaks located at $-$122.8 km~s$^{-1}$ and $-$93.0 km~s$^{-1}$, as reported by \citet{1989A&A...217..157Z}. However, more recent observations detected only one peak at $-$120.0 km~s$^{-1}$ \citep{2004ApJS..155..595D,2012A&A...537A...5W,2016MNRAS.461.3259G,2018ApJS..239...15Q}. Additionally, the main line (1665 MHz) was detected with two peaks at $-$127.0 km~s$^{-1}$ and $-$116.0 km~s$^{-1}$ \citep{1989A&A...217..157Z,2016MNRAS.461.3259G}, and a single peak around $-$116 km~s$^{-1}$ was observed by \citet{2004ApJS..155..595D,2018ApJS..239...15Q}. The most blue-shifted component of the main line (1665 MHz) \citep{1989A&A...217..157Z,2016MNRAS.461.3259G} lies outside the velocity range of the satellite line. Similarly, the most blue-shifted component at $-$125.0 km~s$^{-1}$ in the main line (1667 MHz) \citep{2004ApJS..155..595D,2012A&A...537A...5W,2016MNRAS.461.3259G,2018ApJS..239...15Q} also exceeds the velocity range of the satellite line, although the difference is minor (about 2 km~s$^{-1}$).
Furthermore, \citet{2015A&A...578A.119G} detected  $\rm H_{2}O$ maser emission with a single component at $-$107.6 km~s$^{-1}$. This source exhibits a bipolar morphology in the optical \citep{2011AJ....141...80M} and is classified as a young planetary nebula  \citep{2007ApJ...666L..33G,2012A&A...547A..40U,2018ApJS..239...15Q}, similar to K 3-35 \citep{2001Natur.414..284M}.

\subsection{IRAS 17501--2830}\label{sec: IRAS 17501-2830}
This OH maser source shows a double-peak profile in the satellite line, but there are no spectra given in the literature to confirm this claim \citep{2018ApJS..239...15Q}. According to the available data, the main line (1667 MHz) exhibits a double-peaked profile at 76.5 km~s$^{-1}$ and 84.1 km~s$^{-1}$, though the spectrum is not available in the literature. No prior spectra for the  $\rm H_{2}O$ maser or SiO maser have been reported.

\subsection{IRAS 18006--1734}\label{sec: IRAS 18006-1734}
Database records indicate that the satellite line shows two peaks at 18.2 km~s$^{-1}$ and 29.5 km~s$^{-1}$. The main line (1667 MHz) displays a typical double-peaked profile at 5 km~s$^{-1}$ and 29 km~s$^{-1}$, as noted by \citet{1993A&AS...98..245D}. No main line (1665 MHz) spectra have been reported.

\subsection{IRAS 18025--2113}\label{sec: IRAS 18025-2113}
According to database records, the satellite line displays two peaks at velocities of 24.7 km~s$^{-1}$ and 38.6 km~s$^{-1}$, while the main line (1665 MHz) shows peaks at $-$5.1 km~s$^{-1}$ and 33.9 km~s$^{-1}$. The main line (1667 MHz) exhibits peaks at $-$6.8 km~s$^{-1}$ and 18.5 km~s$^{-1}$. However, no literature was found confirming these velocities through spectral observations. 
The  $\rm H_{2}O$ maser emission line has a single peak at approximately 17 km~s$^{-1}$ \citep{1978AJ.....83.1206K}, which falls outside the velocity range of the satellite line. However, the  $\rm H_{2}O$ maser line profile shows multiple peaks ranging from 10.0 km~s$^{-1}$ to 25.0 km~s$^{-1}$, as reported by \citet{2010ApJS..188..209K}. The SiO \textit{v} = 1, \textit{J} = 1$-$0 maser line shows a single peak at 17.5 km~s$^{-1}$ with a broad linewidth  \citep[about 20 km~s$^{-1}$,][]{2010ApJS..188..209K}. In addition, the SiO \textit{v} = 2, \textit{J} = 1$-$0 maser line shows two peaks at 17.5 km~s$^{-1}$ and 20.0 km~s$^{-1}$, also with a broad linewidth (about 20 km~s$^{-1}$), as reported by \citet{2010ApJS..188..209K}. This star is classified as a red supergiant  \citep{1975AJ.....80.1011H}.

\subsection{IRAS 18033--2111}\label{sec: IRAS 18033-2111}
According to the database records, the satellite line shows two peaks at velocities of $-$66.0 km~s$^{-1}$ and $-$32.6 km~s$^{-1}$. However, this could not be confirmed in the literature. The main line (1667 MHz) exhibits four emission peaks at $-$124.4 km~s$^{-1}$, $-$107.0 km~s$^{-1}$, $-$65.0 km~s$^{-1}$, and $-$32.0 km~s$^{-1}$, as reported by \citet{1993A&AS...98..245D}. Based solely on the database information, a significant velocity excess is suggested. However, since no spectra of the satellite line supporting the database values were found in the literature, we conservatively consider the velocity excess as undetected in the present study.

\subsection{OH 11.54+0.1}\label{sec: OH 11.54+0.1}
The satellite line profile is predominantly characterized by a double-peaked profile with peaks at 20 km~s$^{-1}$ and 65 km~s$^{-1}$ \citep{2004ApJS..155..595D,2012A&A...537A...5W}. The main line (1665 MHz) profile displays a single peak around 60 km~s$^{-1}$, as noted by \citet{1991ApJS...75.1323D,2004ApJS..155..595D}. The main line (1667 MHz) profile exhibits two peaks around 13 km~s$^{-1}$ and 68 km~s$^{-1}$ \citep{1991ApJS...75.1323D,2004ApJS..155..595D}. The most blue-shifted component of the main line (1667 MHz)  exceeds the velocity of the satellite line.

\subsection{IRAS 18095+2704}\label{sec: IRAS 18095+2704}
The satellite line profile shows a single peak at $-$12.7 km~s$^{-1}$, as reported by \citet{2012A&A...537A...5W}, while two emission peaks at $-$17.5 km~s$^{-1}$ and 0 km~s$^{-1}$ were observed by \citet{1988ApJS...66..183E}. According to database records, the main line (1665 MHz)  displays two peaks at $-$27.1 km~s$^{-1}$ and 0.3 km~s$^{-1}$, and the main line (1667 MHz)  exhibits peaks at $-$26.1 km~s$^{-1}$ and $-$17.4 km~s$^{-1}$. However, no spectra confirming these findings for the OH main lines have been found. This star is classified as a pre-PN \citep{2018AJ....156..300H}.

\subsection{IRAS 18195--2804}\label{sec: IRAS 18195-2804}
According to database records, the satellite line displays two peaks at velocities of 97.7 km~s$^{-1}$ and 131.2 km~s$^{-1}$. However, no corresponding spectra have been reported in the literature. The main line (1667 MHz)  shows a double-peaked profile at velocities of approximately 241 km~s$^{-1}$ and 258 km~s$^{-1}$ \citep{1993A&AS...98..245D}, though the velocity difference between the two peaks is narrower than typically observed in AGB stars.

\subsection{G20.635+0.102}\label{sec: G20.635+0.102}
The satellite line shows a double-peaked profile at velocities around 97 km~s$^{-1}$ and 102 km~s$^{-1}$ \citep{2019A&A...628A..90B}. The main line (1665 MHz) exhibits a distinct peak at 112.5 km~s$^{-1}$, as reported by \citet{2019A&A...628A..90B}, while the database indicates an additional peak at 103.5 km~s$^{-1}$. The presence of the 112.5 km~s$^{-1}$ component in the main line (1665 MHz) profile extends the velocity range beyond that of the satellite line line. The velocity difference between the two peaks in the satellite line profile is less than 10 km~s$^{-1}$. No previous $\rm H_{2}O$ maser or SiO maser spectra have been reported.

\subsection{IRAS 18284--0946}\label{sec: IRAS 18284-0946}
The satellite line profile is characterized by a double-peaked profile, with peaks at 66 km~s$^{-1}$ and 78 km~s$^{-1}$ \citep{2019A&A...628A..90B}. The velocity difference between the two peaks in the satellite  line profile is about 12 km~s$^{-1}$. Additionally, the main line (1665 MHz) spectrum shows two peaks at 85.5 km~s$^{-1}$ and 94.5 km~s$^{-1}$ \citep{2019A&A...628A..90B}, which extend beyond the velocity range of the satellite line. No previous  $\rm H_{2}O$ maser or SiO maser spectra have been reported.

\subsection{IRAS 18349+1023}\label{sec: IRAS 18349+1023}
The satellite line shows a typical double-peaked profile at approximately $-$47 km~s$^{-1}$ and $-$18 km~s$^{-1}$ \citep{1993ApJS...89..189C}. However, another peak at 4 km~s$^{-1}$ was reported by \citet{1989A&AS...78..399T}, though no spectra were included in the report. The main line (1665 MHz)  spectrum shows multiple peaks ranging from $-$48 km~s$^{-1}$ to $-$17 km~s$^{-1}$, with a similar profile for the main line (1667 MHz). According to database information, both the two main lines reveal an additional, larger velocity component at $-$62.7 km~s$^{-1}$ and $-$69.3 km~s$^{-1}$, respectively. 
The GHz $\rm H_{2}O$ maser emission line shows three peaks at approximately $-$44 km~s$^{-1}$, $-$36 km~s$^{-1}$, and $-$31 km~s$^{-1}$, all within the velocity range of the satellite  line \citep{1996A&AS..116..117E}. However, the $\rm H_{2}O$ maser spectrum reported by \citet{2010ApJS..188..209K} shows a strong single peak at $-$20 km~s$^{-1}$. 
The SiO \textit{v} = 1 and 2, \textit{J} = 1$-$0 maser lines show a single peak at $-$32.5 km~s$^{-1}$, as reported by \citet{1977MNRAS.180..415B,2010ApJS..188..209K}, with a linewidth of 7.5 km~s$^{-1}$. This object, known as V1111 Oph, is classified as an O-rich AGB star (i.e., mira variable) with a period of 500 days \citep[e.g.,][]{2001MNRAS.326..490O,2016ApJ...817..115C}. Notably, \citet{2001MNRAS.326..490O} identified it as a dust-enshrouded AGB star undergoing significant mass loss (about $10^{-5}$ M$_{\odot}$ yr$^{-1}$).

\subsection{IRAS 18451--0332}\label{sec: IRAS 18451-0332}
The satellite line spectrum reveals multiple peaks within the velocity range of 114.0 km~s$^{-1}$ to 138.0 km~s$^{-1}$ \citep{1977A&AS...28..199J}. The main line (1665 MHz) exhibits a typical double-peaked profile at 117.0 km~s$^{-1}$ and 136.0 km~s$^{-1}$ \citep{1991ApJS...75.1323D}. The main line (1667 MHz)  spectrum shows multiple peaks spanning from 107 km~s$^{-1}$ to 140 km~s$^{-1}$ \citep{1991ApJS...75.1323D}. The satellite line profile is enclosed within that of the main line (1667 MHz).

\subsection{IRAS 19024+0044}\label{sec: IRAS 19024+0044} The satellite line spectrum shows two emission peaks at approximately 35 km~s$^{-1}$ and 62 km~s$^{-1}$, as reported by \citet{2004ApJS..155..595D}. Similarly, in the main line (1665 MHz) spectrum, two emission peaks are observed around 40 km~s$^{-1}$ and 62 km~s$^{-1}$ \citep{2004ApJS..155..595D}. The main line (1667 MHz) spectrum also shows two peaks, at roughly 25 km~s$^{-1}$ and 75 km~s$^{-1}$ \citep{2004ApJS..155..595D}. The satellite line  profile is enclosed within the main line (1667 MHz) profile. This source has been identified as a post-AGB star by \citet{2006A&A...458..173S}. No $\rm H_{2}O$ and SiO maser spectra have been previously reported for this source.

\subsection{IRAS 19083+0851}\label{sec: IRAS 19083+0851} The satellite line spectrum reveals three peaks around 17 km~s$^{-1}$, 37 km~s$^{-1}$, and 75 km~s$^{-1}$, as reported by \citet{1993ApJS...89..189C}. The main line (1665 MHz) shows multiple peaks in the LHC spectrum, ranging from 2 km~s$^{-1}$ to 17 km~s$^{-1}$, while the RHC spectrum presents peaks from 2 km~s$^{-1}$ to 20.5 km~s$^{-1}$ \citep{1997ApJS..109..489L}. In the main line (1667 MHz), both LHC and RHC spectra exhibit three peaks near 2 km~s$^{-1}$, 20 km~s$^{-1}$ (faint in the LHC), and 35 km~s$^{-1}$ \citep{1997ApJS..109..489L}. The outermost blue-shifted components of the two main lines deviate significantly from the satellite line. The $\rm H_{2}O$ 22 GHz maser emission displays two peaks at approximately 7 km~s$^{-1}$ and 58 km~s$^{-1}$ \citep{2001PASJ...53..517T}. The 7.3 km~s$^{-1}$ peak lies outside the velocity range of the 1612 MHz OH maser. As a result, this object has been classified as a WF candidate by \citet{2024ApJS..270...13F}. Additionally, \citet{1996A&AS..116..117E} reported multiple peaks between 44 km~s$^{-1}$ and 70 km~s$^{-1}$ in the $\rm H_{2}O$ maser spectrum. The profile resembles that of the known WF, IRAS 18286--0959 \citep{2011ApJ...741...94Y}. SiO \textit{v} = 1 and \textit{v} = 2, \textit{J} = 1$-$0 maser lines were detected by \citet{2003PASJ...55..203N}, with both lines showing a single peak at around 57 km~s$^{-1}$ and a large linewidth of approximately 20 km~s$^{-1}$, but with a weak second peak  on the redshifted side of the main peak by about 5--6 km~s$^{-1}$. Such broad linewidths are typically observed in RSGs \citep{2010PASJ...62..391D, 2012A&A...541A..36V}. The object was classified as an RSG star by \citet{1996A&AS..116..117E,2021MNRAS.505.6051J}.

\subsection{IRAS 19178+1206}\label{sec: IRAS 19178+1206} According to available data, the satellite line exhibits two peaks at 45.9 km~s$^{-1}$ and 49.6 km~s$^{-1}$. However, only one velocity component at 45.9 km~s$^{-1}$ is evident in the spectrum reported by \citet{1993ApJS...89..189C}. For the main line (1667 MHz), \citet{1997ApJS..109..489L} observed two peaks at 25.3 km~s$^{-1}$ and 56.3 km~s$^{-1}$, which deviate from the 45.9 km~s$^{-1}$ peak of the satellite line. No previous $\rm H_{2}O$ maser spectra have been reported. In the SiO \textit{v} = 2, \textit{J} = 1$-$0 maser line, a clear peak appears at around 40 km~s$^{-1}$, while the SiO \textit{v} = 1, \textit{J} = 1$-$0 maser line exhibits no distinct peaks, as reported by \citet{2003PASJ...55..229N}.

\subsection{IRAS 19229+1708}\label{sec: IRAS 19229+1708} The satellite line spectrum exhibits two peaks at 14.6 km~s$^{-1}$ and 18.7 km~s$^{-1}$, based on available data. However, only one peak, at around 15 km~s$^{-1}$, was detected in the spectrum reported by \citet{1990ApJ...362..634L,2019A&A...628A..90B}. The main line (1667 MHz) spectrum displays two emission peaks at 23.1 km~s$^{-1}$ and 27.9 km~s$^{-1}$ \citep{1997ApJS..109..489L}. The $\rm H_{2}O$ maser spectrum reveals three distinct peaks at approximately 30 km~s$^{-1}$, 40 km~s$^{-1}$, and 50 km~s$^{-1}$, as documented by \citet{2001PASJ...53..517T}. Additionally, \citet{1996A&AS..116..117E} observed a series of peaks between 28 km~s$^{-1}$ and 49 km~s$^{-1}$, a finding corroborated by the studies of \citet{2001A&A...368..845V} and \citet{2013ApJ...769...20Y}. The velocity range of the  $\rm H_{2}O$ maser lies outside the range of the satellite line. The detection of the SiO \textit{v} = 1 and \textit{v} = 2, \textit{J} = 1$-$0 maser lines were reported by \citet{2003PASJ...55..203N}. The $v=1$ line shows three peaks within a velocity width of about 20 km~s$^{-1}$ centered around 40 km~s$^{-1}$. The $v=2$ line is also detected at a similar velocity, but the emission intensity is relatively weaker, and the fine structure is not resolved.

\subsection{IRAS 19271+1354}\label{sec: IRAS 19271+1354} The satellite line spectrum reveals a single peak at around 45 km~s$^{-1}$, as reported by \citet{1993ApJS...89..189C}. The main line (1665 MHz) spectrum exhibits two peaks, around 45 km~s$^{-1}$ and 65 km~s$^{-1}$, with the profile predominantly showing one peak near 65 km~s$^{-1}$ \citep{1997ApJS..109..489L}. The 65 km~s$^{-1}$ components in the two main lines deviate significantly from the satellite line peak. The  $\rm H_{2}O$ maser spectrum, documented by \citet{1996A&AS..116..117E}, shows three peaks at approximately 50 km~s$^{-1}$, 52 km~s$^{-1}$, and 55 km~s$^{-1}$. In contrast, the $\rm H_{2}O$ maser spectrum reported by \citet{2013ApJ...769...20Y} shows two peaks at about 55 km~s$^{-1}$ and 68 km~s$^{-1}$. Furthermore, \citet{2001A&A...368..845V,2017ApJS..232...13C} documented a strong peak at around 54 km~s$^{-1}$. These peaks fall outside the velocity range of the satellite line. SiO \textit{v} = 1 and \textit{v} = 2, \textit{J} = 1$-$0 maser lines were detected by \citet{2003PASJ...55..203N,2017ApJS..232...13C}, both exhibiting single peaks around 52 km~s$^{-1}$. \citet{2013ApJ...769...20Y} suggested that this object might have begun developing asymmetry within the inner envelope, but it was not classified as a WF candidate due to the detection of only a single OH satellite line peak.

\subsection{IRAS 19319+2214}\label{sec: IRAS 19319+2214} The satellite line shows two peaks at approximately 10 km~s$^{-1}$ and 30 km~s$^{-1}$, with two symmetric peaks around 20 km~s$^{-1}$ between them \citep{1990ApJ...362..634L,2004AJ....127..501L}. The main line (1667 MHz) exhibits two peaks at 11.1 km~s$^{-1}$ and 37.5 km~s$^{-1}$, while the main line (1665 MHz) shows a single peak around 10 km~s$^{-1}$, within the velocity range of the satellite line \citep{2004AJ....127..501L}. The  $\rm H_{2}O$ maser line shows three peaks at around 10 km~s$^{-1}$, 15 km~s$^{-1}$, and 20 km~s$^{-1}$ \citep{1996A&AS..116..117E}. However, \citet{2015A&A...578A.119G} reported a spectrum with multiple peaks ranging from 12.5 km~s$^{-1}$ to 32.5 km~s$^{-1}$, outside the velocity range of the satellite line, though the red-shifted deviation is minor (2.5 km~s$^{-1}$). This object is also classified as a WF candidate by \citet{2024ApJS..270...13F}, and its $\rm H_{2}O$ maser profile is similar to that of the known WF, IRAS 18286--0959 \citep{2011ApJ...741...94Y}. No SiO maser spectra have been reported. the satellite lines, discovered by \citet{1990ApJ...362..634L} and \citet{2004AJ....127..501L}, exhibit a unique profile that differs from typical OH/IR stars, strongly suggesting the existence of two shells in the CSE, as proposed by \citet{2004AJ....127..501L}. It is likely that IRAS 19319+2214 is in a common envelope (CE) phase, which is associated with the formation of WFs. For example, \citet{2018MNRAS.480.4991G} suggested that IRAS 15103$-$5754, a known WF PN and a triple stellar system, underwent a CE phase during its evolution. This object has been classified as a post-AGB star based on maser and IR observations \citep{2009A&A...501.1207R,2012A&A...545A..20R,2021MNRAS.505.6051J}, with near-infrared observations by \citet{2012A&A...545A..20R} aligning with the slight brightness increase observed by \citet{2021MNRAS.505.6051J} after 2002.

\subsection{IRAS 19344+0016}\label{sec: IRAS 19344+0016} The satellite line spectrum shows two peaks at approximately 128 km~s$^{-1}$ and 145 km~s$^{-1}$ \citep{1988ApJS...66..183E}. The main line (1665 MHz) spectrum shows two peaks at 147.4 km~s$^{-1}$ and 155.1 km~s$^{-1}$, and the main line (1667 MHz) displays two peaks at 144.9 km~s$^{-1}$ and 145.8 km~s$^{-1}$ \citep{1997ApJS..109..489L}. The  $\rm H_{2}O$ maser line has a single peak at around 133 km~s$^{-1}$ \citep{1996A&AS..116..117E}, which falls within the velocity range of the satellite line. SiO \textit{v} = 1 and \textit{v} = 2, \textit{J} = 1$-$0 maser lines have been detected by \citet{2017ApJS..232...13C}, with both profiles showing single peaks at around 135 km~s$^{-1}$.

\subsection{IRAS 19386+0155}\label{sec: IRAS 19386+0155} The satellite line shows multiple peaks within the velocity range of approximately 4 km~s$^{-1}$ to 42 km~s$^{-1}$ \citep{2000ApJ...533..959L}. 
Similarly, the main lines exhibit multiple peaks in the velocity ranges, from approximately 4 km~s$^{-1}$ to 41 km~s$^{-1}$ in the 1665 MHz and 8 km~s$^{-1}$ to 40 km~s$^{-1}$ in the 1667 MHz spectra, respectively \citep{2000ApJ...533..959L}. This source was classified as a post-AGB star by \citet{2006A&A...458..173S}. No $\rm H_{2}O$ or SiO maser spectra have been reported previously.

\subsection{IRAS 19566+3423}\label{sec: IRAS 19566+3423} the satellite line spectrum displays multiple peaks within the velocity range from $-50.3$ km~s$^{-1}$ to $-37.2$ km~s$^{-1}$, as observed by \citet{1994ApJS...93..549L}. It also has two prominent peaks at  $-45.4$ km~s$^{-1}$ and $-41.2$ km~s$^{-1}$ \citep{1989ApJ...344..350L}. The main line (1665 MHz and 1667 MHz) spectra show several peaks, spanning from $-54.5$ km~s$^{-1}$ to $-32.8$ km~s$^{-1}$ and $-55.7$ km~s$^{-1}$ to $-39.9$ km~s$^{-1}$, respectively \citep{1997ApJS..109..489L}. The outermost blue-shifted component of the main line (1667 MHz) deviates slightly from the satellite line by more than 5 km~s$^{-1}$. The $\rm H_{2}O$ maser emission shows a strong single peak at $-39.0$ km~s$^{-1}$ \citep{2001PASJ...53..517T}, while \citet{1996A&AS..116..117E} reported two peaks at $-45.0$ km~s$^{-1}$ and $-41.0$ km~s$^{-1}$. The $\rm H_{2}O$ maser spectra do not cover a larger velocity range than the satellite line. No SiO maser spectra have been detected previously. This object is classified as a post-AGB star, specifically a protoplanetary nebula, as described by \citet{2012ApJS..203...16S}. Its classification is supported by its redder near/mid-IR color. Notably, it has experienced a significant brightness increase of 0.43 mag per year over a period of about 7 years, likely due to the dilution of its CSE  \citep{2021MNRAS.505.6051J}, making it an excellent candidate for follow-up observations to study the AGB to post-AGB transition.

\subsection{IRAS 20127+2430}\label{sec: IRAS 20127+2430} The satellite line spectrum shows a single peak at 13.7 km~s$^{-1}$, as observed by \citet{1988ApJS...66..183E}, while \citet{1993ApJS...89..189C} detected two peaks, one at 13.7 km~s$^{-1}$ and another faint peak at 39.1 km~s$^{-1}$. The main line (1665 MHz) spectrum shows two peaks at 8.7 km~s$^{-1}$ and 19.7 km~s$^{-1}$ \citep{1997ApJS..109..489L}, while the 1667 MHz spectrum reveals three peaks at 8.3 km~s$^{-1}$, 12 km~s$^{-1}$, and 19.7 km~s$^{-1}$. The outermost blue-shifted component of the main line (1667 MHz) deviates by more than 5 km~s$^{-1}$. The  $\rm H_{2}O$ maser emission, as detailed by \citet{2017ApJS..232...13C}, displays two peaks at approximately 11 km~s$^{-1}$ and 15 km~s$^{-1}$, with the blue-shifted end deviating by 2.7 km~s$^{-1}$ from the velocity range of the satellite line. No SiO maser spectra have been reported to date.

\subsection{IRAS 20165+3413}\label{sec: IRAS 20165+3413} The satellite line spectrum exhibits several peaks within the velocity range of 13.1 km~s$^{-1}$ to 15 km~s$^{-1}$ \citep{1994ApJS...93..549L}. However, the main line (1665 MHz) spectrum lacks noticeable peaks, while the main line (1667 MHz) spectrum shows two peaks at 1.5 km~s$^{-1}$ and 6.6 km~s$^{-1}$ \citep{1994ApJS...93..549L}. The outermost blue-shifted component in the main line (1667 MHz) clearly deviates from the satellite line. No $\rm H_{2}O$ and SiO maser spectra have been reported previously.

\subsection{IRAS 20460+3253}\label{sec: IRAS 20460+3253} The satellite line spectrum exhibits two peaks at around $-10$ km~s$^{-1}$ and $-8$ km~s$^{-1}$ \citep{1990ApJ...362..634L}, while the main line (1665 MHz) spectrum shows multiple peaks ranging from $-8.8$ km~s$^{-1}$ to $-2.5$ km~s$^{-1}$, similar to the profile observed in the main line (1667 MHz) \citep{1997ApJS..109..489L}. The $-2.5$ km~s$^{-1}$ components in both the two main lines deviate noticeably from the peak in the satellite line. The  $\rm H_{2}O$ maser line displays two peaks at approximately $-8$ km~s$^{-1}$ and $-3$ km~s$^{-1}$, outside the velocity range of the satellite line emission line \citep{1996A&AS..116..117E}. The red-shifted end of the $\rm H_{2}O$ maser line deviates by 5.5 km~s$^{-1}$ from the velocity range of the OH maser line. The SiO \textit{v} = 1, \textit{J} = 1$-$0 maser line exhibits a single peak at $-4.8$ km~s$^{-1}$, which is consistent with the SiO \textit{v} = 2, \textit{J} = 1$-$0 maser line, as reported by \citet{2017ApJS..232...13C}.

\subsection{IRAS 22180+3225}\label{sec: IRAS 22180+3225} The satellite line spectrum displays multiple peaks ranging from $-22.5$ km~s$^{-1}$ to $-12.6$ km~s$^{-1}$, as documented by \citet{1993ApJS...89..189C}. Meanwhile, the main line (1665 MHz) spectrum shows two distinct peaks at $-20.3$ km~s$^{-1}$ and $-17.6$ km~s$^{-1}$ \citep{1997ApJS..109..489L}, within the velocity range of the satellite line. Additionally, the main line (1667 MHz) spectrum reveals three peaks at approximately $-33$ km~s$^{-1}$, $-20$ km~s$^{-1}$, and $-13$ km~s$^{-1}$ \citep{1997ApJS..109..489L}. The outermost blue-shifted component in the main line (1667 MHz) clearly deviates from the satellite line. The  $\rm H_{2}O$ maser line shows a solitary peak at $-17.5$ km~s$^{-1}$ \citep{1996A&AS..116..117E}, confirmed by \citet{2013ApJ...769...20Y}. Similarly, the SiO \textit{v} = 1, \textit{J} = 1$-$0 maser line presents a single peak at $-18$ km~s$^{-1}$ \citep{2017ApJS..232...13C}.

\subsection{IRAS 22525+6033}\label{sec: IRAS 22525+6033} The satellite line spectrum shows two emission peaks at $-65.1$ km~s$^{-1}$ and $-56.4$ km~s$^{-1}$ \citep{1990A&A...233..112S}. The 1665 MHz spectrum displays multiple peaks ranging from $-62.5$ km~s$^{-1}$ to $-46$ km~s$^{-1}$ \citep{1990A&A...233..112S, 2004A&A...420..217E}, resembling the profile of the main line (1667 MHz), as noted by \citet{2004A&A...420..217E}. The $-46$ km~s$^{-1}$ components in both the two main lines deviate significantly from the peak in the satellite line. The  $\rm H_{2}O$ maser line shows multiple peaks ranging from approximately $-65$ km~s$^{-1}$ to $-45$ km~s$^{-1}$ \citep{2013ApJ...769...20Y,1990A&AS...84..179C,2010ApJS..188..209K}, outside the velocity range of the satellite line emission line. The SiO \textit{v} = 1, \textit{J} = 1$-$0 maser exhibits a single peak at $-56$ km~s$^{-1}$ \citep{1990A&A...231..431A}, consistent with observations by \citet{2012A&A...541A..36V}.

\subsection{IRAS 22556+5833}\label{sec: IRAS 22556+5833} The satellite line shows a typical double-peaked profile with velocities at $-60$ km~s$^{-1}$ and $-45.0$ km~s$^{-1}$, as reported by \citet{2017ARep...61...16A}. The main line (1665 MHz) spectrum exhibits multiple peaks ranging from $-58.0$ km~s$^{-1}$ to $-45$ km~s$^{-1}$, while the main line (1667 MHz) profile also displays several features spanning from $-55.0$ km~s$^{-1}$ to $-47$ km~s$^{-1}$ \citep{2017ARep...61...16A}. The  $\rm H_{2}O$ maser line has a complex shape with multiple peaks within the velocity range of $-45.0$ km~s$^{-1}$ to $-46.0$ km~s$^{-1}$ \citep{2010ApJS..188..209K}. \citet{2017ARep...61...16A} also detected many peaks spanning from $-60.0$ km~s$^{-1}$ to $-40.0$ km~s$^{-1}$. Both the SiO \textit{v} = 1 and \textit{v} = 2, \textit{J} = 1$-$0 maser lines show a single peak at $-55.0$ km~s$^{-1}$ \citep{2010ApJS..188..209K}.

\section{Infrared photometric data for known WFs and RNRs}\label{appendix: Infrared photometric data for known WFs and RNRs}

\begin{figure}
\centering
\includegraphics[width=0.8\textwidth, angle=0]{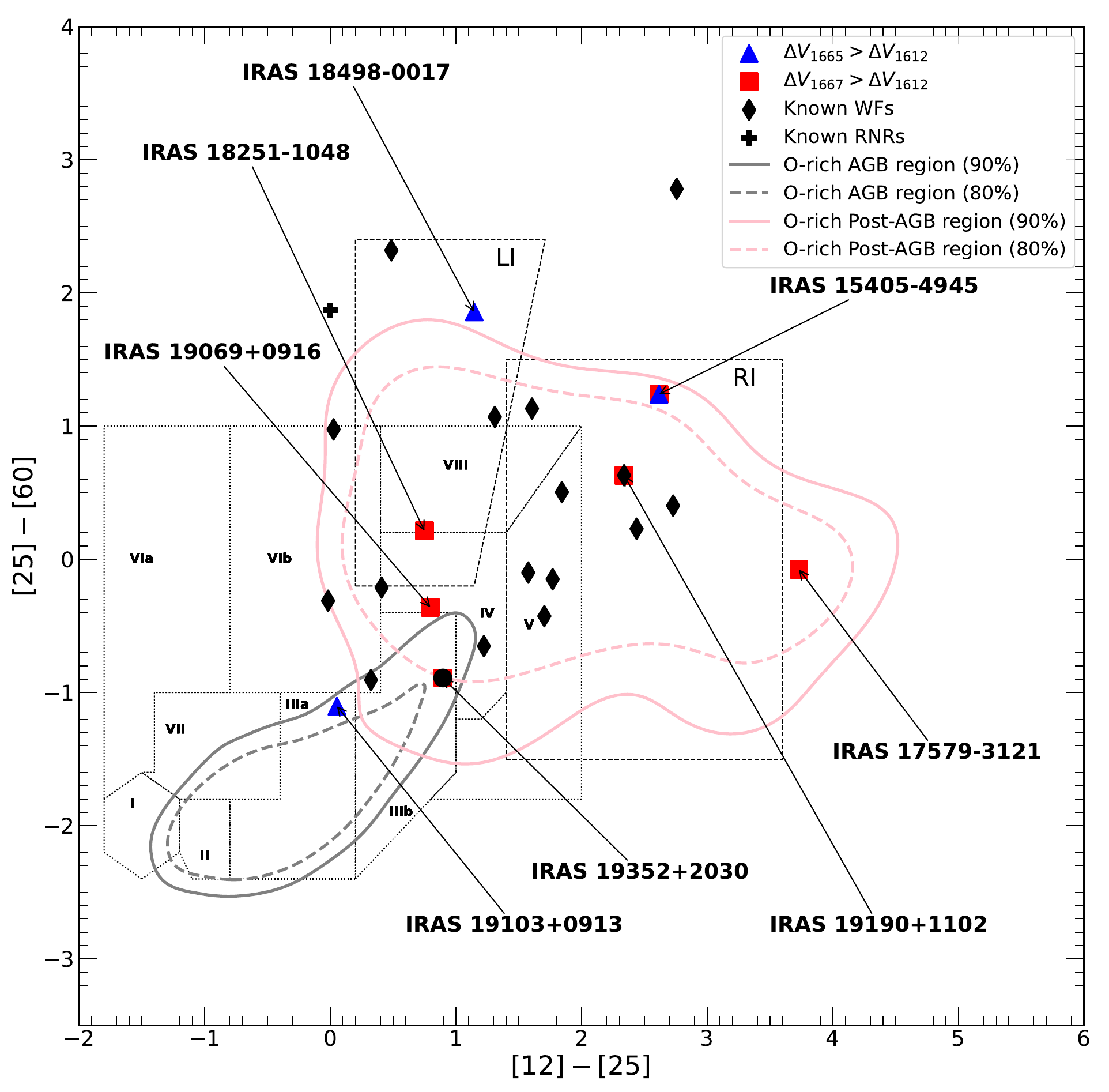}
\caption{Comparison of the IRAS colors of the eight sources with detected velocity excess and the known WFs and RNRs. The known WFs are represented by the black diamonds, and the known RNRs are represented by the black crosses. All other notations are consistent with those in Figure~\ref{Fig: [New_IRAS_two_color_diagram_of_AGB_and_postAGB]}.}
\label{Fig: [New_IRAS_two_color_diagram_of_known_WFs_RNRs]}
\end{figure}

\begin{figure}
\centering
\includegraphics[width=0.8\textwidth, angle=0]{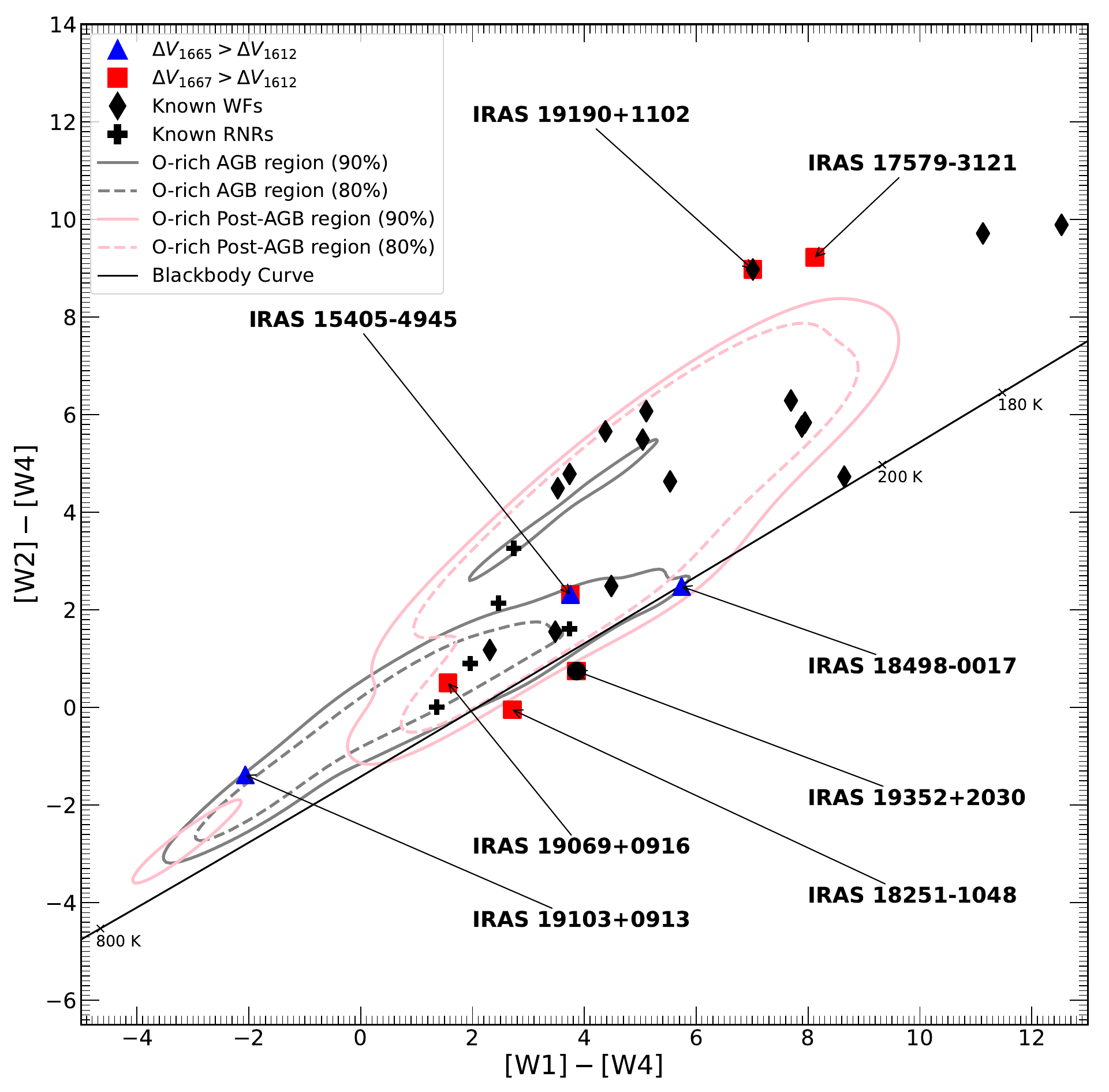}
\caption{Comparison of the WISE colors of the eight sources with detected velocity excess and the known WFs and RNRs. The known WFs are represented by the black diamonds, and the known RNRs are represented by the black crosses. All other notations are consistent with those in Figure~\ref{Fig: [New_IRAS_two_color_diagram_of_AGB_and_postAGB]}.}
\label{Fig: [New_WISE_two_color_diagram_of_known_WFs_RNRs]}
\end{figure}

\begin{figure}
\centering
\includegraphics[width=0.8\textwidth, angle=0]{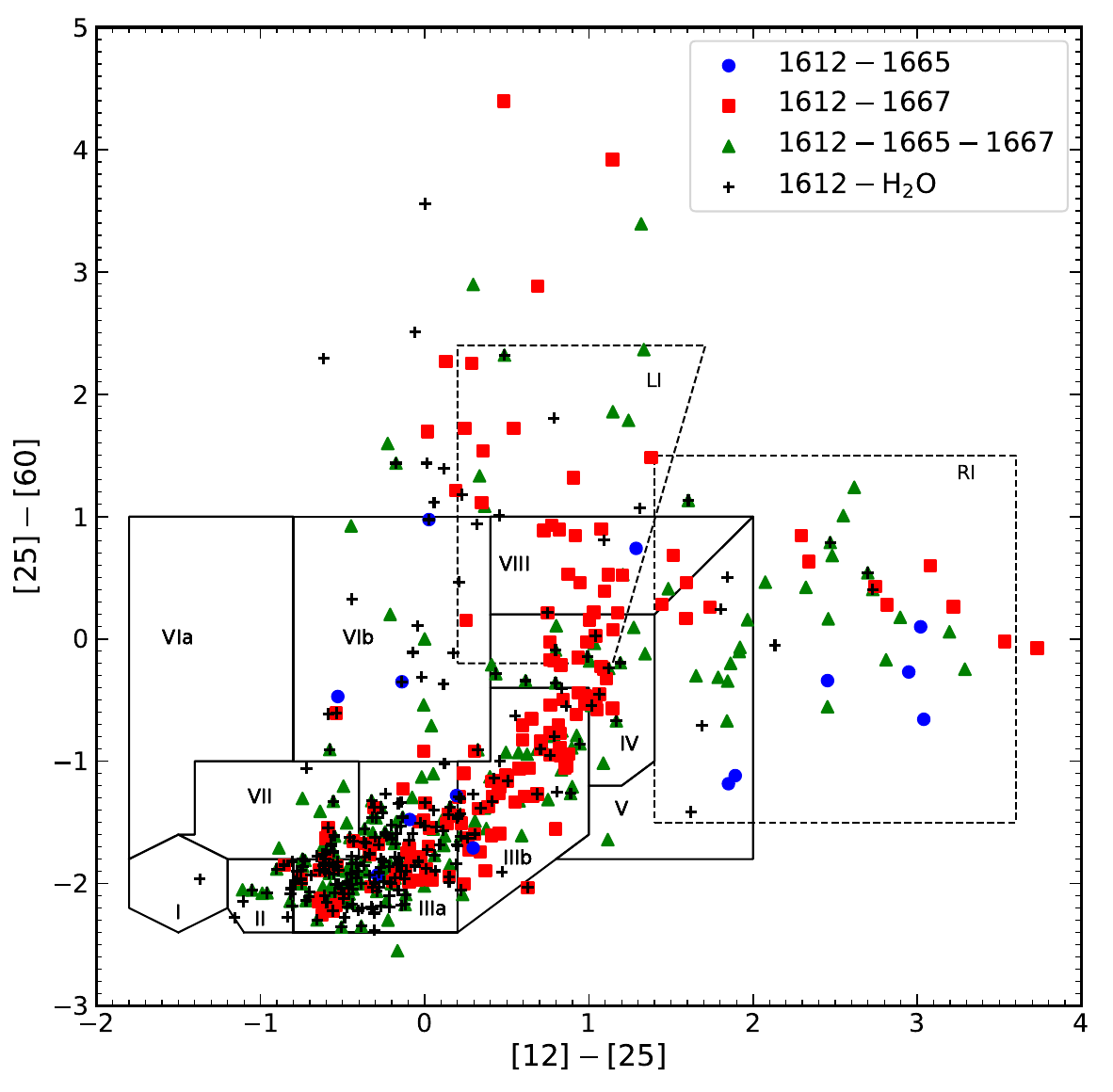}
\caption{IRAS two-color diagram showing the distribution of sources where the line combinations indicated in the legend were detected simultaneously. '1612--1665' (blue circles) represents sources where both the satellite line and main line (1665 MHz) were detected simultaneously (with no detection of the 1667 MHz main line). '1612--1667' (red squares) represents sources where both the satellite line MHz and main line (1667 MHz) were detected simultaneously (with no detection of the 1665 MHz main line). '1612--1665--1667 pair' (green triangles) represents sources where all three OH lines were detected. '1612--$\rm H_{2}O$'  (black crosses) represents sources where both the satellite line and the H$_2$O maser were detected. The color regions are the same as in Figure~\ref{Fig: [New_IRAS_two_color_diagram_of_AGB_and_postAGB]}.}
\label{Fig: [OH maser distribution]}
\end{figure}

\begin{figure}
\centering
\includegraphics[width=0.8\textwidth, angle=0]{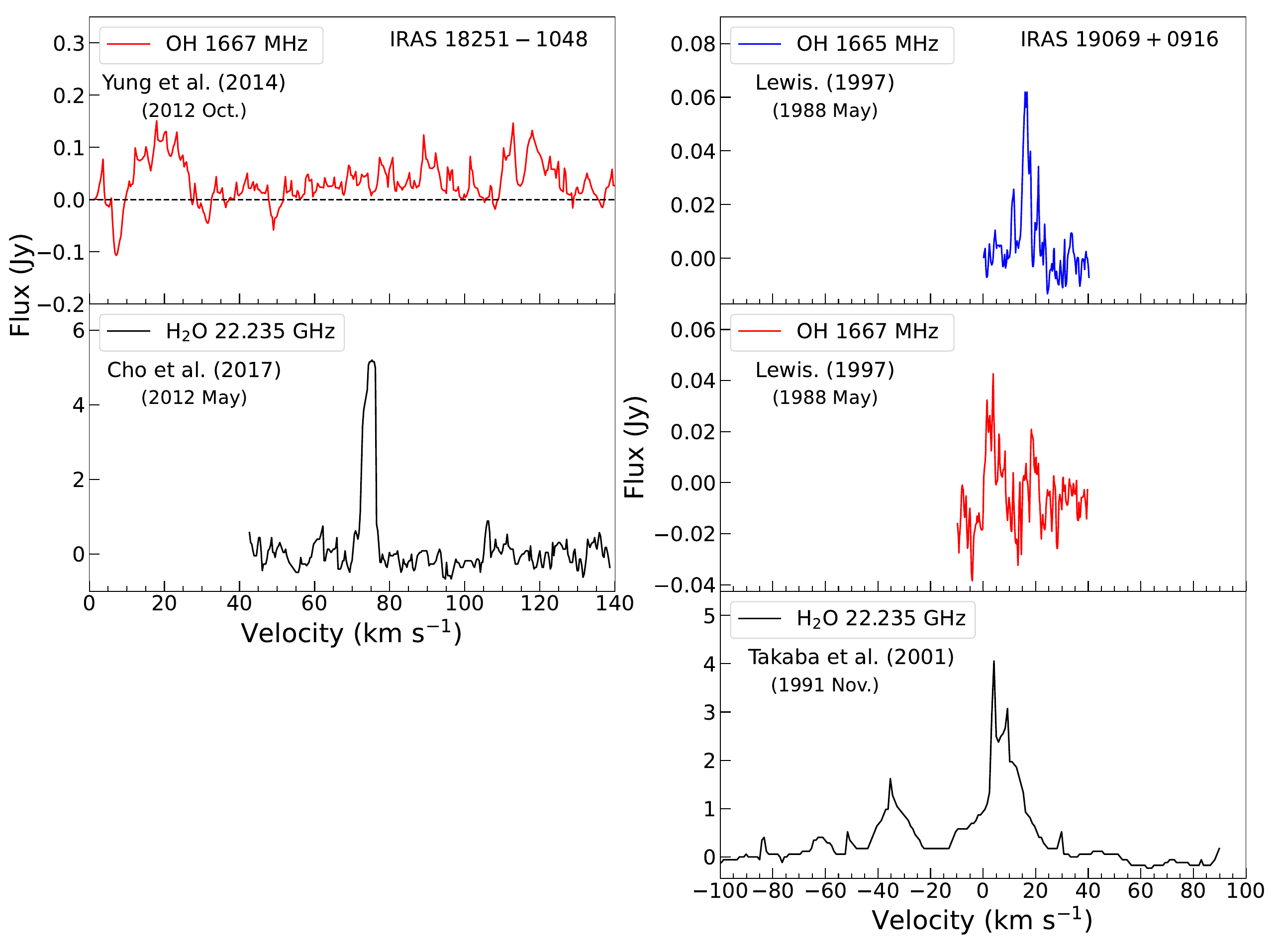}
\caption{Left panel: Spectra of the main line (1667 MHz) and  H$_2$O maser line for IRAS 18251--1048. The main line (1667 MHz) spectrum is cited from \citet{2014ApJ...794...81Y}, and the H$_2$O spectrum is cited from \citet{2017ApJS..232...13C}. Right panel: Spectra of the two main lines and the H$_2$O line for IRAS 19069+0916. The main line spectra are cited from \citet{1997ApJS..109..489L}, while the H$_2$O spectrum is cited from \citet{2001PASJ...53..517T}. The references for the spectra and observation times are provided below the legend.}
\label{Fig: Spectra_IRAS18251-1048_and_IRAS19069+0916}
\end{figure}

\begin{figure}
\centering
\includegraphics[width=0.7\textwidth, angle=0]{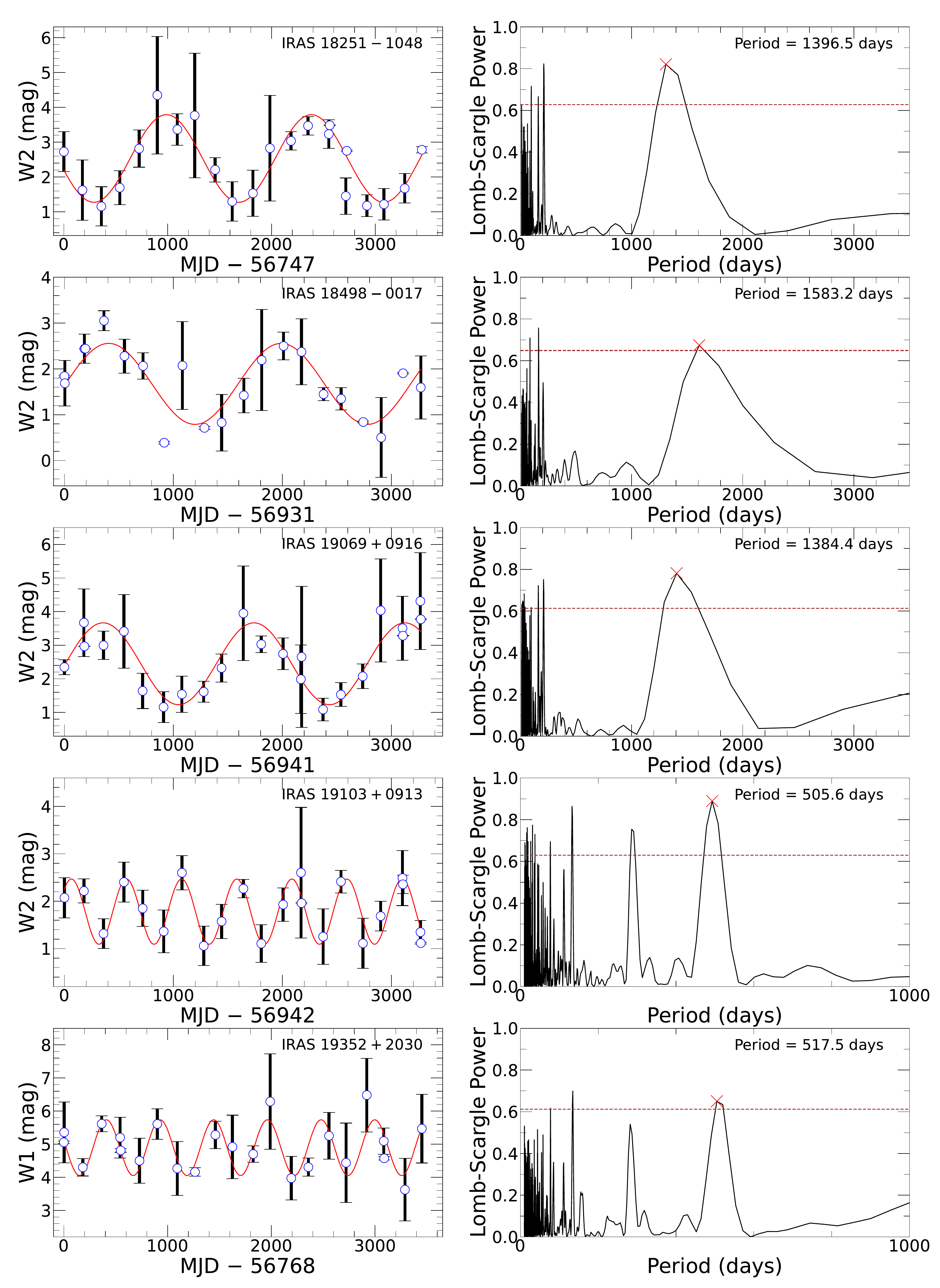}
\caption{WISE light curves (left panel) and periodic analysis results (right panel) for the sources listed in Table~\ref{Tab: [Sources showing true velocity deviations in the OH 1665/1667 MHz line]} for which periodicity was detected. The red line in each WISE light curve represents the best-fit sinusoid based on the Lomb–Scargle periodogram. The corresponding fit results are: --1.26sin(2$\pi$(ft+0.04))+2.53, 0.88sin(2$\pi$(ft--0.01))+1.67, 1.22sin(2$\pi$(ft--0.01))+2.45, 0.69sin(2$\pi$(ft+0.12))+1.78, and --0.69sin(2$\pi$(ft--0.05))+1.46, where $f$ is the frequency equal to 1/period. The black vertical lines represent the error bars (see the main text for details). In the Lomb–Scargle periodogram, the dashed brown horizontal line indicates the periodogram level corresponding to a maximum peak false alarm probability of 2\%, and the red X mark denotes the selected peak.}
\label{Fig: [WISE light curves with periodicity]}
\end{figure}

\begin{figure}
\centering
\includegraphics[width=0.4\textwidth, angle=0]{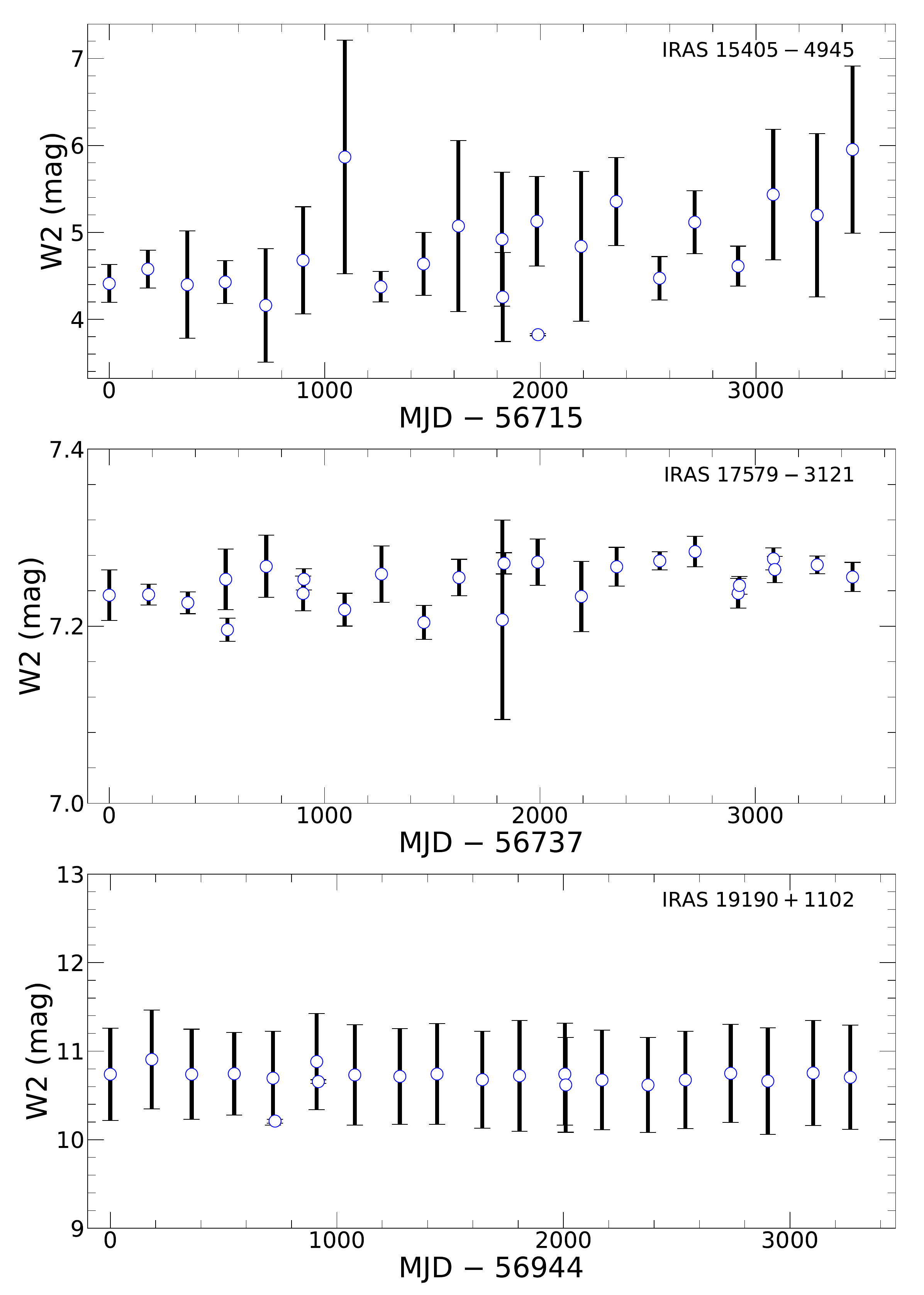}
\caption{WISE light curves for the sources listed in Table~\ref{Tab: [Sources showing true velocity deviations in the OH 1665/1667 MHz line]} for which no periodicity was detected.}
\label{Fig: [WISE light curves without periodicity}
\end{figure}

For comparison, the known WFs and RNRs are plotted on infrared two-color diagrams in Figures~\ref{Fig: [New_IRAS_two_color_diagram_of_known_WFs_RNRs]} and \ref{Fig: [New_WISE_two_color_diagram_of_known_WFs_RNRs]}. A summary of the known WFs and RNRs, along with the collected infrared photometric data, is provided in Table~\ref{Tab: [WISE, IRAS and 2MASS Infrared photometric data of the known WFs and RNRs]} for reference.


\input{WISE_Position_information_of_the_selected_sources}
\input{Velocity_information_of_the_selected_sources}

\input{WISE_and_IRAS_and_2MASS_information_of_the_selected_sources}
\input{Sources_excluded_from_further_analysis} 
\input{Eye_inspection_objects_velocity_excess}

\input{Detection_rate_of_vel_excess_sources}

\input{Known_WFs_RNRs_infrared_photometric_data_wise_iras_2mass}

\input{OH_maser_status_of_the_known_WF_samples}

\clearpage
\begin{figure}
		\centering
  		\includegraphics[width=0.8\textwidth, angle=0]{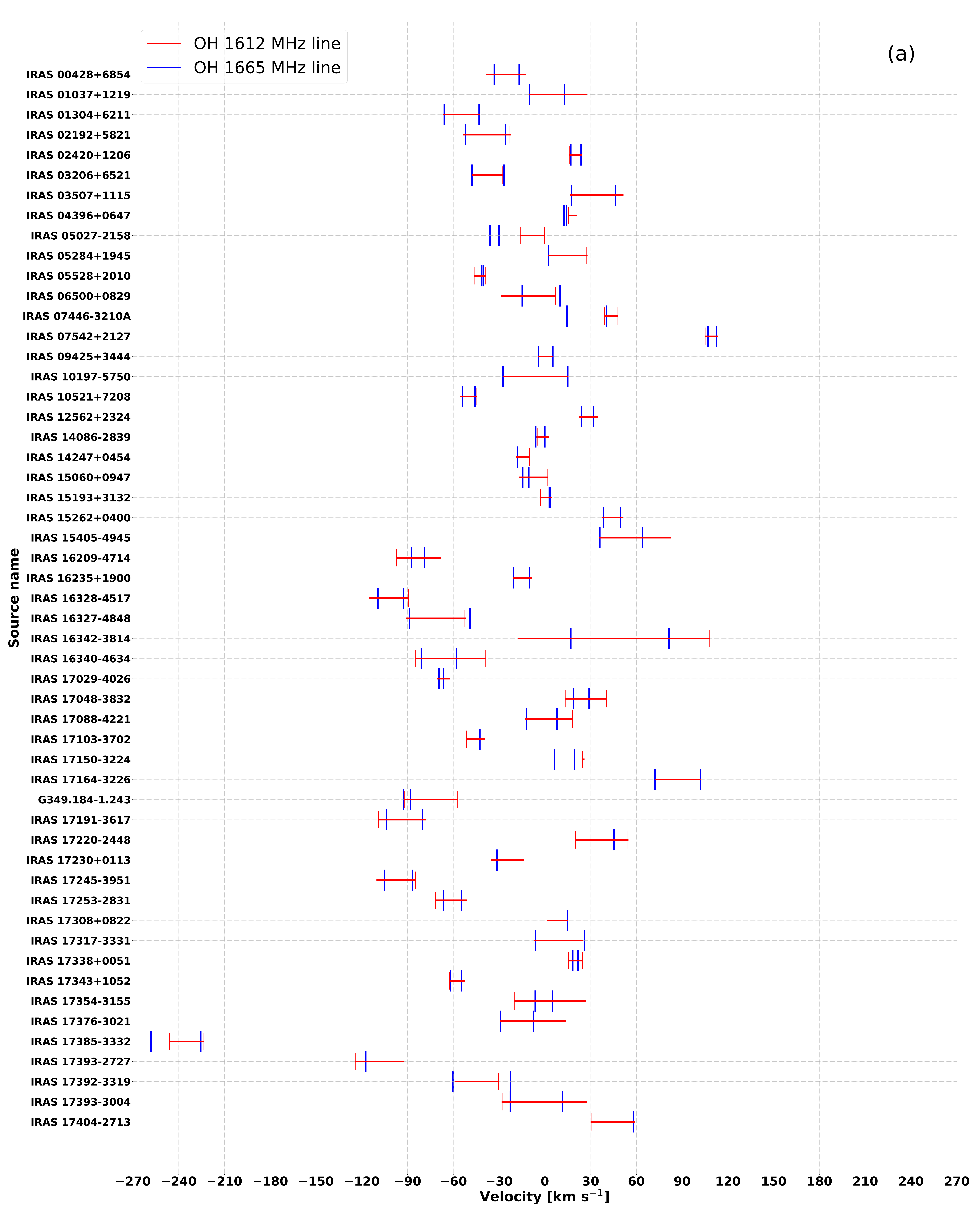}
\caption{Velocity range comparison for the 213 OH maser sources detected in both the satellite line and the main line (1665 MHz). The red line represents the velocity range of the satellite line, while the two blue vertical lines represent the velocity range of the main line (1665 MHz). The sources are arranged in ascending order of the right ascension. Since all sources cannot be displayed in a single figure, the remaining ones are shown separately in Figures~\ref{Fig: b[comparison of 1612 OH maser with that of 1665 MHz-2]}, \ref{Fig: c[comparison of 1612 OH maser with that of 1665 MHz-3]}, and \ref{Fig: d[comparison of 1612 OH maser with that of 1665 MHz-4]}.}
	\label{Fig: a[comparison of 1612 OH maser with that of 1665 MHz]}
\end{figure}


\begin{figure}
    \centering
  	\includegraphics[width=0.8\textwidth, angle=0]{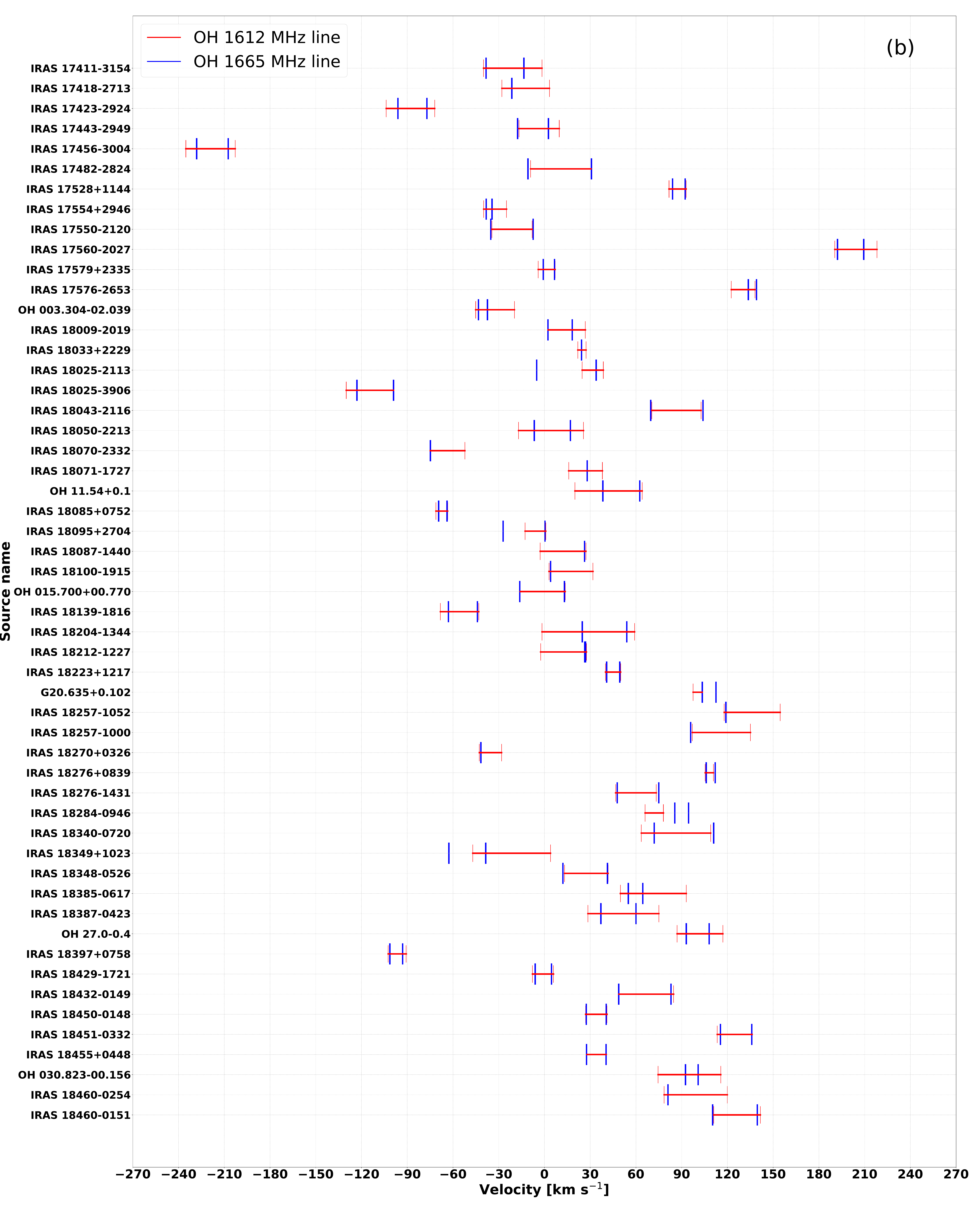}
   \caption{Continuation of Fig.\ref{Fig: a[comparison of 1612 OH maser with that of 1665 MHz]}. The notation is the same as in Fig.~\ref{Fig: a[comparison of 1612 OH maser with that of 1665 MHz]}.}\label{Fig: b[comparison of 1612 OH maser with that of 1665 MHz-2]}
\end{figure}


\begin{figure}
    \centering
  	\includegraphics[width=0.8\textwidth, angle=0]{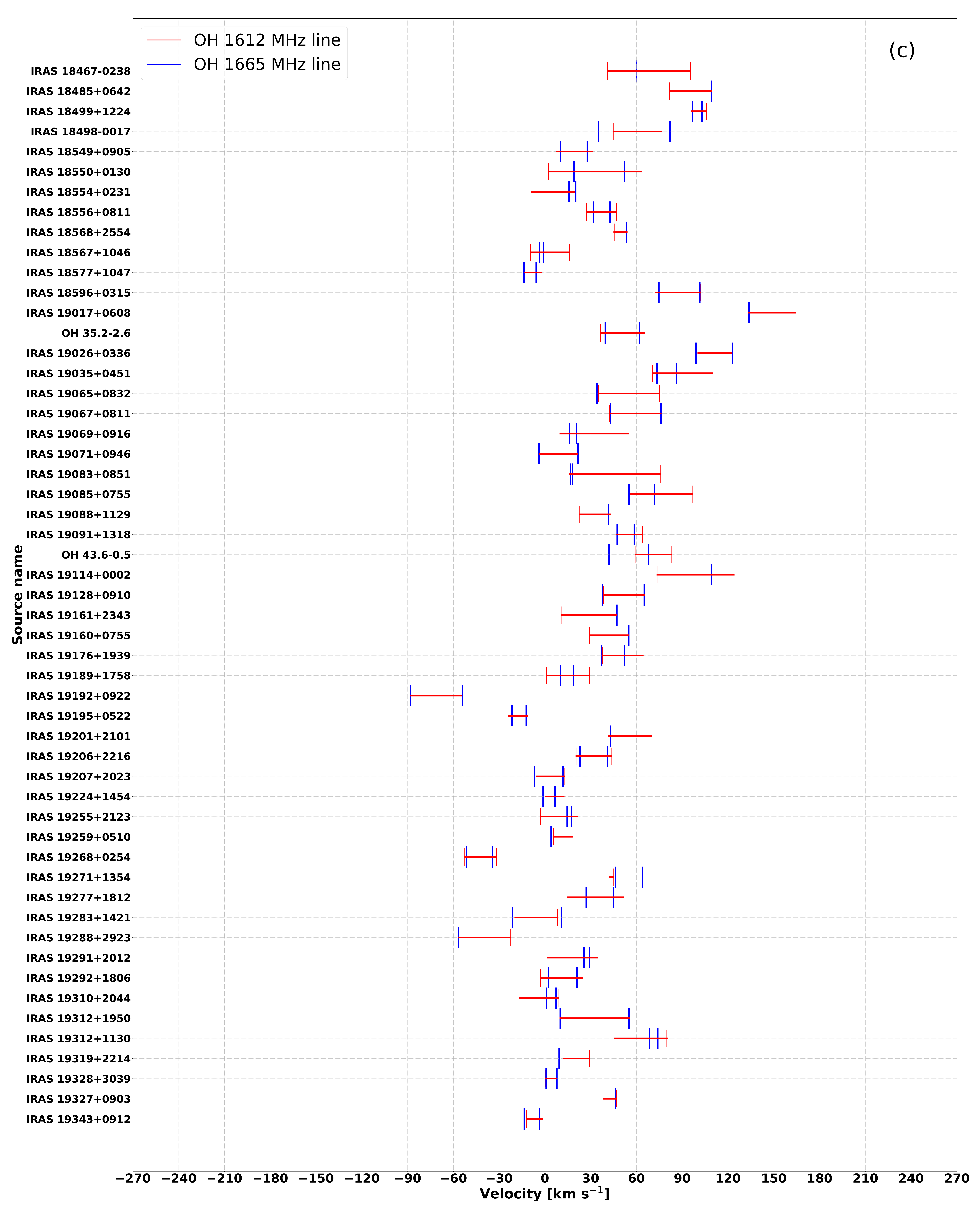}
   \caption{Continuation of Fig.~\ref{Fig: b[comparison of 1612 OH maser with that of 1665 MHz-2]}. The notation is the same as in Fig.\ref{Fig: a[comparison of 1612 OH maser with that of 1665 MHz]}.}\label{Fig: c[comparison of 1612 OH maser with that of 1665 MHz-3]}
\end{figure}


\begin{figure}
    \centering
  	\includegraphics[width=0.8\textwidth, angle=0]{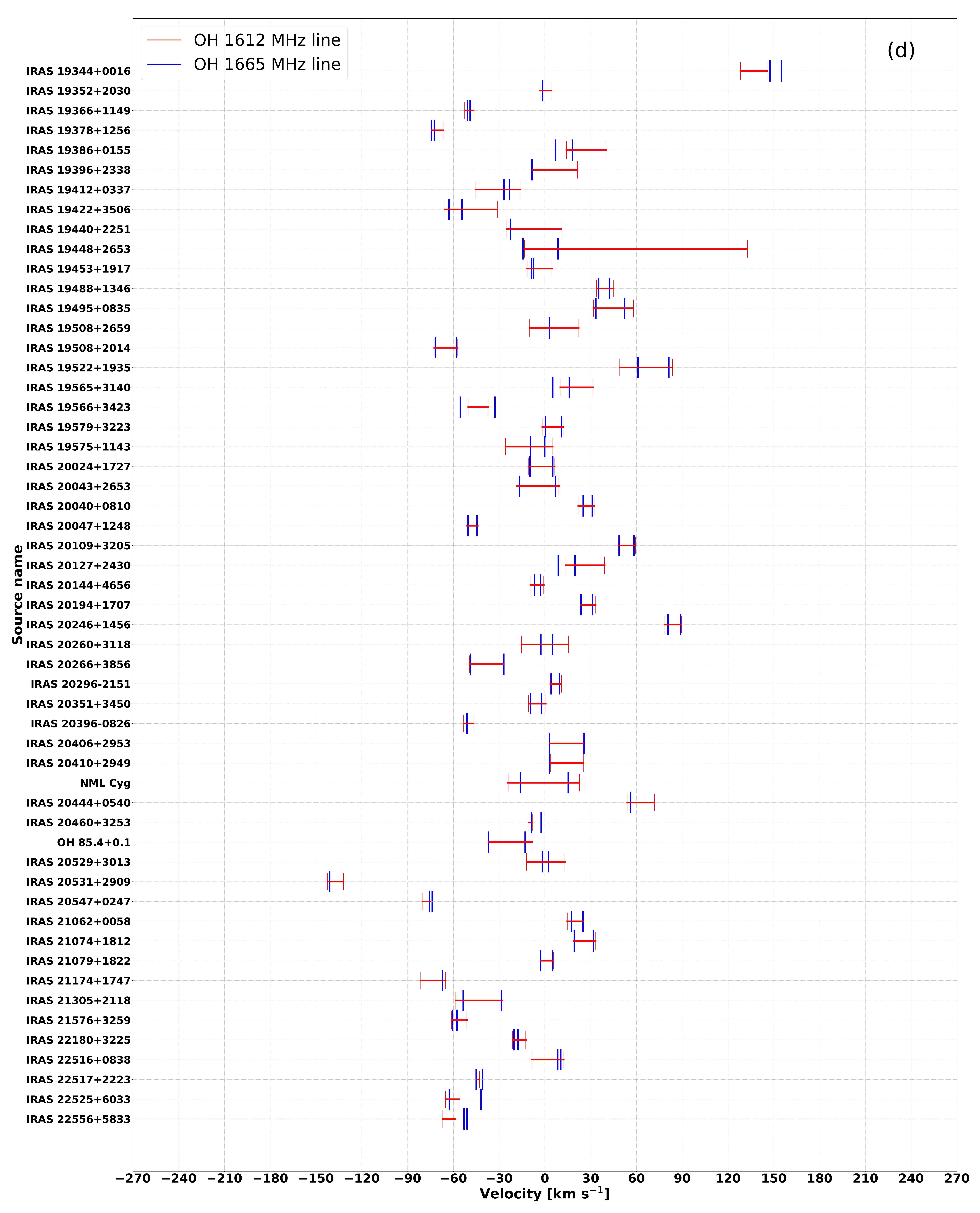}
   \caption{Continuation of Fig.~\ref{Fig: c[comparison of 1612 OH maser with that of 1665 MHz-3]}. The notation is the same as in Fig.\ref{Fig: a[comparison of 1612 OH maser with that of 1665 MHz]}.}\label{Fig: d[comparison of 1612 OH maser with that of 1665 MHz-4]}
\end{figure}

\clearpage
\FloatBarrier
 
\begin{figure}
		\centering
  		\includegraphics[width=0.8\textwidth, angle=0]{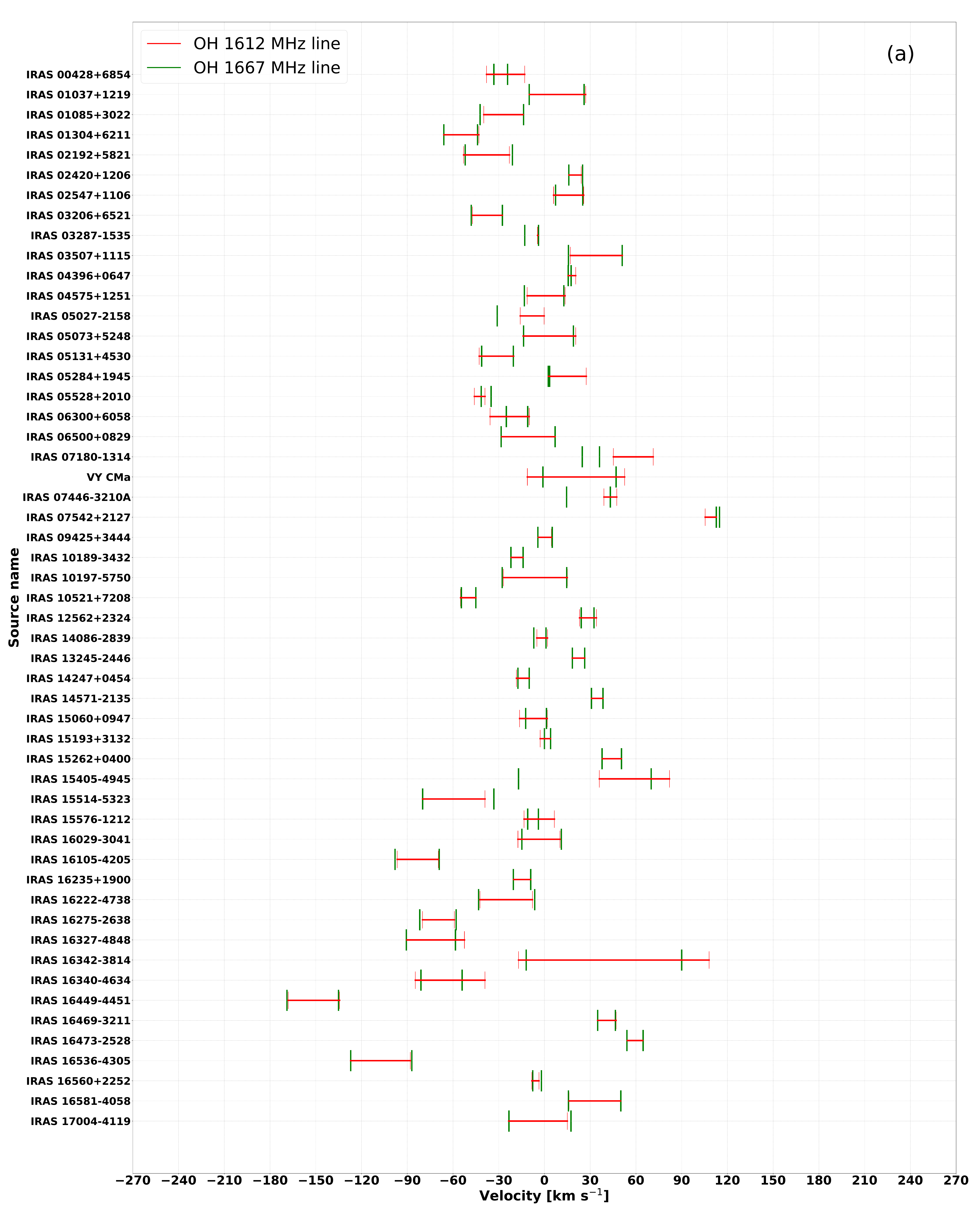}
    	\caption{Velocity range comparison for the 361 OH maser sources detected in both the satellite line and the main line (1667 MHz). The red line represents the velocity range of the satellite line, while the green vertical lines represent the velocity range of the main line (1667 MHz). The sources are arranged in ascending order of the right ascension. Since all sources cannot be displayed in a single figure, the remaining ones are shown separately in Figures~\ref{Fig: b[comparison of 1612 OH maser with that of 1667 MHz-2]}, \ref{Fig: c[comparison of 1612 OH maser with that of 1667 MHz-3]}, \ref{Fig: d[comparison of 1612 OH maser with that of 1667 MHz-4]}, \ref{Fig: e[comparison of 1612 OH maser with that of 1667 MHz-5]}, \ref{Fig: f[comparison of 1612 OH maser with that of 1667 MHz-6]}, and \ref{Fig: g[comparison of 1612 OH maser with that of 1667 MHz-7]}.}
	\label{Fig: a[comparison of 1612 OH maser with that of 1667 MHz]}
\end{figure}
\begin{figure}
    \centering
  	\includegraphics[width=0.8\textwidth, angle=0]{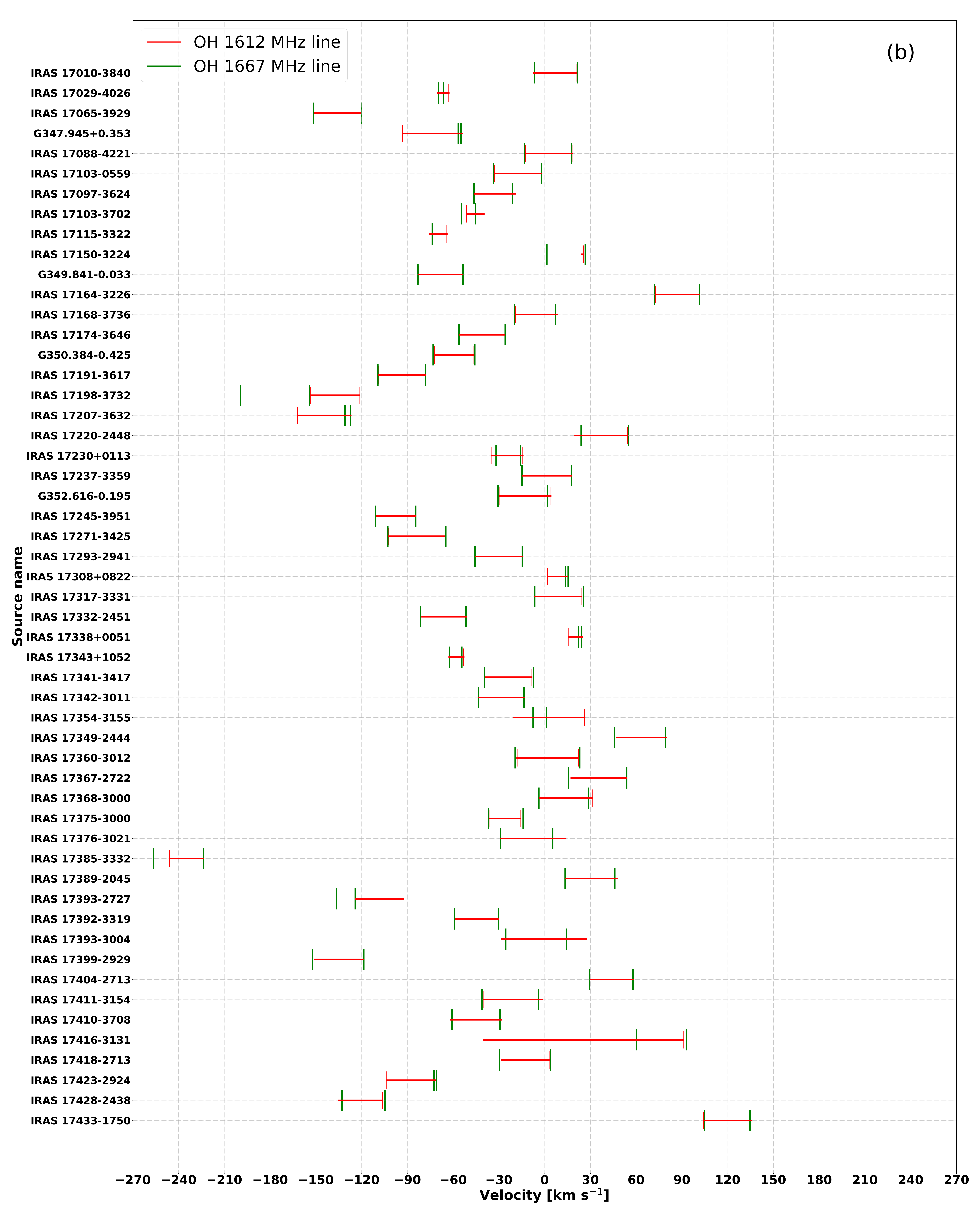}
     \caption{Continuation of Fig.~\ref{Fig: a[comparison of 1612 OH maser with that of 1667 MHz]}. The notation is the same as in Fig.\ref{Fig: a[comparison of 1612 OH maser with that of 1667 MHz]}.}\label{Fig: b[comparison of 1612 OH maser with that of 1667 MHz-2]}
\end{figure}
\begin{figure}
    \centering
  	\includegraphics[width=0.8\textwidth, angle=0]{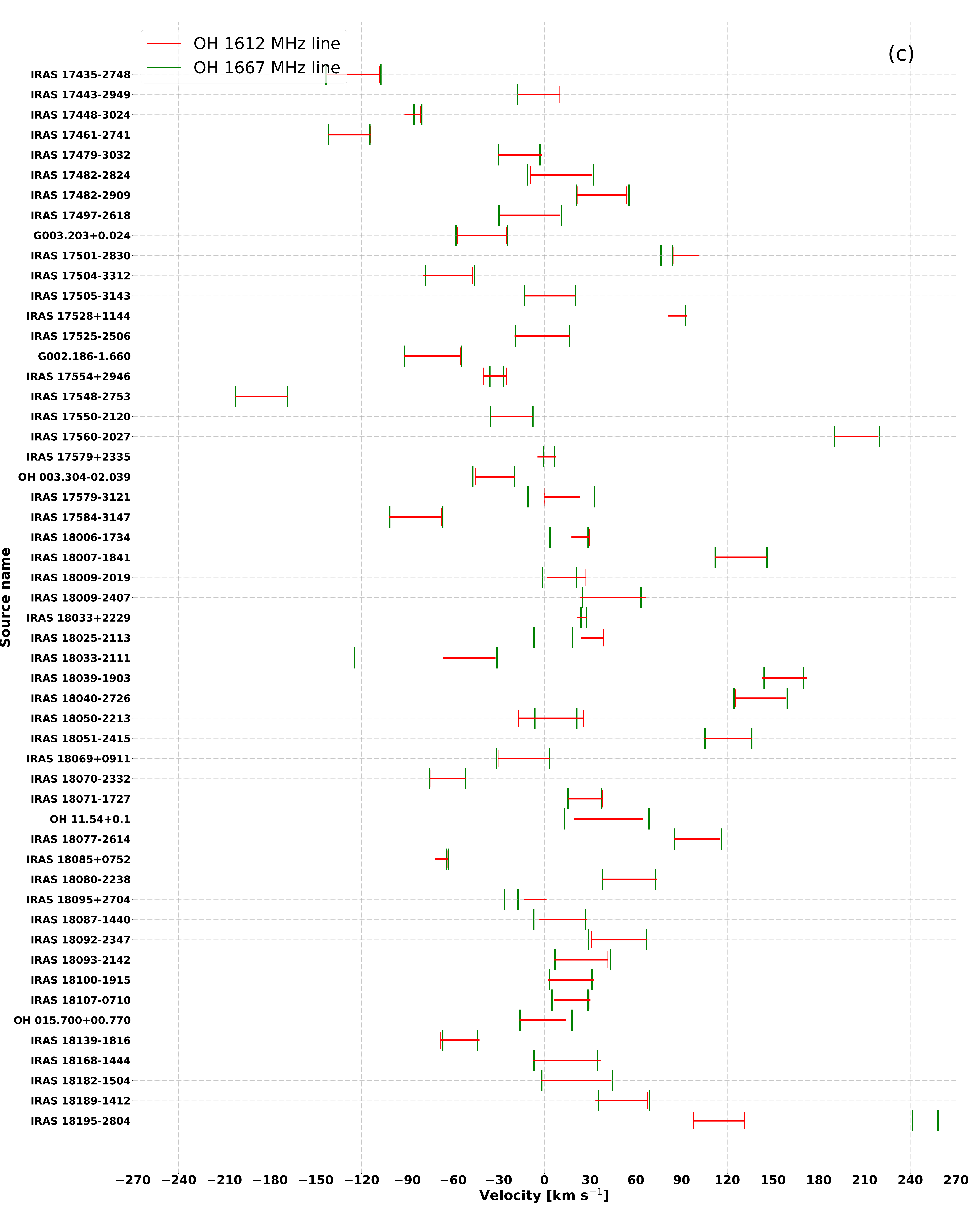}
   \caption{Continuation of Fig.~\ref{Fig: b[comparison of 1612 OH maser with that of 1667 MHz-2]}. The notation is the same as in Fig.\ref{Fig: a[comparison of 1612 OH maser with that of 1667 MHz]}.}\label{Fig: c[comparison of 1612 OH maser with that of 1667 MHz-3]}
\end{figure}
\begin{figure}
    \centering
  	\includegraphics[width=0.8\textwidth, angle=0]{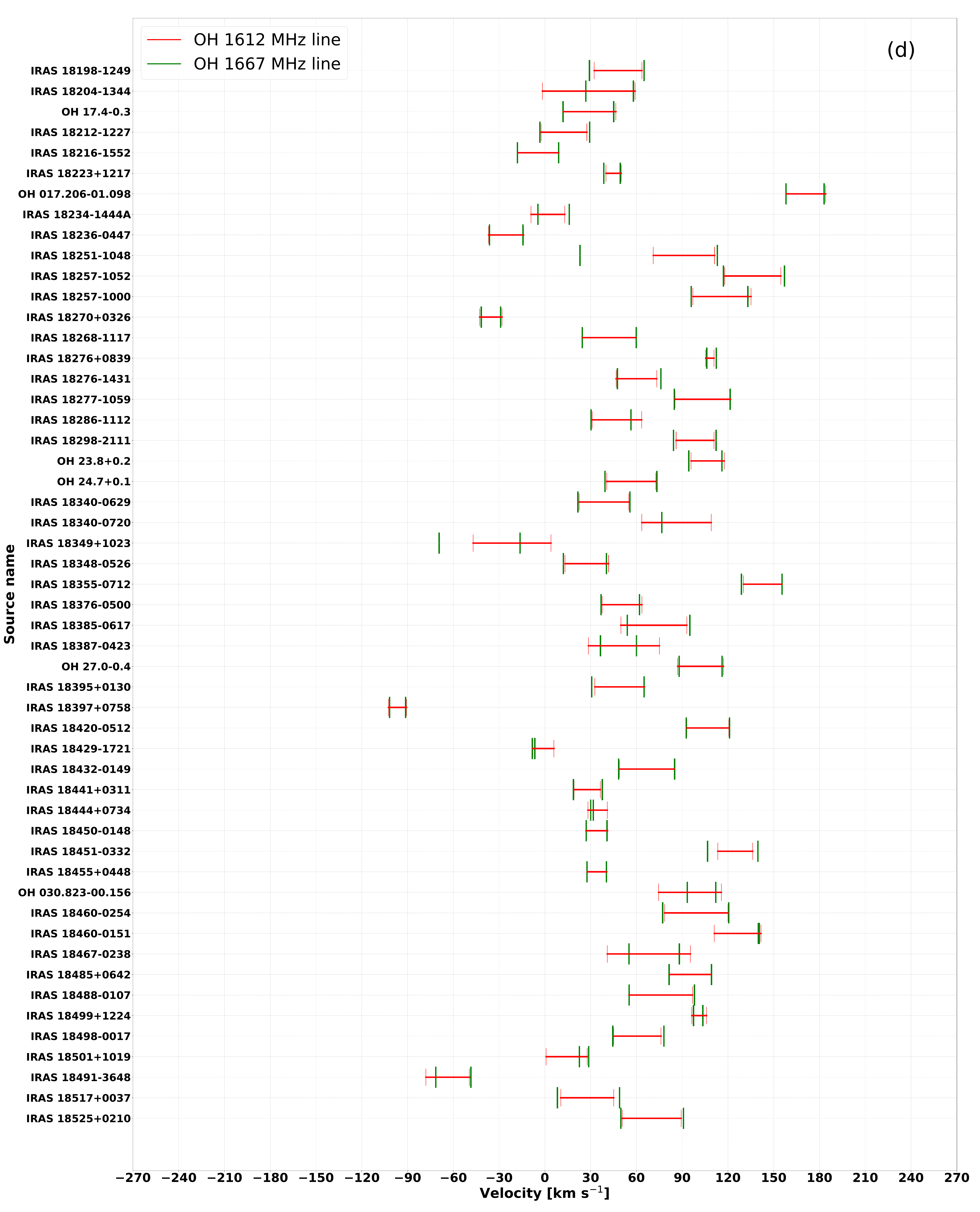}
   \caption{Continuation of Fig.~\ref{Fig: c[comparison of 1612 OH maser with that of 1667 MHz-3]}. The notation is the same as in Fig.\ref{Fig: a[comparison of 1612 OH maser with that of 1667 MHz]}.}\label{Fig: d[comparison of 1612 OH maser with that of 1667 MHz-4]}
\end{figure}
\begin{figure}
    \centering
  	\includegraphics[width=0.8\textwidth, angle=0]{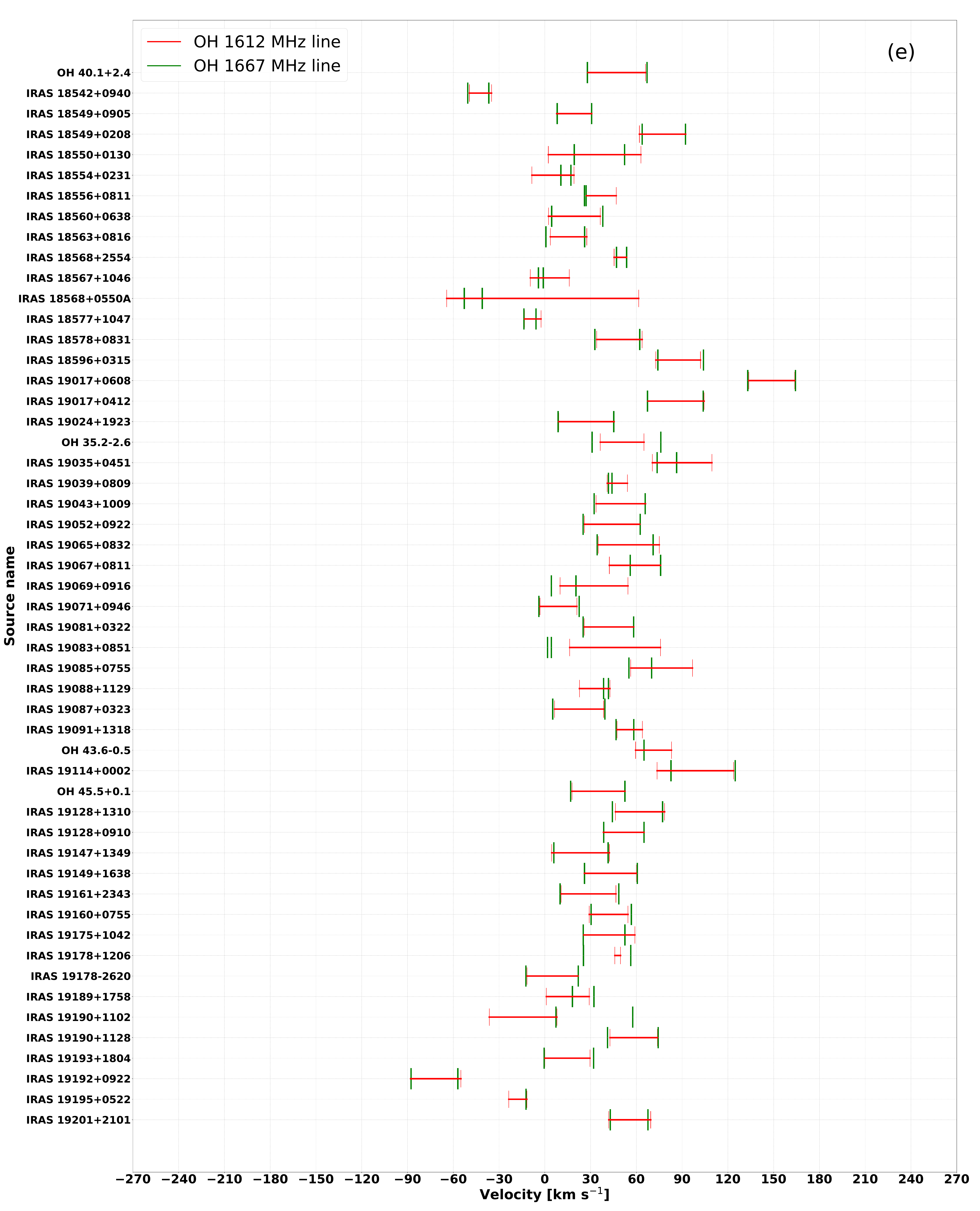}
   \caption{Continuation of Fig.~\ref{Fig: d[comparison of 1612 OH maser with that of 1667 MHz-4]}. The notation is the same as in Fig.\ref{Fig: a[comparison of 1612 OH maser with that of 1667 MHz]}.}\label{Fig: e[comparison of 1612 OH maser with that of 1667 MHz-5]}
\end{figure}
\begin{figure}
    \centering
  	\includegraphics[width=0.8\textwidth, angle=0]{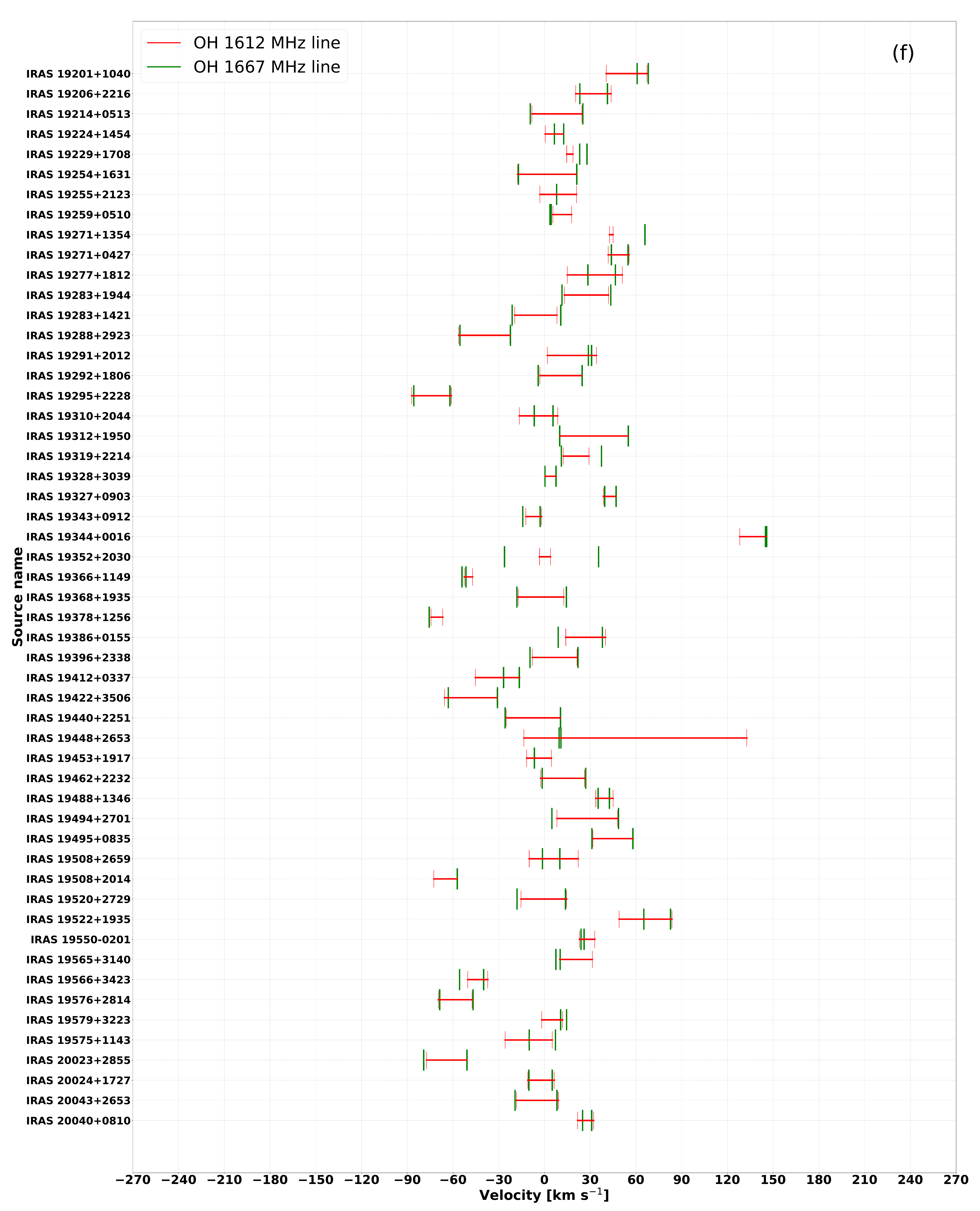}
   \caption{Continuation of Fig.~\ref{Fig: e[comparison of 1612 OH maser with that of 1667 MHz-5]}. The notation is the same as in Fig.\ref{Fig: a[comparison of 1612 OH maser with that of 1667 MHz]}.}\label{Fig: f[comparison of 1612 OH maser with that of 1667 MHz-6]}
\end{figure}
\begin{figure}
    \centering
  	\includegraphics[width=0.8\textwidth, angle=0]{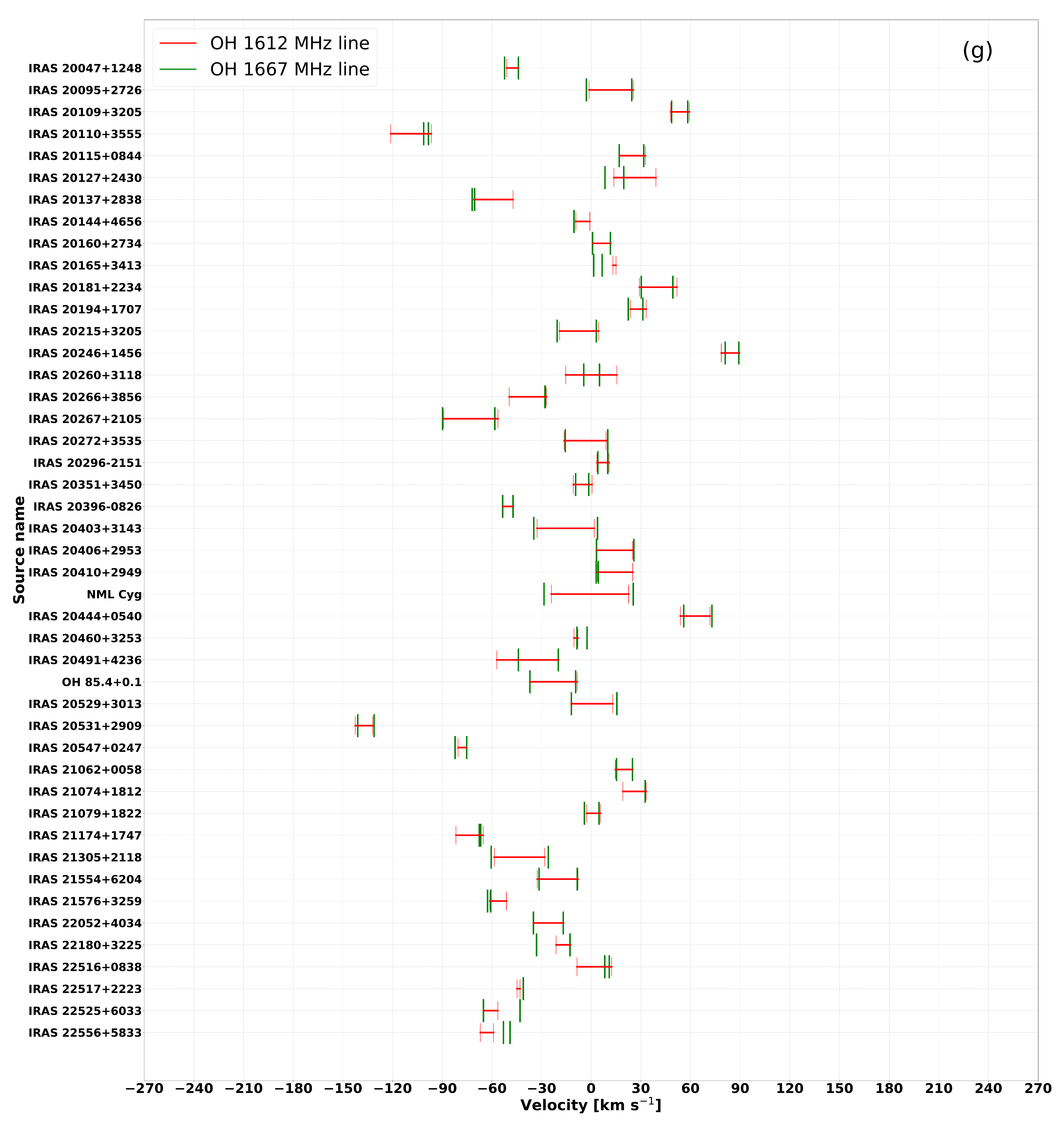}
   \caption{Continuation of Fig.~\ref{Fig: f[comparison of 1612 OH maser with that of 1667 MHz-6]}. The notation is the same as in Fig.\ref{Fig: a[comparison of 1612 OH maser with that of 1667 MHz]}.}\label{Fig: g[comparison of 1612 OH maser with that of 1667 MHz-7]}
\end{figure}

\bibliography{ms}{}
\bibliographystyle{aasjournal}

\end{CJK*}
\end{document}

%% file: WISE_Position_information_of_the_selected_sources.tex
\begin{deluxetable}{cccccr} 
		\caption{WISE coordinates of the OH maser sources used in the analysis.}\label{Tab: [WISE position of the selected sources]}
\tablehead{ \multirow{2}{*}{Source name} & \multicolumn{2}{c}{Catalog position}   &\multicolumn{2}{c}{WISE position} &\multirow{2}{*}{angDist}\\
			& R.A. & Dec.& R.A. & Dec.& \\
			& (J2000) &(J2000)  &(J2000) &(J2000) &arcsec }
\startdata
			IRAS 00428$+$6854 & 00 46 00.1 & $+$69 10 53.4&00 46 00.2 &$+$69 10 53.9&0.54 \\
   			IRAS 01037$+$1219&01 06 26.0 &$+$12 35 53.0&01 06 26.0&$+$12 35 53.2&0.39\\
   			IRAS 01085$+$3022 &01 11 15.9 & $+$30 38 06.0&01 11 15.9&$+$30 38 06.4 &0.78\\
			IRAS 01304$+$6211 &01 33 50.5 &$+$62 26 47.0&01 33 51.2&$+$62 26 53.4 &8.02\\
   			IRAS 02192$+$5821 & 02 22 51.1 & $+$58 35 08.0&02 22 51.7 &$+$58 35 11.5&5.93 \\
\enddata
\tablecomments{This table shows the first five rows as a sample. The full dataset is available in the electronic table attached to this paper. ``angDist'': The angular separation between the locations of the circumstellar OH maser sources in the database and their corresponding WISE counterparts.}
\end{deluxetable}

%% file: Velocity_information_of_the_selected_sources.tex
	
\begin{deluxetable}{crrrrrrrrr} 
		\caption{Velocity information of the OH maser sources used in the analysis.}\label{Tab: [Velocity information of the selected the OH maser sources with WISE counterparts]}
\tablehead{	   \multirow{2}{*}{Source name} & \multicolumn{3}{c}{1612 MHz} & \multicolumn{3}{c}{1665 MHz} & \multicolumn{3}{c}{1667 MHz}\\
			\cline{2-4} \cline{5-7}\cline{8-10}
    & \multicolumn{1}{c}{$V_{\rm b}$} & \multicolumn{1}{c}{$V_{\rm r}$} & \multicolumn{1}{c}{$\Delta V_{1612}$} & \multicolumn{1}{c}{$V_{\rm b}$} & \multicolumn{1}{c}{$V_{\rm r}$} & \multicolumn{1}{c}{$\Delta V_{1665}$} & \multicolumn{1}{c}{$V_{\rm b}$} & \multicolumn{1}{c}{$V_{\rm r}$} & \multicolumn{1}{c}{$\Delta V_{1667}$} \\
    & \multicolumn{1}{c}{(km s$^{-1}$)} & \multicolumn{1}{c}{(km s$^{-1}$)} & \multicolumn{1}{c}{(km s$^{-1}$)} & \multicolumn{1}{c}{(km s$^{-1}$)} & \multicolumn{1}{c}{(km s$^{-1}$)} & \multicolumn{1}{c}{(km s$^{-1}$)} & \multicolumn{1}{c}{(km s$^{-1}$)} & \multicolumn{1}{c}{(km s$^{-1}$)} & \multicolumn{1}{c}{(km s$^{-1}$)}  }
\startdata
			IRAS 00428$+$6854  &$-$38.0 &$-$13.0  &25.0&$-$33.2&$-$16.9 &16.3&$-$33.2&$-$24.2  &9.0\\
   			IRAS 01037$+$1219 &$-$9.9&27.0&36.9 &$-$10.1&12.8&22.9  &$-$10.0&26.0  &36.0\\
   			IRAS 01085$+$3022 &$-$39.8&$-$14.0 &25.8 &&  &&$-$42.2&$-$13.7  &28.5\\
			IRAS 01304$+$6211 &$-$66.0&$-$43.0  &23.0&$-$66.0&$-$43.1 &22.9&$-$66.0&$-$43.9 &22.1\\
   			IRAS 02192$+$5821  &$-$53.0&$-$23.0  &30.0&$-$52.0&$-$26.0&26.0  &$-$52.0&$-$21.0  &31.0\\
\enddata
\tablecomments{This table shows the first five rows as a sample. The full dataset is available in the electronic table attached to this paper. $V_{\rm b}$ and $V_{\rm r}$ represent the peak velocities on the blueshifted and redshifted sides provided by each database, respectively. $\Delta V_{1612}$, $\Delta V_{1665}$, and $\Delta V_{1667}$ represent the values of $V_{\rm r} - V_{\rm b}$.}
\end{deluxetable}

%% file: WISE_and_IRAS_and_2MASS_information_of_the_selected_sources.tex
 \begin{deluxetable}{crrrrcrrrcrrrc} 
			\caption{Infrared photometric data of OH maser sources used in the analysis.}\label{Tab: [WISE, IRAS and 2MASSS information of the selected sources]}
\tablehead{		\multirow{2}{*}{Source name}  & \multicolumn{4}{c}{AllWISE} &\multirow{2}{*}{qph} & \multicolumn{3}{c}{IRAS}&\multirow{2}{*}{QF} & \multicolumn{3}{c}{2MASS}&\multirow{2}{*}{QFl}\\
				\cline{2-5} \cline{7-9} \cline{11-13}
      & \multicolumn{1}{c}{W1} & \multicolumn{1}{c}{W2} & \multicolumn{1}{c}{W3} & \multicolumn{1}{c}{W4} & & \multicolumn{1}{c}{$F_{\rm 12}$} & \multicolumn{1}{c}{$F_{\rm 25}$} & \multicolumn{1}{c}{$F_{\rm 60}$} & & \multicolumn{1}{c}{J} & \multicolumn{1}{c}{H} & \multicolumn{1}{c}{K} & \\
    & \multicolumn{1}{c}{mag} & \multicolumn{1}{c}{mag} & \multicolumn{1}{c}{mag} & \multicolumn{1}{c}{mag} & & \multicolumn{1}{c}{Jy} & \multicolumn{1}{c}{Jy} & \multicolumn{1}{c}{Jy} & & \multicolumn{1}{c}{mag} & \multicolumn{1}{c}{mag} & \multicolumn{1}{c}{mag} &}
\startdata
				IRAS 00428$+$6854  &$-$0.283 & $-$0.776 &$-$0.756 &$-$2.147&UUBA &71.1&50.9&9.9&333 &4.764&3.251&2.180&DDD\\
    			IRAS 01037$+$1219 &$-$1.548&1.325&$-$2.606&$-$3.477&UUUA&1160.0&968.0&215.0&333 &7.437&4.641&2.217&ACD\\
       			IRAS 01085$+$3022 &0.123 &$-$0.249 &$-$1.27 &$-$2.652 &UUCA&165.0&122.0&18.9&333 &4.710&3.356&2.293&DDD\\
				IRAS 01304$+$6211 &3.829 &$-$0.393 &$-$1.031 &$-$2.948 &BUBA&289.0&456.0&194.0&333 &16.747&11.232&7.097&CAA\\
				IRAS 02192$+$5821 & $-$2.232 & $-$2.526 &$-$1.995 &$-$3.044 &UUCA&339.0&233.0&40.6&333 &2.949&1.845&1.123&DDD\\
\enddata
\tablecomments{This table shows the first five rows as a sample. The full dataset is available in the electronic table attached to this paper. ``qph'': The photometric quality flag for the AllWISE fluxes. The four characters correspond to the W1, W2, W3, and W4 bands, in that order. A, B, C, and D represent different quality levels from high to low. U indicates an upper limit on magnitude, and X means that a profile-fit measurement was not possible for this band at this location. In such cases, the magnitude value will be "null" for this band. ``QF'': The quality flag for the IRAS flux density. The three characters correspond to the 12, 25, and 60 micron bands. A value of 3 indicates high quality, 2 represents moderate quality, and 1 represents an upper limit. ``QFI'': The photometric quality flag for the 2MASS fluxes. The meaning of the notation is the same as for qph.}
\end{deluxetable}

%% file: Sources_excluded_from_further_analysis.tex
\begin{deluxetable}{ cccrrrrrr} 
		\caption{OH maser sources excluded from the analysis.}\label{Tab: [Sources without the WISE counterparts]}
\tablehead{	\multirow{2}{*}{Source name} & \multicolumn{2}{c}{Catalog position} & \multicolumn{2}{c}{1612 MHz} & \multicolumn{2}{c}{1665 MHz} & \multicolumn{2}{c}{1667 MHz} \\
			\cline{4-5} \cline{6-7} \cline{8-9}
  & \multicolumn{1}{c}{R.A.} & \multicolumn{1}{c}{Dec.} & \multicolumn{1}{c}{$V_{\rm b}$} & \multicolumn{1}{c}{$V_{\rm r}$} & \multicolumn{1}{c}{$V_{\rm b}$} & \multicolumn{1}{c}{$V_{\rm r}$} & \multicolumn{1}{c}{$V_{\rm b}$} & \multicolumn{1}{c}{$V_{\rm r}$} \\
    & \multicolumn{1}{c}{(J2000)} & \multicolumn{1}{c}{(J2000)} & \multicolumn{1}{c}{(km s$^{-1}$)} & \multicolumn{1}{c}{(km s$^{-1}$)} & \multicolumn{1}{c}{(km s$^{-1}$)} & \multicolumn{1}{c}{(km s$^{-1}$)} & \multicolumn{1}{c}{(km s$^{-1}$)} & \multicolumn{1}{c}{(km s$^{-1}$)} }
\startdata
		IRAS 05506+2414$^{\dagger}$ & 05 53 43.6 & $+$24 14 46.0 & 0.5 & 10.1 & 1.0 &9.8&$-$7.8&$-$5.7 \\
  		IRAS 19059$-$2219$^{\ast}$ & 19 08 54.6 & $-$22 14 20.0 & 8.0 &35.0 & &&8.0&36.0\\
            IRAS 19244$+$1115$^{\ast}$ & 19 26 47.6 & $+$11 21 15.0 & 45.4 &107.1 & 46.6&99.3&40.5&108.5\\
		IRAS 20077$-$0625$^{\ast}$ & 20 10 27.4 & $-$06 16 16.0 & $-$32.0 &$-$7.2 & &&$-$31.0&$-$2.9 \\
		IRAS 22177$+$5936$^{\ast}$ & 22 19 27.8 & $+$59 51 22.0 & $-$40.5 &$-$9.8 & $-$41.2&$-$11.4&$-$41.4&$-$9.0 \\
\enddata

\tablecomments{Objects marked with $^{\dagger}$ are YSO candidates, while those marked with $^{\ast}$ have counterparts in WISE images but lack data in the PSC.}
\end{deluxetable}

%% file: Eye_inspection_objects_velocity_excess.tex


			


    

\begin{deluxetable}{ cccccc} 
\caption{List of circumstellar OH maser sources with confirmed velocity excess.}\label{Tab: [Sources showing true velocity deviations in the OH 1665/1667 MHz line]}
\tablehead{	\multirow{3}{*}{Source name}  & \multicolumn{1}{c}{Spectral type} & \multicolumn{4}{c}{Velocity deviation} \\
\cline{3-6}
&(1612 MHz)& \multicolumn{2}{c}{Blue-shifted side} & \multicolumn{2}{c}{Red-shifted side}  \\
&& 1665 MHz & 1667 MHz & 1665 MHz & 1667 MHz  \\
&& ($\rm km~s^{-1}$) & ($\rm km~s^{-1}$) & ($\rm km~s^{-1}$) & ($\rm km~s^{-1}$)  }
\startdata
IRAS 15405$-$4945 &Irr& 26.0 & 53.0 &--&--\\
IRAS 17579$-$3121 &$\rm D_{+}$&--&12.0 &--& --\\
IRAS 18251$-$1048 &D&--&50.0&--& 5.0 \\
IRAS 18498$-$0017 &D&--&--&7.0 &--\\
IRAS 19069$+$0916 &D&--&8.0 &--&--\\
IRAS 19103$+$0913 &D&18.0 &--&--&--\\
IRAS 19190$+$1102 &$\rm D_{+}$&--&--&--&49.7 \\
IRAS 19352$+$2030 &Irr& --&--&--&31.0\\
\enddata
\tablecomments{Spectral type:``D'' denotes a typical double-peaked line profile. ``$\rm D_{+}$'' refers to a typical double-peak with some minor additional features. ``Irr'' indicates irregular profiles, excluding those classified as ``$\rm D_{+}$''.}
\end{deluxetable}

%% file: Detection_rate_of_vel_excess_sources.tex
	

\begin{deluxetable}{lcccccc} 
	    \caption{Detection rate of velocity excess}\label{Tab: [Detection rate of velocity excess sources]}
	
	\tablehead{Line Combination &  \multicolumn{2}{c}{Number of sources}  &  \multicolumn{2}{c}{Velocity excess} &  \multicolumn{2}{c}{Detectin rate}\\						
		& AGB & post-AGB & AGB & post-AGB & in AGB &in Post-AGB }	
	\startdata
        1612--1665--1667 & 82 & 75 & 0 & 1 & 0\% & 1.3\% \\
        1612--1665 & 3 & 8 & 1 & 1 & 33.3\%  & 12.5\%  \\
        1612--1667 & 37 & 106 & 0 & 4 & 0\%  & 3.8\%  \\
        1612--H$_2$O & 135 & 58 & 6 & 5 & 4.4\%  & 8.6\%  \\
	\enddata
\end{deluxetable}

%% file: Known_WFs_RNRs_infrared_photometric_data_wise_iras_2mass.tex
\begin{deluxetable}{ccrrrrcrrrcrrrc} 
\caption{Infrared photometric data of the known WFs and RNRs.}\label{Tab: [WISE, IRAS and 2MASS Infrared photometric data of the known WFs and RNRs]}
\tablehead{	\multirow{2}{*}{Source name} &\multirow{2}{*}{Class} & \multicolumn{4}{c}{AllWISE} &\multirow{2}{*}{qph} & \multicolumn{3}{c}{IRAS}&\multirow{2}{*}{QF} & \multicolumn{3}{c}{2MASS}&\multirow{2}{*}{QFl}\\
		\cline{3-6} \cline{8-10} \cline{12-14}
	     & & \multicolumn{1}{c}{W1} & \multicolumn{1}{c}{W2} & \multicolumn{1}{c}{W3} & \multicolumn{1}{c}{W4} & & \multicolumn{1}{c}{$F_{\rm 12}$} & \multicolumn{1}{c}{$F_{\rm 25}$} & \multicolumn{1}{c}{$F_{\rm 60}$} & & \multicolumn{1}{c}{J} & \multicolumn{1}{c}{H} & \multicolumn{1}{c}{K} &\\
    & &\multicolumn{1}{c}{mag} & \multicolumn{1}{c}{mag} & \multicolumn{1}{c}{mag} & \multicolumn{1}{c}{mag} & & \multicolumn{1}{c}{Jy} & \multicolumn{1}{c}{Jy} & \multicolumn{1}{c}{Jy} & & \multicolumn{1}{c}{mag} & \multicolumn{1}{c}{mag} & \multicolumn{1}{c}{mag} & }
\startdata
	IRAS 15103$-$5754 &WF &8.287&8.617&5.976&0.842&ABBA&10.8&102.0&126.0&333 &12.743&11.661&11.195&AAA\\
	IRAS 15445$-$5449 &WF  &7.977&8.619&5.545&$-$0.320&BAAA&6.9&87.2&1130.0&231&16.439&14.396&13.459&CAE\\
	IRAS 15544$-$5332 &WF  &5.912&4.142&2.267&$-$0.318&ABAA&4.6&15.5&41.5&333 &14.087&10.522&8.086& AAA\\
    IRAS 16342$-$3814 &WF  &7.195 & 7.607 & & $-$0.461 &AAXA & 16.2 & 200.0 & 290.0 & 333 & 16.104 & 15.528 & 15.558 & ABU\\
    IRAS 16552$-$3050 &WF &15.09&13.038&3.254&0.042&AAAA&2.5&10.5&9.6&333 &16.402&15.359&14.770&BUU\\
    IRAS 17291$-$2147 &WF &9.403&7.864&3.383&$-$0.051&AAAA&2.5&12.2&8.2&333 &13.507&12.188&11.008&AAA\\
    IRAS 18043$-$2116 &WF & 12.312 & 9.567 & 3.549 &0.447& BAAA & 6.6 & 6.8 & 16.6& 133 & 14.546 & 13.404 & 13.042& AAA\\
    IRAS 18113$-$2503 &WF &11.205&11.530&7.758&2.177&AAAA&2.9&14.8&12.9&333 &13.024&12.229&11.901&AAA\\
    IRAS 18139$-$1816 &WF & 6.599 &4.034&0.661&$-$0.802&ABAA&11.6& 16.9& 13.9& 333 &17.041& 15.725& 11.639& UUA\\
    IRAS 18286$-$0959 &WF &9.283&9.090&5.676&0.319&AABA&24.9&24.5&18.4&231 &12.843&10.784&9.947&AAA\\
    IRAS 18450$-$0148 &WF &13.879&10.598&0.451&$-$2.574&BACA&23.7&104.0&295.0&331 &16.361&14.988&14.120&BBA\\
    IRAS 18455+0448 &WF &7.799&5.167&1.135&$-$0.604&AAAA&9.4&12.6&5.5&333 &15.068&13.294&11.305&AAA\\
    IRAS 18460$-$0151 &WF &11.488&6.932&0.707&$-$1.080&AAAA&20.9&32.7&277.0&331 &13.806&11.778&10.790&AAA\\
    IRAS 18596+0315 &WF &11.429&8.659&2.559&$-$0.380&AAAA&2.6&14.2&22.6&333 &16.501&14.945&14.311&UAU\\
    IRAS 19134+2131 &WF &11.132&9.089&1.887&$-$0.483&AAAA&5.1&15.6&8.6&333 &16.543&14.926&13.464&BAA\\
    IRAS 19190+1102 &WF &13.638&14.961&8.237&2.704&BCBA&1.6&13.7&24.5&333 &15.997&15.238&15.068&AAB\\
    V4332 Sgr &RNR &7.304&5.609&4.507&1.426&AAAA&&&& &12.100&11.605&10.992&AAA\\
    V838 Mon &RNR &4.676&3.711&$-$0.134&$-$1.711&BBAA&0.25&0.25&1.4&113 &13.873&13.510&13.333&AAA\\
    OGLE-2002-BLG-360 &RNR &6.220&4.228&1.944&0.937&ABAA&&&& &14.072&12.645&11.251&EAA\\
    V1309 Sco &RNR &8.801&6.037&2.730&1.146&AAAA&&&&&13.282&12.373&12.099&AAA\\
    CK Vul &RNR &13.413&13.294&11.618&6.752&AAUA&&&&&15.462&14.506&14.173&AAA\\
\enddata
\tablecomments{For the quality flags of the photometric data (qph, QH, QFl), refer to the footnote of Table~\ref{Tab: [WISE, IRAS and 2MASSS information of the selected sources]}.}
\end{deluxetable}

%% file: OH_maser_status_of_the_known_WF_samples.tex
\begin{deluxetable}{ccrrrrrr} 
    \caption{OH maser status of the known WF samples.}\label{Tab: [OH maser status of the known WF samples]}
    \tablehead{
        \multirow{2}{*}{Source name} & \multicolumn{1}{c}{Spectral type} & \multicolumn{2}{c}{1612 MHz} & \multicolumn{2}{c}{1665 MHz} & \multicolumn{2}{c}{1667 MHz} \\
        \cline{3-4} \cline{5-6} \cline{7-8}
        &(1612 MHz) & \multicolumn{1}{c}{$V_{\rm b}$} & \multicolumn{1}{c}{$V_{\rm r}$} & \multicolumn{1}{c}{$V_{\rm b}$} & \multicolumn{1}{c}{$V_{\rm r}$} & \multicolumn{1}{c}{$V_{\rm b}$} & \multicolumn{1}{c}{$V_{\rm r}$} \\
        & & \multicolumn{1}{c}{(km s$^{-1}$)} & \multicolumn{1}{c}{(km s$^{-1}$)} & \multicolumn{1}{c}{(km s$^{-1}$)} & \multicolumn{1}{c}{(km s$^{-1}$)} & \multicolumn{1}{c}{(km s$^{-1}$)} & \multicolumn{1}{c}{(km s$^{-1}$) }
    }
    \startdata
    IRAS 15103$-$5754 & S & $-$47.0 & & & & & \\
    IRAS 15445$-$5449 & Irr & $-$140.2  & & $-$139.0 &  &  &   \\
    IRAS 15544$-$5332 & Irr & $-$118.3 & $-$104.8 & $-$107.2 & & $-$123.2 &    \\
    IRAS 16342$-$3814 & Irr & $-$17.0 & 108.0 & 17.0 & 81.3 & $-$12.0 & 90.0  \\
    IRAS 16552$-$3050 &  &  &  &  &  &  &  \\
    IRAS 17291$-$2147 &  &  &  &  &  &  &  \\
    IRAS 18043$-$2116 & Irr & 70.4 & 102.7 & 69.7 & 104.0 &  &   \\
    IRAS 18113$-$2503 &  \\
    IRAS 18139$-$1816 & D & $-$68.3 & $-$43.1 & $-$63.0 & $-$44.0& $-$66.7 & $-$44.0   \\
    IRAS 18286$-$0959 & Irr  & $-$0.5 & 72.6 & 20.5 &  &  & \\
    IRAS 18450$-$0148 &  D  & 27.0 & 41.0 & 27.4 & 40.5 & 27.1 & 40.7 \\
    IRAS 18455$+$0448 & D & 27.8 & 40.5 & 27.6 & 40.4 &27.6  &40.3   \\
    IRAS 18460$-$0151 & D & 111.0 & 141.6 & 110.3 & 139.6 & 139.9 & 140.5 \\
    IRAS 18596$+$0315 & D & 72.7 & 102.0 & 74.6 & 101.5 & 74.1 & 104.0  \\
    IRAS 19134$+$2131 &  &  &  &  &  &  &   \\
    IRAS 19190$+$1102 & D+ & $-$36.4 & 7.9 & &&7.3 & 57.6   \\
    \enddata
    \tablecomments{Spectral type: “S” denotes a single-peaked line profile, and other notations are the same as Table~\ref{Tab: [Sources showing true velocity deviations in the OH 1665/1667 MHz line]}. For details, refer to the footnotes of Table~\ref{Tab: [Sources showing true velocity deviations in the OH 1665/1667 MHz line]} and the main text. The parameters of the OH maser lines were obtained from the references listed in Table 5 of  \citet{2024ApJS..270...13F}. For IRAS 18455$+$0448, two peaks were observed in 1988, but the peak at 40.5 km s$^{-1}$ disappeared by 2000. For IRAS 18460$-$0151, the observed peak at 141.6 km s$^{-1}$ was notably weak. }
\end{deluxetable}